\documentclass[arps,prd,onecolumn,noshowpacs,nofootinbib]{revtex4}

\usepackage{epsfig}
\input{epsf.sty}
\usepackage{color}
\usepackage{graphicx}
\usepackage[large]{subfigure}
\usepackage[amssymb]{SIunits}
\usepackage{epstopdf}
\usepackage{url}
\usepackage{aas_macros}
\usepackage{natbib}

\newcommand\beq{\begin{equation}}
\newcommand\eeq{\end{equation}}
\newcommand\beqn{\begin{eqnarray}}
\newcommand\eeqn{\end{eqnarray}}
\newcommand\be{\begin{eqnarray}}
\newcommand\ee{\end{eqnarray}}

\newcommand\lsim{\mathrel{\rlap{\lower4pt\hbox{\hskip1pt$\sim$}}
        \raise1pt\hbox{$<$}}}
\newcommand\gsim{\mathrel{\rlap{\lower4pt\hbox{\hskip1pt$\sim$}}
        \raise1pt\hbox{$>$}}}

\newcommand\fnl{$f_{\mathrm{NL}}\,\,$}

\newcommand\Mnu{$M_{\nu}\,\,$}

\newcommand{\ex}[1]{\langle #1 \rangle}
\newcommand{\aSZ}{\hat a^{SZ}}

\input{epsf}

\begin{document}
\title{Cosmology from the Thermal Sunyaev-Zel'dovich Power Spectrum: Primordial non-Gaussianity and Massive Neutrinos}
\author{J.\ Colin Hill$^{1}$ and Enrico Pajer$^{2}$}
\affiliation{$^{1}$Department of Astrophysical Sciences, Princeton
  University, Princeton, NJ 08544 \\
$^{2}$Department of Physics, Princeton University, Princeton, NJ 08544\\
jch@astro.princeton.edu, epajer@princeton.edu}
\date{\today}

\begin{abstract}
We carry out a comprehensive analysis of the possible constraints on cosmological and astrophysical parameters achievable with measurements of the thermal Sunyaev-Zel'dovich (tSZ) power spectrum from upcoming full-sky CMB observations, with a particular focus on one-parameter extensions to the $\Lambda$CDM standard model involving local primordial non-Gaussianity (described by \fnl) and massive neutrinos (described by \Mnu).  We include all of the relevant physical effects due to these additional parameters, including the change to the halo mass function and the scale-dependent halo bias induced by local primordial non-Gaussianity.  We use the halo model to compute the tSZ power spectrum and provide a pedagogical derivation of the one- and two-halo terms in an appendix.  We model the pressure profile of the intracluster medium (ICM) using a parametrized fit that agrees well with existing observations, and include uncertainty in the ICM modeling by including the overall normalization and outer logarithmic slope of the profile as free parameters.  We compute forecasts for Planck, PIXIE, and a cosmic variance (CV)-limited experiment, using multifrequency subtraction to remove foregrounds and implementing two masking criteria based on the ROSAT and eROSITA cluster catalogs to reduce the significant CV errors at low multipoles.  We find that Planck can detect the tSZ power spectrum with $>30\sigma$ significance, regardless of the masking scenario.  However, neither Planck or PIXIE is likely to provide competitive constraints on \fnl from the tSZ power spectrum due to CV noise at low-$\ell$ overwhelming the unique signature of the scale-dependent bias.  A future CV-limited experiment could provide a $3\sigma$ detection of \fnl$\simeq 37$, which is the WMAP9 maximum-likelihood result.  The outlook for neutrino masses is more optimistic: Planck can reach levels comparable to the current upper bounds $\lesssim 0.3$ eV with conservative assumptions about the ICM; stronger ICM priors could allow Planck to provide $1-2\sigma$ evidence for massive neutrinos from the tSZ power spectrum, depending on the true value of the sum of the neutrino masses.  We also forecast a $<10$\% constraint on the outer slope of the ICM pressure profile using the unmasked Planck tSZ power spectrum.
\end{abstract}

\maketitle
\section{Introduction}
\label{sec:intro}

The thermal Sunyaev-Zel'dovich (tSZ) effect is a spectral distortion of the cosmic microwave background (CMB) that arises due to the inverse Compton scattering of CMB photons off hot electrons that lie between our vantage point and the surface of last scattering~\cite{Sunyaev-Zeldovich1970}.  The vast majority of these hot electrons are located in the intracluster medium (ICM) of galaxy clusters, and thus the tSZ signal is dominated by contributions from these massive objects.  The tSZ effect has been used for many years to study individual clusters in pointed observations (e.g., \cite{Reeseetal2011,Plaggeetal2012,Lancasteretal2011,AMIetal2011}) and in recent years has been used as a method with which to find and characterize massive clusters in blind millimeter-wave surveys~\cite{Marriageetal2011,Williamsonetal2011,Plancketal2011,Hasselfieldetal2013, Reichardtetal2013}.  Moreover, recent years have brought the first detections of the angular power spectrum of the tSZ effect through its contribution to the power spectrum in arcminute-resolution maps of the microwave sky made by the Atacama Cosmology Telescope (ACT)\footnote{{\tt http://www.princeton.edu/act/ }} and the South Pole Telescope (SPT)\footnote{{\tt http://pole.uchicago.edu/ }}~\cite{Dunkleyetal2011,Sieversetal2013, Reichardtetal2012,Storyetal2012}.  In addition, three-point statistics of the tSZ signal have been detected within the past year: first, the real-space skewness was detected in ACT data~\cite{Wilsonetal2012} using methods first anticipated by~\cite{Rubino-Martin-Sunyaev2003}, and second, the Fourier-space bispectrum was very recently detected in SPT data~\cite{Crawfordetal2013}.  The amplitudes of these measurements were shown to be consistent in the SPT analysis, despite observing different regions of sky and using different analysis methods.  Note that the tSZ signal is highly non-Gaussian since it is dominated by contributions from massive collapsed objects in the late-time density field; thus, higher-order tSZ statistics contain significant information beyond that found in the power spectrum.  Furthermore, the combination of multiple different $N$-point tSZ statistics provides an avenue to extract tighter constraints on cosmological parameters and the astrophysics of the ICM than the use of the power spectrum alone, through the breaking of degeneracies between ICM and cosmological parameters~\cite{Hill-Sherwin2013,Bhattacharyaetal2012}.  The recent SPT bispectrum detection used such methods in order to reduce the error bar on the tSZ power spectrum amplitude by a factor of two~\cite{Crawfordetal2013}.

Thus far, tSZ power spectrum detections have been limited to measurements or constraints on the power at a single multipole (typically $\ell = 3000$) because ACT and SPT do not have sufficient frequency coverage to fully separate the tSZ signal from other components in the microwave sky using its unique spectral signature.  However, this situation will shortly change with the imminent release of full-sky temperature maps from the Planck satellite\footnote{{\tt http://planck.esa.int}}.  Planck has nine frequency channels that span the spectral region around the tSZ null frequency $\approx 218$ GHz.  Thus, it should be possible to separate the tSZ signal from other components in the sky maps to high accuracy, allowing for a measurement of the tSZ power spectrum over a wide range of multipoles, possibly $\sim 100 \lesssim \ell \lesssim 1500$, as we demonstrate in this paper.  The proposed Primordial Inflation Explorer (PIXIE) experiment~\cite{Kogutetal2011} will also be able to detect the tSZ power spectrum at high significance, as its wide spectral coverage and high spectral resolution will allow for very accurate extraction of the tSZ signal.  However, its angular resolution is much lower than Planck's, and thus the tSZ power spectrum will be measured over a much smaller range of multipoles ($\ell \lesssim 200$).  However, the PIXIE data (after masking using X-ray cluster catalogs --- see below) will permit tSZ measurements on large angular scales ($\ell \lesssim 100 $) that are essentially inaccessible to Planck due to its noise levels; these multipoles are precisely where one would expect the signature of scale-dependent bias induced by primordial non-Gaussianity to arise in the tSZ power spectrum.  Assessing the amplitude and detectability of this signature is a primary motivation for this paper.

The tSZ power spectrum has been suggested as a potential cosmological probe by a number of authors over the past few decades (e.g., \cite{Cole-Kaiser1988,Komatsu-Kitayama1999,Komatsu-Seljak2002}).  Nearly all studies in the last decade have focused on the small-scale tSZ power spectrum ($\ell \gtrsim 1000$) due to its role as a foreground in high angular resolution CMB measurements, and likely because without multi-frequency information, the tSZ signal has only been able to be isolated by looking for its effects on small scales (e.g., using an $\ell$-space filter to upweight tSZ-dominated small angular scales in CMB maps).  Much of this work was driven by the realization that the tSZ power spectrum is a very sensitive probe of the amplitude of matter density fluctuations, $\sigma_8$~\cite{Komatsu-Seljak2002}.  The advent of multi-frequency data promises measurements of the large-scale tSZ power spectrum very shortly, and thus we believe it is timely to reassess its value as a cosmological probe, including parameters beyond $\sigma_8$ and including a realistic treatment of the uncertainties due to modeling of the ICM.  We build on the work of~\cite{Komatsu-Kitayama1999} to compute the full angular power spectrum of the tSZ effect, including both the one- and two-halo terms, and moving beyond the Limber/flat-sky approximations where necessary.

Our primary interest is in assessing constraints from the tSZ power spectrum on currently unknown parameters beyond the $\Lambda$CDM standard model: the amplitude of local primordial non-Gaussianity, \fnl, and the sum of the neutrino masses, $M_{\nu} \equiv \sum m_{\nu}$.  The values of these parameters are currently unknown, and determining their values is a key goal of modern cosmology.

Primordial non-Gaussianity is one of the few known probes of the physics of inflation.  Models of single-field, minimally-coupled slow-roll inflation predict negligibly small deviations from Gaussianity in the initial curvature perturbations~\cite{Acquavivaetal2003,Maldacena2003}.  In particular, a detection of a non-zero bispectrum amplitude in the so-called ``squeezed'' limit ($k_1 \ll k_2, k_3$) would falsify essentially all single-field models of inflation~\cite{Maldacena2003,Creminelli-Zaldarriaga2004}.  This type of non-Gaussianity can be parametrized using the ``local'' model, in which \fnl describes the lowest-order deviation from Gaussianity~\cite{Salopek-Bond1990,Ganguietal1994,Komatsu-Spergel2001}:
\be
\label{eq.fNLlocmodel}
\Phi(\vec{x}) = \Phi_G(\vec{x}) + f_{\mathrm{NL}} \left( \Phi_G^2(\vec{x}) - \langle \Phi_G^2 \rangle \right) + \cdots \,,
\ee
where $\Phi$ is the primordial potential and $\Phi_G$ is a Gaussian field.  Note that $\Phi = \frac{3}{5} \zeta$, where $\zeta$ is the initial adiabatic curvature perturbation.  Local-type non-Gaussianity can be generated in multi-field inflationary scenarios, such as the curvaton model~\cite{PNG1,PNG2,PNG3}, or by non-inflationary models for the generation of perturbations, such as the new ekpyrotic/cyclic scenario~\cite{Ekp1,Ekp2,Ekp3}.  It is perhaps most interesting when viewed as a method with which to rule out single-field inflation, however.  Current constraints on \fnl are consistent with zero~\cite{Bennettetal2012,Giannantonioetal2013}, but the errors will shrink significantly very soon with the imminent CMB results from Planck.  We review the effects of \fnl$\neq 0$ on the large-scale structure of the universe in Section~\ref{sec:LSS}.

In contrast to primordial non-Gaussianity, massive neutrinos are certain to exist at a level that will be detectable within the next decade or so; neutrino oscillation experiments have precisely measured the differences between the squared masses of the three known species, leading to a lower bound of $\approx 0.05$ eV on the total summed mass~\cite{McKeown-Vogel2004}.  The remaining questions surround their detailed properties, especially their absolute mass scale.  The contribution of massive neutrinos to the total energy density of the universe today can be expressed as
\be
\label{eq.Omeganu}
\Omega_{\nu} \approx \frac{M_{\nu}}{93.14 \, h^2 \,\, \mathrm{eV}} \approx 0.0078 \frac{M_{\nu}}{0.1 \,\, \mathrm{eV}} \,,
\ee
where $M_{\nu}$ is the sum of the masses of the three known neutrino species.  Although this contribution appears to be small, massive neutrinos can have a significant influence on the small-scale matter power spectrum.  Due to their large thermal velocities, neutrinos free-stream out of gravitational potential wells on scales below their free-streaming scale~\cite{Lesgourgues-Pastor2006,Abazajianetal2011}.  This suppresses power on scales below the free-streaming scale.  Current upper bounds from various cosmological probes assuming a flat $\Lambda$CDM+\Mnu model are in the range \Mnu$\lesssim 0.3$ eV~\cite{Sieversetal2013,Vikhlininetal2009,Mantzetal2010b,dePutteretal2012}, although a $3\sigma$ detection near this mass scale was recently claimed in~\cite{Houetal2012}.  Should the true total mass turn out to be near $0.3$ eV, its effect on the tSZ power spectrum may be marginally detectable using the Planck data even for fairly conservative assumptions about the ICM physics, as we show in this paper.  With stronger ICM priors, Planck could achieve a $\sim 2-3\sigma$ detection for masses at this scale, using only the primordial CMB temperature power spectrum and the tSZ power spectrum.  We review the effects of massive neutrinos on the large-scale structure of the universe in Section~\ref{sec:LSS}.

In addition to the effects of both known and currently unknown cosmological parameters, we also model the effects of the physics of the ICM on the tSZ power spectrum.  This subject has attracted intense scrutiny in recent years after the early measurements of tSZ power from ACT~\cite{Dunkleyetal2011} and SPT~\cite{Luekeretal2010} were significantly lower than the values predicted from existing ICM pressure profile models~(e.g.,~\cite{Komatsu-Seljak2001}) in combination with WMAP5 cosmological parameters.  Subsequent ICM modeling efforts have ranged from fully analytic approaches~(e.g., \cite{Shawetal2010}) to cosmological hydrodynamics simulations~(e.g.,~\cite{Battagliaetal2010,Battagliaetal2012}), with other authors adopting semi-analytic approaches between these extremes~(e.g., \cite{Tracetal2011,Bodeetal2012}), in which dark matter-only $N$-body simulations are post-processed to include baryonic physics according to various prescriptions.  In addition, recent SZ and X-ray observations have continued to further constrain the ICM pressure profile from data, although these results are generally limited to fairly massive, nearby systems~(e.g.,~\cite{Arnaudetal2010,Plancketal2013}).  We choose to adopt a parametrized form of the ICM pressure profile known as the Generalized NFW (GNFW) profile, with our fiducial parameter values chosen to match the constrained pressure profile fit from hydrodynamical simulations in~\cite{Battagliaetal2012}.  In order to account for uncertainty in the ICM physics, we free two of the parameters in the pressure profile (the overall normalization and the outer logarithmic slope) and treat them as additional parameters in our model.  This approach is discussed in detail in Section~\ref{sec:ICM}.  We use this profile to compute the tSZ power spectrum following the halo model approach, for which we provide a complete derivation in Appendix~\ref{appendix}.

In addition to a model for the tSZ signal, we must compute the tSZ power spectrum covariance matrix in order to forecast parameter constraints.  There are two important issues that must be considered in computing the expected errors or signal-to-noise ratio (SNR) for a measurement of the tSZ power spectrum.  First, we must assess how well the tSZ signal can actually be separated from the other components in maps of the microwave sky, including the primordial CMB, thermal dust, point sources, and so on.  Following~\cite{THEO} and \cite{CHT}, we choose to implement a multi-frequency subtraction technique that takes advantage of the unique spectral signature of the tSZ effect, and also takes advantage of the current state of knowledge about the frequency- and multipole-dependence of the foregrounds.  Although there are other approaches to this problem, such as using internal linear combination techniques to construct a Compton-$y$ map from the individual frequency maps in a given experiment (e.g., \cite{Remazeillesetal2013, Remazeillesetal2011,Leachetal2008}), we find this method to be fairly simple and robust.  We describe these calculations in detail in Section~\ref{sec:exp}.

Second, we must account for the extreme cosmic variance induced in the large-angle tSZ power spectrum by massive clusters at low redshifts.  The one-halo term from these objects dominates the angular trispectrum of the tSZ signal, even down to very low multipoles~\cite{Cooray2001}.  The trispectrum represents a large non-Gaussian contribution to the covariance matrix of the tSZ power spectrum~\cite{Komatsu-Seljak2002}, which is especially problematic at low multipoles.  However, the trispectrum can be greatly suppressed by masking massive, low-redshift clusters using existing X-ray, optical, or SZ catalogs~\cite{Komatsu-Kitayama1999}.  This procedure can greatly increase the SNR for the tSZ power spectrum at low multipoles.  For constraints on \fnl it also has the advantage of enhancing the relative importance of the two-halo term compared to the one-halo, thus showing greater sensitivity to the scale-dependent bias at low-$\ell$.  Moreover, even in a Gaussian cosmology, the inclusion of the two-halo term slightly changes the shape of the tSZ power spectrum, which likely helps break degeneracies amongst the several parameters which effectively only change the overall amplitude of the one-halo term; the relative enhancement of the two-halo term due to masking should help further in this regard.  We consider two masking scenarios motivated by the flux limits of the cluster catalogs from all-sky surveys performed with the ROSAT\footnote{{\tt http://www.dlr.de/en/rosat}} X-ray telescope and the upcoming eROSITA\footnote{Extended ROentgen Survey with an Imaging Telescope Array, http://www.mpe.mpg.de/erosita/ } X-ray telescope.  These scenarios are detailed in Section~\ref{sec:covmasking}; by default all calculations and figures are computed for the unmasked scenario unless they are labelled otherwise.

Earlier studies have investigated the consequences of primordial non-Gaussianity for the tSZ power spectrum~\cite{Sadehetal2007,Roncarellietal2010}, though we are not aware of any calculations including the two-halo term (and hence the scale-dependent bias) or detailed parameter constraint forecasts.  We are also not aware of any previous work investigating constraints on massive neutrinos from the tSZ power spectrum, although previous authors have computed their signature~\cite{Shimonetal2011}.  Other studies have investigated detailed constraints on the primary $\Lambda$CDM parameters from the combination of CMB and tSZ power spectrum measurements~\cite{Taburetetal2010}.  Many authors have investigated constraints on \fnl and \Mnu from cluster counts, though the results depend somewhat on the cluster selection technique and mass estimation method.  Considering SZ cluster count studies only, \cite{Shimonetal2011} and~\cite{Shimonetal2012} investigated constraints on \Mnu from a Planck-derived catalog of SZ clusters (in combination with CMB temperature power spectrum data).  The earlier paper found a $1\sigma$ uncertainty of $\Delta M_{\nu} \approx 0.28$ eV while the later paper found $\Delta M_{\nu} \approx 0.06-0.12$ eV; the authors state that the use of highly degenerate nuisance parameters degraded the results in the former study.  In either case, the result is highly sensitive to uncertainties in the halo mass function, as the clusters included are deep in the exponential tail of the mass function.  We expect that our results using the tSZ power spectrum should be less sensitive to uncertainties in the tail of the mass function, as the power spectrum is dominated at most angular scales by somewhat less massive objects ($10^{13}-10^{14} M_{\odot}/h$)~\cite{Komatsu-Seljak2002}.  Finally, a very recent independent study~\cite{Mak-Pierpaoli2013} found $\Delta M_{\nu} \approx 0.3-0.4$ eV for Planck SZ cluster counts (with CMB temperature power spectrum information added), although they estimated that this bound could be improved to $\Delta M_{\nu} \approx 0.08$ eV with the inclusion of stronger priors on the ICM physics.

Our primary findings are as follows:
\begin{itemize}
\item The tSZ power spectrum can be detected with a total SNR $>30$ using the imminent Planck data up to $\ell=3000$, regardless of masking;
\item The tSZ power spectrum can be detected with a total SNR between $\approx 6$ and 22 using the future PIXIE data up to $\ell=300$, with the result being sensitive to the level of masking applied to remove massive, nearby clusters;
\item Adding the tSZ power spectrum information to the forecasted constraints from the Planck CMB temperature power spectrum and existing $H_0$ data is unlikely to significantly improve constraints on the primary cosmological parameters, but may give interesting constraints on the extensions we consider:
\begin{itemize}
\item If the true value of \fnl is near the WMAP9 ML value of $\approx 37$, a future CV-limited experiment combined with eROSITA-masking could provide a $3\sigma$ detection, completely independent of the primordial CMB temperature bispectrum; alternatively, PIXIE could give $1-2\sigma$ evidence for such a value of \fnl with this level of masking;
\item If the true value of \Mnu is near 0.1 eV, the Planck tSZ power spectrum with eROSITA masking can provide upper limits competitive with the current upper bounds on \Mnu; with stronger external constraints on the ICM physics, Planck with eROSITA masking could provide $1-2\sigma$ evidence for massive neutrinos from the tSZ power spectrum, depending on the true neutrino mass;
\end{itemize}
\item Regardless of the cosmological constraints, Planck will allow for a very tight constraint on the logarithmic slope of the ICM pressure profile in the outskirts of galaxy clusters, and may also provide some information on the overall normalization of the pressure profile (which sets the zero point of the $Y-M$ relation).
\end{itemize}

The remainder of this paper is organized as follows.  In Section~\ref{sec:LSS}, we describe our models for the halo mass function and halo bias, as well as the effects of primordial non-Gaussianity and massive neutrinos on large-scale structure.  In Section~\ref{sec:tSZPS}, we describe our halo model-based calculation of the tSZ power spectrum, including the relevant ICM physics.  We also demonstrate the different effects of each parameter in our model on the tSZ power spectrum.  In Section~\ref{sec:exp}, we consider the extraction of the tSZ power spectrum from the other components in microwave sky maps via multifrequency subtraction techniques.  Having determined the experimental noise levels, in Section~\ref{sec:cov} we detail our calculation of the covariance matrix of the tSZ power spectrum, and discuss the role of masking massive nearby clusters in reducing the low-$\ell$ cosmic variance.  In Section~\ref{sec:forecast}, we use our tSZ results to forecast constraints on cosmological and astrophysical parameters from a variety of experimental set-ups and masking choices.  We also compute the expected SNR of the tSZ power spectrum detection for each possible scenario.  We discuss our results and conclude in Section~\ref{sec:discussion}.  Finally, in Appendix~\ref{appendixPlanck}, we provide a brief comparison between our forecasts and the Planck tSZ power spectrum results that were publicly released while this manuscript was under review~\cite{Plancketal2013b}. 

The WMAP9+eCMB+BAO+$H_0$ maximum-likelihood parameters~\cite{Hinshawetal2012} define our fiducial model (see Section~\ref{sec:params} for details). All masses are quoted in units of $M_{\odot}/h$, where $h \equiv H_0/(100 \, \mathrm{km} \, \mathrm{s}^{-1} \, \mathrm{Mpc}^{-1})$ and $H_0$ is the Hubble parameter today.  All distances and wavenumbers are in comoving units of $\mathrm{Mpc}/h$.  All tSZ observables are computed at $\nu = 150$ GHz, since ACT and SPT have observed the tSZ signal at (or very near) this frequency, where the tSZ effect leads to a temperature decrement in the CMB along the line-of-sight (LOS) to a galaxy cluster.


\section{Modeling Large-Scale Structure}
\label{sec:LSS}
In order to compute statistics of the tSZ signal, we need to model the comoving number density of halos as a function of mass and redshift (the halo mass function) and the bias of halos with respect to the underlying matter density field as a function of mass and redshift.  Moreover, in order to extract constraints on \fnl and \Mnu from the tSZ power spectrum, we must include the effects of these parameters on large-scale structure.  We describe our approach to these computations in the following.


\subsection{Halo Mass Function}
\label{sec:HMF}
We define the mass of a dark matter halo by the spherical overdensity (SO) criterion: $M_{\delta,c}$ ($M_{\delta,d}$) is the mass enclosed within a sphere of radius $r_{\delta,c}$ ($r_{\delta,d}$) such that the enclosed density is $\delta$ times the critical (mean matter) density at redshift $z$.  To be clear, $c$ subscripts refer to masses referenced to the critical density at redshift $z$, $\rho_{cr}(z) = 3H^2(z)/8\pi G$ with $H(z)$ the Hubble parameter at redshift $z$, whereas $d$ subscripts refer to masses referenced to the mean matter density at redshift $z$, $\bar{\rho}_m(z) \equiv \bar{\rho}_m$ (this quantity is constant in comoving units).

We will generally work in terms of a particular SO mass, the virial mass, which we denote as $M$.  The virial mass is the mass enclosed within a radius $r_{vir}$~\cite{Bryan-Norman1998}:
\be
\label{eq.rvir}
r_{vir} = \left( \frac{3 M}{4 \pi \Delta_{cr}(z) \rho_{cr}(z)} \right)^{1/3} \,,
\ee
where $\Delta_{cr}(z) = 18 \pi^2 + 82(\Omega(z)-1) - 39(\Omega(z)-1)^2$ and $\Omega(z) = \Omega_m(1+z)^3/(\Omega_m(1+z)^3+\Omega_{\Lambda})$.  For many calculations, we need to convert between $M$ and various other SO masses (e.g., $M_{200c}$ or $M_{200d}$).  We use the NFW density profile~\cite{NFW1997} and the concentration-mass relation from~\cite{Duffyetal2008} in order to do these conversions, which require solving the following non-linear equation for $r_{\delta,c}$ (or $r_{\delta,d}$):
\be
\label{eq.rvir2}
\int_0^{r_{\delta,c}} 4 \pi r'^2 \rho_{\mathrm{NFW}} (r', M, c_{vir}) dr' = \frac{4}{3} \pi r_{\delta,c}^3 \rho_{cr}(z) \delta
\ee
where $c_{vir} \equiv r_{vir}/r_{NFW}$ is the concentration parameter ($r_{NFW}$ is the NFW scale radius) and we replace the critical density $\rho_{cr}(z)$ with the mean matter density $\bar{\rho}_m$ in this equation in order to obtain $r_{\delta,d}$ instead of $r_{\delta,c}$.  After solving Eq.~(\ref{eq.rvir2}) to find $r_{\delta,c}$, we calculate $M_{\delta,c}$ via $M_{\delta,c} = \frac{4}{3} \pi r_{\delta,c}^3 \rho_{cr}(z) \delta$.

The halo mass function, $dn(M,z)/dM$ describes the comoving number density of halos per unit mass as a function of redshift.  We employ the approach developed from early work by Press and Schechter~\cite{Press-Schechter1974} and subsequently refined by many other authors (e.g., \cite{Sheth-Tormen1999,Shethetal2001,Jenkinsetal2001,Tinkeretal2008}):
\beqn
\frac{dn(M,z)}{dM} & = & \frac{\bar{\rho}_m}{M} \frac{d \ln (\sigma^{-1}(M,z))}{dM} f(\sigma(M,z)) \nonumber \\
   & = & -\frac{\bar{\rho}_m}{2M^2} \frac{R(M)}{3\sigma^2(M,z)} \frac{d\sigma^2(M,z)}{dR(M)} f(\sigma(M,z)) \,,
\label{eq.dndM}
\eeqn
where $\sigma^2(M,z)$ is the variance of the linear matter density field smoothed with a (real space) top-hat filter on a scale $R(M) = \left( \frac{3M}{4\pi\bar{\rho}_m} \right)^{1/3}$ at redshift $z$:
\be
\label{eq.sigmaMz}
\sigma^2(M,z) = \frac{1}{2\pi^2} \int k^3 \, P_{\mathrm{lin}}(k,z) \, W^2(k,R(M)) \, d\ln k \,,
\ee
where $P_{\mathrm{lin}}(k,z)$ is the linear theory matter power spectrum at wavenumber $k$ and redshift $z$.  Note that the window function $W(k,R)$ is a top-hat filter in real space, which in Fourier space is given by
\be
\label{eq.WkR}
W(k,R) = \frac{3}{x^2} \left( \frac{\sin x}{x} - \cos x \right) \,,
\ee
where $x \equiv kR$.  In Eq.~(\ref{eq.dndM}), the function $f(\sigma(M,z))$ is known as the halo multiplicity function.  It has been measured to increasingly high precision from large $N$-body simulations over the past decade~\cite{Jenkinsetal2001,Warrenetal2006,Tinkeretal2008,Bhattacharyaetal2011}.  However, many of these calibrated mass functions are specified in terms of the friends-of-friends (FOF) mass rather than the SO mass, hindering their use in analytic calculations such as ours.  For this reason, we use the parametrization and calibration from~\cite{Tinkeretal2008}, where computations are performed in terms of the SO mass with respect to the mean matter density, $M_{\delta,d}$, for a variety of overdensities.  The halo multiplicity function in this model is parametrized by
\be
f(\sigma(M,z)) = A \left[ \left( \frac{\sigma}{b} \right)^{-a} + 1 \right] e^{-c/\sigma^2}
\label{eq.fsigmaTink}
\ee
where $\left\{A, a, b, c\right\}$ are (redshift- and overdensity-dependent) parameters fit from simulations.  We use the values of these parameters appropriate for the $M_{200,d}$ halo mass function from~\citep{Tinkeretal2008} with the redshift-dependent parameters given in their Eqs.~(5)--(8); we will hereafter refer to this as the Tinker mass function.  Note that the authors of that study caution against extrapolating their parameters beyond the highest redshift measured in their simulations ($z=2.5$) and recommend setting the parameters equal to their $z=2.5$ values at higher redshifts; we adopt this recommendation in our calculations.  Also, note that our tSZ power spectrum calculations in Section~\ref{sec:tSZPS} are phrased in terms of the virial mass $M$, and thus we compute the Jacobian $dM_{200,d}/dM$ using the procedure described in Eq.~(\ref{eq.rvir2}) in order to convert the Tinker mass function $dn/dM_{200,d}$ to a virial mass function $\frac{dn}{dM} = \frac{dn}{dM_{200,d}} \frac{dM_{200,d}}{dM}$.

We compute the smoothed matter density field in Eq.~(\ref{eq.sigmaMz}) by first obtaining the linear theory matter power spectrum from CAMB\footnote{{\tt http://camb.info/}} at $z_{in} = 30$ and subsequently rescaling $\sigma^2(M,z)$ by $D^2(z)$, where $D(z)$ is the linear growth factor. We normalize $D(z)$ by requiring that $D(z) \rightarrow 1/(1+z)$ deep in the matter-dominated era (e.g., at $z_{in}$).  The resulting $\sigma^2(M,z)$ is then used to compute the mass function in Eq.~(\ref{eq.dndM}).

Note that we assume the mass function to be known to high enough precision that the parameters describing it can be fixed; in other words, we do not consider $\left\{A, a, b, c\right\}$ to be free parameters in our model.  These parameters are certainly better constrained at present than those describing the ICM pressure profile (see Section~\ref{sec:ICM}), and thus this assumption seems reasonable for now.  However, precision cosmological constraints based on the mass function should in principle consider variations in the mass function parameters in order to obtain robust results, as has been done in some recent X-ray cluster cosmology analyses~\cite{Mantzetal2010a}.  However, we leave the implications of these uncertainties for tSZ statistics as a topic for future work.

\subsubsection{Effect of Primordial non-Gaussianity}
\label{sec:HMFNG}

The influence of primordial non-Gaussianity on the halo mass function has been studied by many authors over the past two decades using a variety of approaches~(e.g., \cite{NGMF1,NGMF2,NGMF3,NGMF4,NGMF5,NGMF6,NGMF7,LoVerdeetal2008,NGMF8}).  The physical consequences of the model specified in Eq.~(\ref{eq.fNLlocmodel}) are fairly simple to understand for the halo mass function, especially in the exponential tail of the mass function where massive clusters are found.  Intuitively, the number of clusters provides information about the tail of the probability distribution function of the primordial density field, since these are the rarest objects in the universe, which have only collapsed recently.  For positive skewness in the primordial density field ($f_{\mathrm{NL}} > 0$), one obtains an increased number of massive clusters at late times relative to the $f_{\mathrm{NL}} = 0$ case, because more regions of the smoothed density field have $\delta > \delta_c$, the collapse threshold ($\approx 1.686$ in the spherical collapse model).  Conversely, for negative skewness in the primordial density field ($f_{\mathrm{NL}} < 0$), one obtains fewer massive clusters at late times relative to the $f_{\mathrm{NL}} = 0$ case, because fewer regions of the smoothed density field are above the collapse threshold.  As illustrated in recent analytic calculations and simulation measurements~\cite{LoVerde-Smith2011,Pillepichetal2010,Grossietal2009,Wagneretal2010,Grossietal2007}, these changes can be quite significant for the number of extremely massive halos ($\sim 10^{15} \, M_{\odot}/h$) in the late-time universe; for example, the $z=0$ abundance of such halos for $f_{\mathrm{NL}} \approx 250$ can be $\approx 1.5-2$ times larger than in a Gaussian cosmology.  These results have been used as a basis for recent studies constraining \fnl by looking for extremely massive outliers in the cluster distribution (e.g., \cite{Hoyleetal2011,Cayonetal2011,Mortonsonetal2012,Enqvistetal2011,Harrison-Hotchkiss2012,Hoyleetal2012}).

We model the effect of \fnl on the halo mass function by multiplying the Tinker mass function by a non-Gaussian correction factor:
\be
\label{eq.NGMF}
\left( \frac{dn}{dM} \right)_{NG} = \frac{dn}{dM} R_{NG}(M,z,f_{\mathrm{NL}}) \,,
\ee
where $dn/dM$ is given by Eq.~(\ref{eq.dndM}).  We use the model for $R_{NG}(M,z,f_{\mathrm{NL}})$ given by Eq.~(35) in~\cite{LoVerde-Smith2011} (the ``log-Edgeworth'' mass function).  In this approach, the density field is approximated via an Edgeworth expansion, which captures small deviations from Gaussianity.  The Press-Schechter approach is then applied to the Edgeworth-expanded density field to obtain an expression for the halo mass function in terms of cumulants of the non-Gaussian density field.  The results of \cite{LoVerde-Smith2011} include numerical fitting functions for these cumulants obtained from $N$-body simulations.  We use both the expression for $R_{NG}(M,z,f_{\mathrm{NL}})$ and the cumulant fitting functions from~\cite{LoVerde-Smith2011} to compute the non-Gaussian correction to the mass function.  This prescription was shown to accurately reproduce the non-Gaussian halo mass function correction factor measured directly from $N$-body simulations in~\cite{LoVerde-Smith2011}, and in particular improves upon the similar prescription derived in~\cite{LoVerdeetal2008} (the ``Edgeworth'' mass function).

Note that we apply the non-Gaussian correction factor $R_{NG}(M,z,f_{\mathrm{NL}})$ to the Tinker mass function in Eq.~(\ref{eq.NGMF}), which is an SO mass function, as mentioned above.  The prescription for computing $R_{NG}(M,z,f_{\mathrm{NL}})$ makes no assumption about whether $M$ is an FOF or SO mass, so there is no logical flaw in this procedure.  However, the comparisons to $N$-body results in~\cite{LoVerde-Smith2011} were performed using FOF halos.  Thus, without having tested the results of Eq.~(\ref{eq.NGMF}) on SO mass functions from simulations, our calculation assumes that the change in the mass function due to non-Gaussianity is quasi-universal, even if the underlying Gaussian mass function itself is not.  This assumption was tested in~\cite{Wagneretal2010} for the non-Gaussian correction factor from~\cite{LoVerdeetal2008} (see Fig.~9 in \cite{Wagneretal2010}) and found to be valid; thus, we choose to adopt it here.  We will refer to the non-Gaussian mass function computed via Eq.~(\ref{eq.NGMF}) using the prescription from~\cite{LoVerde-Smith2011} as the LVS mass function.

\subsubsection{Effect of Massive Neutrinos}
\label{sec:HMFMnu}

It has long been known that massive neutrinos suppress the amplitude of the matter power spectrum on scales below their free-streaming scale, $k_{fs}$~\cite{Lesgourgues-Pastor2006}:
\be
\label{eq.kfs}
k_{fs} \approx 0.082 \frac{H(z)}{H_0 (1+z)^2} \left( \frac{M_{\nu}}{0.1 \,\, \mathrm{eV}} \right) \,\, h/\mathrm{Mpc} \,.
\ee
Neutrinos do not cluster on scales much smaller than this scale (i.e., $k > k_{fs}$), as they are able to free-stream out of small-scale gravitational potential wells.  This effect leads to a characteristic decrease in the small-scale matter power spectrum of order $\Delta P/P \approx -8 \Omega_{\nu}/\Omega_m$ in linear perturbation theory~\cite{Lesgourgues-Pastor2006,Abazajianetal2011}.  Nonlinear corrections increase this suppression to $\Delta P/P \approx -10 \Omega_{\nu}/\Omega_m$ for modes with wavenumbers $k \sim 0.5-1 \,\, \mathrm{Mpc}/h$~\cite{Abazajianetal2011}.

The neutrino-induced suppression of the small-scale matter power spectrum leads one to expect that the number of massive halos in the low-redshift universe should also be decreased.  Several papers in recent years have attempted to precisely model this change in the halo mass function using both $N$-body simulations and analytic theory~\cite{Brandbygeetal2010,Marullietal2011,Ichiki-Takada2012}.  In~\cite{Brandbygeetal2010}, $N$-body simulations are used to show that massive neutrinos do indeed suppress the halo mass function, especially for the largest, latest-forming halos (i.e., galaxy clusters).  Moreover, the suppression is found to arise primarily from the suppression of the initial transfer function in the linear regime, and not due to neutrino clustering effects in the $N$-body simulations.  This finding suggests that an analytic approach similar to the Press-Schecter theory should work for massive neutrino cosmologies as well, and the authors subsequently show that a modified Sheth-Tormen formalism~\cite{Sheth-Tormen1999} gives a good fit to their simulation results.  Similar $N$-body simulations are examined in~\cite{Marullietal2011}, who find generally similar results to those in~\cite{Brandbygeetal2010}, but also point out that the effect of $M_{\nu} > 0$ on the mass function cannot be adequately represented by simply rescaling $\sigma_8$ to a lower value in an analytic calculation without massive neutrinos.  Finally, \cite{Ichiki-Takada2012} study the effect of massive neutrinos on the mass function using analytic calculations with the spherical collapse model.  Their results suggest that an accurate approximation is to simply input the $M_{\nu}$-suppressed linear theory (cold$+$baryonic-only) matter power spectrum computed at $z_{in}$ to a $\Lambda$CDM-calibrated mass function fit (note that a similar procedure was used in some recent X-ray cluster-based constraints on \Mnu~\cite{Vikhlininetal2009}).  The net result of this suppression can be quite significant at the high-mass end of the mass function; for example, $M_{\nu} = 0.1$ eV leads to a factor of $\sim 2$ decrease in the abundance of $10^{15} \,\, M_{\odot}/h$ halos at $z=1$ as compared to a massless-neutrino cosmology~\cite{Ichiki-Takada2012}.  We follow the procedure used in~\cite{Ichiki-Takada2012} in our work, although we input the suppressed linear theory matter power spectrum to the Tinker mass function rather than that of \cite{Bhattacharyaetal2011}, as was done in \cite{Ichiki-Takada2012}.  We will refer to the \Mnu-suppressed mass function computed with this prescription as the IT mass function.


\subsection{Halo Bias}
\label{sec:bias}

Dark matter halos are known to cluster more strongly than the underlying matter density field; they are thus biased tracers.  This bias can depend on scale, mass, and redshift~(e.g., \cite{BBKS,Mo-White1996,Smithetal2007}).  We define the halo bias $b(k,M,z)$ by
\be
\label{eq.halobias}
b(k,M,z) = \sqrt{\frac{P_{hh}(k,M,z)}{P(k,z)}}\,,
\ee
where $P_{hh}(k,M,z)$ is the power spectrum of the halo density field and $P(k,z)$ is the power spectrum of the matter density field.  Knowledge of the halo bias is necessary to model and extract cosmological information from the clustering of galaxies and galaxy clusters.  For our purposes, it will be needed to compute the two-halo term in the tSZ power spectrum, which requires knowledge of $P_{hh}(k,M,z)$.

In a Gaussian cosmology, the halo bias depends on mass and redshift but is independent of scale for $k \lesssim 0.05 \,\, \mathrm{Mpc}/h$, i.e. on large scales~(e.g., \cite{Tinkeretal2010}).  We compute this linear Gaussian bias, $b_G(M,z)$, using the fitting function in Eq.~(6) of \cite{Tinkeretal2010} with the parameters appropriate for $M_{200,d}$ SO masses (see Table 2 in \cite{Tinkeretal2010}).  This fit was determined from the results of many large-volume $N$-body simulations with a variety of cosmological parameters and found to be quite accurate.  We will refer to this prescription as the Tinker bias model.

Although the bias becomes scale-dependent on small scales even in a Gaussian cosmology, it becomes scale-dependent on large scales in the presence of local primordial non-Gaussianity, as first shown in~\cite{Dalaletal2008}.  The scale-dependence arises due to the coupling of long- and short-wavelength density fluctuations induced by local \fnl$\neq 0$.  We model this effect as a correction to the Gaussian bias described in the preceding paragraph:
\be
\label{eq.btot}
b(k,M,z) = b_G(M,z) + \Delta b_{NG}(k,M,z) \,,
\ee
where the non-Gaussian correction is given by~\cite{Dalaletal2008}
\be
\label{eq.bNG}
\Delta b_{NG}(k,M,z) = 2 \delta_c \left( b_G(M,z) - 1 \right) \frac{f_{\mathrm{NL}}}{\alpha(k,z)} \,.
\ee
Here, $\delta_c = 1.686$ (the spherical collapse threshold) and
\be
\label{eq.alpha}
\alpha(k,z) = \frac{2 k^2 T(k) D(z) c^2}{3 \Omega_m H_0^2}
\ee
relates the linear density field to the primordial potential via $\delta(k,z) = \alpha(k,z) \Phi(k)$.  Note that $T(k)$ is the linear matter transfer function, which we compute using CAMB.  Since the original derivation in~\cite{Dalaletal2008}, the results in Eqs.~(\ref{eq.bNG}) and~(\ref{eq.alpha}) have subsequently been confirmed by other authors~\cite{NGbias1,NGbias2,NGbias3} and tested extensively on $N$-body simulations~(e.g., \cite{NGbias4,Dalaletal2008,Pillepichetal2010,Smithetal2012}).  The overall effect is a steep increase in the large-scale bias of massive halos, which is even larger for highly biased tracers like galaxy clusters.  We will refer to this effect simply as the scale-dependent halo bias.

The influence of massive neutrinos on the halo bias has been studied far less thoroughly than that of primordial non-Gaussianity.  Recent $N$-body simulations analyzed in~\cite{Marullietal2011} indicate that massive neutrinos lead to a nearly scale-independent increase in the large-scale halo bias.  This effect arises because of the mass function suppression discussed in Section~\ref{sec:HMFMnu}: halos of a given mass are rarer in an \Mnu$>0$ cosmology than in a massless neutrino cosmology (for fixed $A_s$), and thus they are more highly biased relative to the matter density field.  However, the amplitude of this change is far smaller than that induced by \fnl$\neq 0$, especially on very large scales.  For example, the results of~\cite{Marullietal2011} indicate an overall increase of $\sim 10$\% in the mean bias of massive halos at $z=1$ for \Mnu$=0.3$ eV as compared to \Mnu$=0$.  Our implementation of the scale-dependent bias due to local \fnl yields a factor of $\sim 100-1000$ increase in the large-scale ($k \sim 10^{-4} \,\, h/\mathrm{Mpc}$) bias of objects in the same mass range at $z=1$ for \fnl$=50$.  Clearly, the effect of \fnl is much larger than that of massive neutrinos, simply because it is so strongly scale-dependent, while \Mnu only leads to a small scale-independent change (at least on large scales; the small-scale behavior may be more complicated).  Moreover, the change in bias due to \Mnu is larger at higher redshifts ($z \gtrsim 1$), whereas most of the tSZ signal originates at lower redshifts.  Lastly, due to the smallness of the two-halo term in the tSZ power spectrum compared to the one-halo term (see Section~\ref{sec:tSZPS}), small variations in the Gaussian bias cause essentially no change in the total signal.  For all of these reasons, we choose to neglect the effect of massive neutrinos on the halo bias in our calculations.


\section{Thermal SZ Power Spectrum}
\label{sec:tSZPS}

The tSZ effect results in a frequency-dependent shift in the CMB temperature observed in the direction of a galaxy group or cluster.  The temperature shift $\Delta T$ 
at angular position $\vec{\theta}$ with respect to the center of a cluster of mass $M$ at redshift $z$ is given by~\cite{Sunyaev-Zeldovich1970}
\be
\label{eq.tSZdef}
\frac{\Delta T(\vec{\theta}, M, z)}{T_{\mathrm{CMB}}} & = & g_{\nu} y(\vec{\theta}, M, z) \\
 &= & g_{\nu} \frac{\sigma_T}{m_e c^2} \int_{\mathrm{LOS}} P_e \left( \sqrt{l^2 + d_A^2 |\vec{\theta}|^2}, M, z \right) dl \,, \nonumber
\ee
where $g_{\nu} = x\,\mathrm{coth}(x/2)-4$ is the tSZ spectral function with $x \equiv h\nu/k_B T_{\mathrm{CMB}}$, $y$ is the Compton-$y$ parameter, $\sigma_T$ is the Thomson scattering cross-section, $m_e$ is the electron mass, and $P_e(\vec{r})$ is the ICM electron pressure at location $\vec{r}$ with respect to the cluster center.  We have neglected relativistic corrections in Eq.~(\ref{eq.tSZdef}) (e.g.,~\cite{Nozawaetal2006}), as these effects are relevant only for the most massive clusters in the universe ($\gtrsim 10^{15} \,\, M_{\odot}/h$).  Such clusters contribute non-negligibly to the tSZ power spectrum at low-$\ell$, and thus our results in unmasked calculations may be slightly inaccurate; however, the optimal forecasts for cosmological constraints arise from calculations in which such nearby, massive clusters are masked (see Section~\ref{sec:forecast}), and thus these corrections will not be relevant.  Therefore, we do not include them in our calculations.

Note that we only consider spherically symmetric pressure profiles in this work, i.e. $P_e(\vec{r}) = P_e(r)$ in Eq.~(\ref{eq.tSZdef}).  The integral in Eq.~(\ref{eq.tSZdef}) is computed along the LOS such that $r^2 = l^2 + d_A(z)^2 \theta^2$, where $d_A(z)$ is the angular diameter distance to redshift $z$ and $\theta \equiv |\vec{\theta}|$ is the angular distance between $\vec{\theta}$ and the cluster center in the plane of the sky (note that this formalism assumes the flat-sky approximation is valid; we provide exact full-sky results for the tSZ power spectrum in Appendix~\ref{appendix}).  In the flat-sky limit, a spherically symmetric pressure profile implies that the temperature decrement (or Compton-$y$) profile is azimuthally symmetric in the plane of the sky, i.e., $\Delta T(\vec{\theta},M,z) = \Delta T(\theta,M,z)$.  Finally, note that the electron pressure $P_e(\vec{r})$ is related to the thermal gas pressure via $P_{th} = P_e (5 X_H+3)/2(X_H+1) = 1.932 P_e$, where $X_H=0.76$ is the primordial hydrogen mass fraction.  We calculate all tSZ power spectra in this paper at $\nu = 150$ GHz, where the tSZ effect is observed as a decrement in the CMB temperature ($g_{150 \, \mathrm{GHz}} = -0.9537$).  We make this choice simply because recent tSZ measurements have been performed at this frequency using ACT and SPT~(e.g., \cite{Wilsonetal2012,Crawfordetal2013,Storyetal2012,Sieversetal2013}), and thus the temperature values in this regime are perhaps more familiar and intuitive.  All of our calculations can be phrased in a frequency-independent manner in terms of the Compton-$y$ parameter, and we will often use ``y'' as a label for tSZ quantities, although they are calculated numerically at $\nu = 150$ GHz.

In the remainder of this section, we outline the halo model-based calculations used to compute the tSZ power spectrum, discuss our model for the gas physics of the ICM, and explain the physical effects of each cosmological and astrophysical parameter on the tSZ power spectrum.


\subsection{Halo Model Formalism}
\label{sec:tSZPShalomodel}

We compute the tSZ power spectrum using the halo model approach~(see~\cite{Cooray-Sheth2002} for a review).  We provide complete derivations of all the relevant expressions in Appendix~\ref{appendix}, first obtaining completely general full-sky results and then specializing to the flat-sky/Limber-approximated case.  Here, we simply quote the necessary results and refer the interested reader to Appendix~\ref{appendix} for the derivations.  Note that we will work in terms of the Compton-$y$ parameter; the results can easily be multiplied by the necessary $g_{\nu}$ factors to obtain results at any frequency.

The tSZ power spectrum, $C_{\ell}^y$, is given by the sum of the one-halo and two-halo terms:
\be
\label{eq.Cell}
C_{\ell}^y = C_{\ell}^{y,1h} + C_{\ell}^{y,2h} \,.
\ee
The exact expression for the one-halo term is given by Eq.~(\ref{eq.yCl1hexact2}):
\be
\label{eq.Cell1hexactquote}
C_{\ell}^{y,1h} = \int \frac{dz}{\chi(z)} \frac{d^2V}{dz d\Omega} \int dM \frac{dn}{dM} \left| \int k \, dk \, J_{\ell+1/2}(k\chi(z)) \tilde{y}_{3D}(k;M,z) \int \frac{c \, dz'}{H(z')(1+z')\sqrt{\chi(z')}} J_{\ell+1/2}(k\chi(z')) \right|^2 \,,
\ee
where $\chi(z)$ is the comoving distance to redshift $z$, $d^2V/dz d\Omega$ is the comoving volume element per steradian, $dn/dM$ is the halo mass function discussed in Section~\ref{sec:HMF}, $\tilde{y}_{3D}(k;M,z)$ is given in Eq.~(\ref{eq.y3Dtwid}), and $J_{\ell+1/2}(x)$ is a Bessel function of the first kind.  In the flat-sky limit, the one-halo term simplifies to the following widely-used expression (given in e.g. Eq.~(1) of~\cite{Komatsu-Seljak2002}), which we derive in Eq.~(\ref{eq.yCl1hflatsky}):
\be
\label{eq.yCl1hflatskyquote}
C_{\ell \gg 1}^{y,1h} \approx \int dz \frac{d^2V}{dz d\Omega} \int dM \frac{dn(M,z)}{dM} \left| \tilde{y}_{\ell}(M,z) \right|^2 \,,
\ee
where 
\be
\tilde{y}_{\ell}(M,z) \approx \frac{4 \pi r_s}{\ell_s^2} \int dx \, x^2 \frac{\sin((\ell+1/2) x/\ell_s)}{(\ell+1/2) x/\ell_s} y_{3D}(x;M,z) \,.
\label{eq.yelltwidquote}
\ee
Here, $r_s$ is a characteristic scale radius (not the NFW scale radius) of the $y_{3D}$ profile given by $y_{3D}(\vec{r}) = \frac{\sigma_T}{m_e c^2} P_e(\vec{r})$ and $\ell_s = a(z)\chi(z)/r_s = d_A(z)/r_s$ is the multipole moment associated with the scale radius.  For the pressure profile from~\cite{Battagliaetal2012} used in our calculations, the natural scale radius is $r_{200,c}$.  In our calculations, we choose to implement the flat-sky result for the one-halo term at all $\ell$ --- see Appendix~\ref{appendix} for a justification of this decision and an assessment of the associated error at low-$\ell$ (the only regime where this correction would be relevant).

The exact expression for the two-halo term is given by Eq.~(\ref{eq.yCl2hexact}):
\be
\label{eq.yCl2hexactquote}
C_{\ell}^{y,2h} = \int dk \, k \, \frac{P_{\mathrm{lin}}(k;z_{in})}{D^2(z_{in})} \left[ \int \frac{dz}{\sqrt{\chi(z)}} \frac{d^2V}{dz d\Omega} J_{\ell+1/2}(k\chi(z)) D(z) \int dM \frac{dn}{dM} b(k,M,z) \tilde{y}_{k\chi(z)}(M,z) \right]^2 \,,
\ee
where $P_{\mathrm{lin}}(k,z_{in})$ is the linear theory matter power spectrum at $z_{in}$ (which we choose to set equal to 30), $b(k,M,z)$ is the halo bias discussed in Section~\ref{sec:bias}, and $\tilde{y}_{k\chi(z)}(M,z)$ refers to the expression for $\tilde{y}_{\ell}(M,z)$ given in Eq.~(\ref{eq.yelltwidquote}) evaluated with $\ell+1/2=k\chi$.  This notation is simply a mathematical convenience; no flat-sky or Limber approximation was used in deriving Eq.~(\ref{eq.yCl2hexact}), and no $\ell$ appears in $\tilde{y}_{k\chi}(M,z)$.  In the Limber approximation~\cite{Limber}, the two-halo term simplifies to the result given in~\cite{Komatsu-Kitayama1999}, which we derive in Eq.~(\ref{eq.yCl2hLimber}):
\be
\label{eq.yCl2hLimberquote}
C_{\ell \gg 1}^{y,2h} \approx \int dz \frac{d^2V}{dz d\Omega} \left[ \int dM \frac{dn(M,z)}{dM} b(k,M,z) \tilde{y}_{\ell}(M,z) \right]^2 P_{\mathrm{lin}}\left(\frac{\ell+1/2}{\chi(z)};z\right) \,.
\ee
We investigate the validity of the Limber approximation in detail in Appendix~\ref{appendix}.  We find that it is necessary to compute the exact expression in Eq.~(\ref{eq.yCl2hexactquote}) in order to obtain sufficiently accurate results at low-$\ell$, where the signature of the scale-dependent bias induced by \fnl is present (looking for this signature is our primary motivation for computing the two-halo term to begin with).  In particular, we compute the exact expression in Eq.~(\ref{eq.yCl2hexactquote}) for $\ell < 50$, while we use the Limber-approximated result in Eq.~(\ref{eq.yCl2hLimberquote}) at higher multipoles.

The fiducial integration limits in our calculations are $0.005 < z < 4$ for all redshift integrals, $5 \times 10^{11} M_{\odot}/h < M < 5 \times 10^{15} M_{\odot}/h$ for all mass integrals, and $10^{-4} \,\, h/\mathrm{Mpc} < k < 3 \,\, h/\mathrm{Mpc}$ for all wavenumber integrals.  We check that extending the wavenumber upper limit further into the nonlinear regime does not affect our results.  Note that the upper limit in the mass integral becomes redshift-dependent in the masked calculations that we discuss below, in which the most massive clusters at low redshifts are removed from the computation.

We use the halo mass functions discussed in Section~\ref{sec:HMF} (Tinker, LVS, and IT) and the bias models discussed in Section~\ref{sec:bias} (Tinker and scale-dependent bias) in Eqs.~(\ref{eq.yCl1hflatskyquote}), (\ref{eq.yCl2hexactquote}), and (\ref{eq.yCl2hLimberquote}).  The only remaining ingredient needed to complete the tSZ power spectrum calculation is a prescription for the ICM electron pressure profile as a function of mass and redshift.  Note that this approach to the tSZ power spectrum calculation separates the cosmology-dependent component (the mass function and bias) from the ICM-dependent component (the pressure profile).  This separation arises from the fact that the small-scale baryonic physics that determines the structure of the ICM pressure profile effectively decouples from the large-scale physics described by the background cosmology and linear perturbation theory.  Thus, it is a standard procedure to constrain the ICM pressure profile from cosmological hydrodynamics simulations~(e.g., \cite{Battagliaetal2012,Bodeetal2012}) or actual observations of galaxy clusters~(e.g., \cite{Arnaudetal2010,Plancketal2013}, which are obtained for a fixed cosmology in either case (at present, it is prohibitively computationally expensive to run many large hydrodynamical simulations with varying cosmological parameters).  Of course, it is also possible to model the ICM analytically and obtain a pressure profile~(e.g., \cite{Komatsu-Seljak2001,Shawetal2010}.  Regardless of its origin (observations/simulations/theory), the derived ICM pressure profile can then be applied to different background cosmologies by using the halo mass function and bias model appropriate for that cosmology in the tSZ power spectrum calculations.  We follow this approach.

Note that because the tSZ signal is heavily dominated by contributions from collapsed objects, the halo model approximation gives very accurate results when compared to direct LOS integrations of numerical simulation boxes (see Figs.~7 and 8 in~\cite{Battagliaetal2012} for direct comparisons).  In particular, the halo model agrees very well with the simulation results for $\ell \lesssim 1000$, which is predominantly the regime we are interested in for this paper (on smaller angular scales effects due to asphericity and substructure become important, which are not captured in the halo model approach).  These results  imply that contributions from the intergalactic medium, filaments, and other diffuse structures are unlikely to be large enough to significantly impact the calculations and forecasts in the remainder of the paper.  Contamination from the Galaxy is a separate issue, which we assume can be minimized to a sufficient level through sky cuts and foreground subtraction (see Section~\ref{sec:exp}).

\begin{figure}
  \begin{center}
    \includegraphics[trim=0cm 0cm 0cm 0cm, clip=true, totalheight=0.5\textheight]{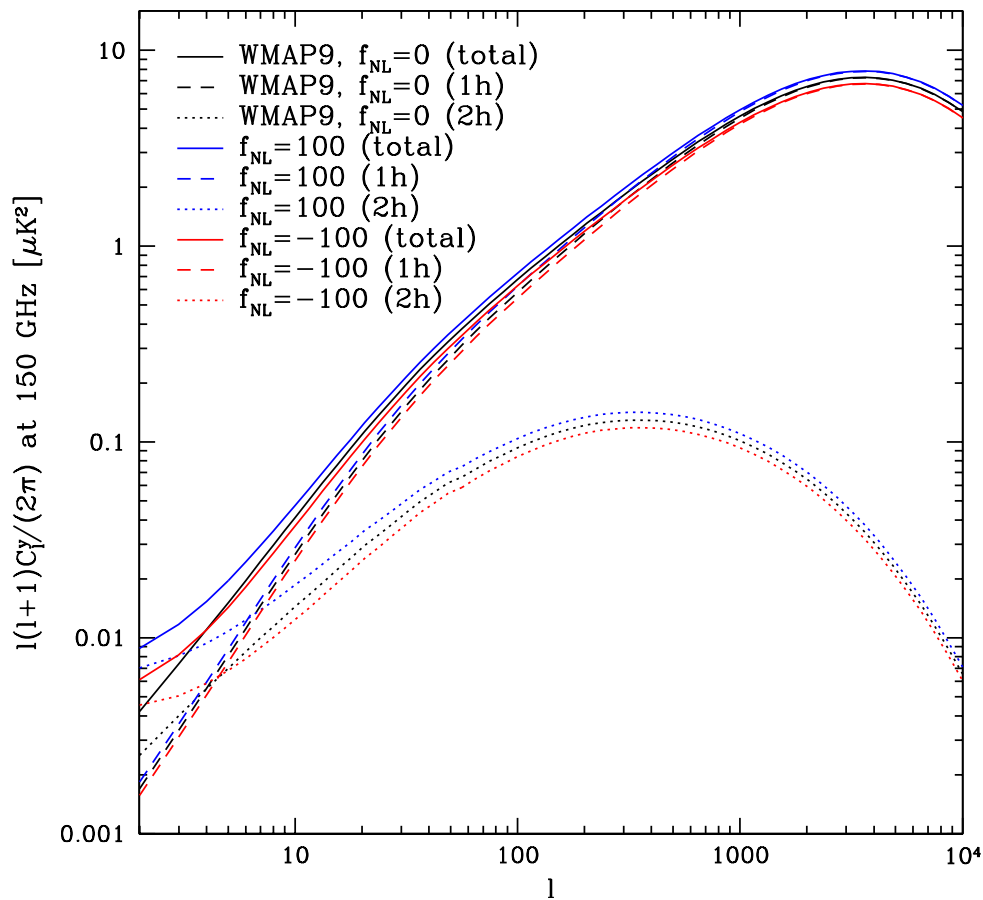}
    \caption{This plot shows the unmasked tSZ power spectrum for our fiducial model (black curves), as specified in Section~\ref{sec:params}, as well as variations with \fnl$=100$ (blue curves) and \fnl$=-100$ (red curves).  \fnl values of this magnitude are highly disfavored by current constraints, but we plot them here to clarify the influence of primordial non-Gaussianity on the tSZ power spectrum.  The effect of \fnl on the one-halo term is simply an overall amplitude shift due to the corresponding increase or decrease in the number of massive clusters in the universe, as described in Section~\ref{sec:HMFNG}.  The effect of \fnl on the two-halo term includes not only an amplitude shift due to the change in the mass function, but also a steep upturn at low-$\ell$ due to the influence of the scale-dependent halo bias, as described in~Section~\ref{sec:bias}.  Note that for \fnl$<0$ the two effects cancel for $\ell \approx 4-5$.  The relative smallness of the two-halo term (compared to the one-halo term) makes the scale-dependent bias signature subdominant for all $\ell$ values except $\ell \lesssim 7-8$.  However, masking of nearby massive clusters suppresses the low-$\ell$ one-halo term in the power spectrum (in addition to decreasing the cosmic variance, as discussed in Section~\ref{sec:cov}), which increases the relative importance of the two-halo term and thus the dependence of the total signal on \fnl at low-$\ell$. \label{fig.fNL1h2h}}
  \end{center}
\end{figure}

\begin{figure}
  \begin{center}
    \includegraphics[trim=0cm 0cm 0cm 0cm, clip=true, totalheight=0.5\textheight]{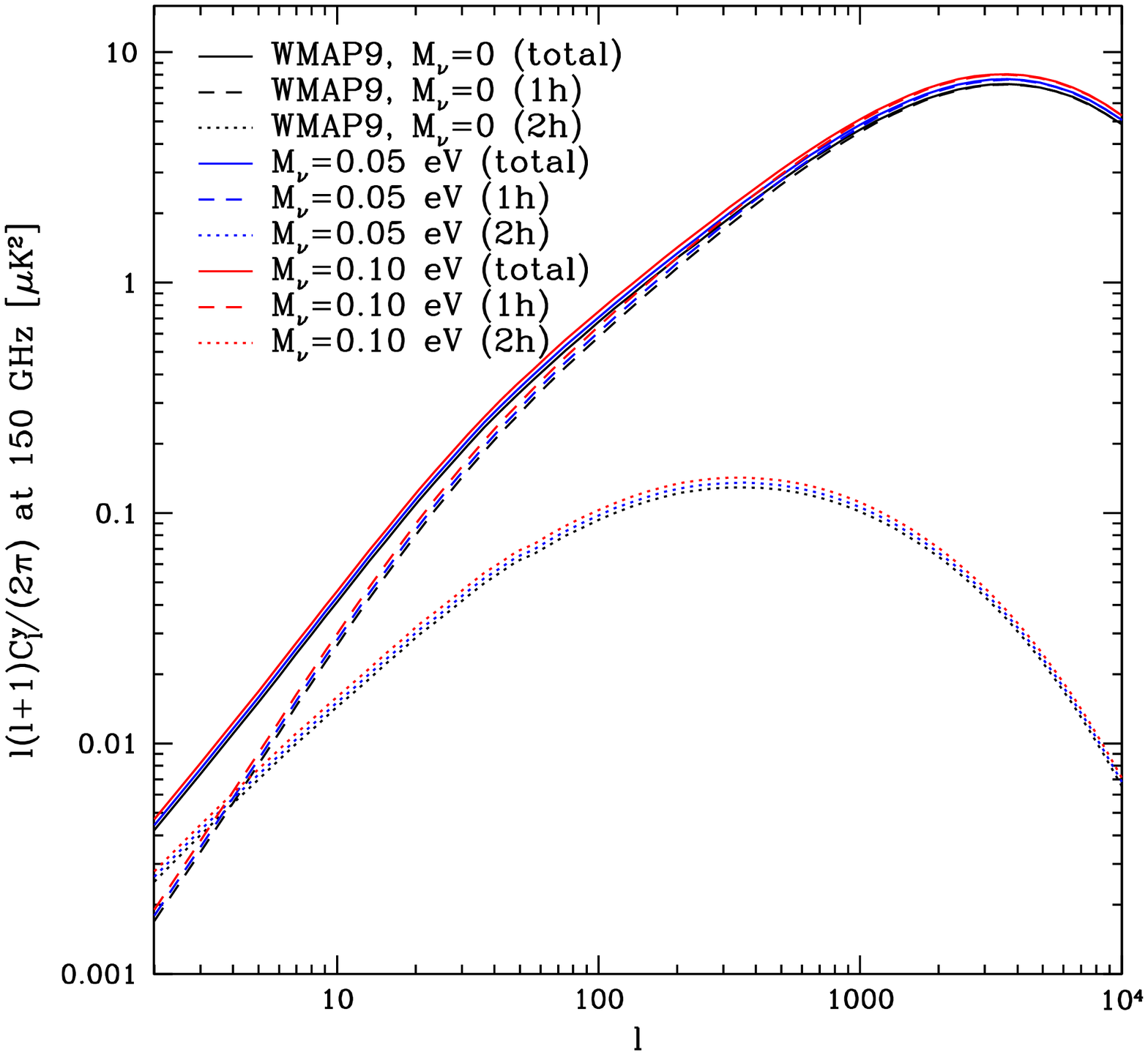}
    \caption{This plot shows the unmasked tSZ power spectrum for our fiducial model (black curves), as specified in Section~\ref{sec:params}, as well as variations with \Mnu$=0.05$ eV (blue curves) and \Mnu$=0.10$ eV (red curves).  \Mnu values of this magnitude are at the lower bound allowed by neutrino oscillation measurements, and thus an effect of at least this magnitude is expected in our universe.  The effect of \Mnu on both the one- and two-halo terms is effectively an overall amplitude shift, although the effect tapers off very slightly at high-$\ell$.  It may be puzzling at first to see an increase in the tSZ power when \Mnu$>0$, but the key fact is that we are holding $\sigma_8$ constant when varying \Mnu (indeed, we hold all of the other parameters constant).  In order to keep $\sigma_8$ fixed despite the late-time suppression of structure growth due to \Mnu$>0$, we must increase the primordial amplitude of scalar perturbations, $A_s$.  Thus, the net effect is an increase in the tSZ power spectrum amplitude, counterintuitive though it may be.  If our $\Lambda$CDM parameter set included $A_s$ rather than $\sigma_8$, and we held $A_s$ constant while \Mnu$>0$, we would indeed find a corresponding decrease in the tSZ power (and in $\sigma_8$, of course).\label{fig.Mnu1h2h}}
  \end{center}
\end{figure}


\subsection{Modeling the ICM}
\label{sec:ICM}

We adopt the parametrized ICM pressure profile fit from~\cite{Battagliaetal2012} as our fiducial model.  This profile is derived from cosmological hydrodynamics simulations described in~\cite{Battagliaetal2010}.  These simulations include (sub-grid) prescriptions for radiative cooling, star formation, supernova feedback, and feedback from active galactic nuclei (AGN).  Taken together, these feedback processes typically decrease the gas fraction in low-mass groups and clusters, as the injection of energy into the ICM blows gas out of the cluster potential.  In addition, the smoothed particle hydrodynamics used in these simulations naturally captures the effects of non-thermal pressure support due to bulk motions and turbulence, which must be modeled in order to accurately characterize the cluster pressure profile in the outskirts.

The ICM thermal pressure profile in this model is parametrized by a dimensionless GNFW form, which has been found to be a useful parametrization by many observational and numerical studies~(e.g.,~\cite{Nagaietal2007,Arnaudetal2010,Plaggeetal2012,Plancketal2013}):
\be
\label{eq.GNFW}
\frac{P_{th}(x)}{P_{200,c}} = \frac{P_0 \left( x/x_c\right)^{\gamma}}{\left[1+\left( x/x_c \right)^{\alpha} \right]^{\beta}} \,\,\,\,\,, x \equiv r/r_{200,c} \,,
\ee
where $P_{th}(x) = 1.932 P_e(x)$ is the thermal pressure profile, $x$ is the dimensionless distance from the cluster center, $x_c$ is a core scale length, $P_0$ is a dimensionless amplitude, $\alpha$, $\beta$, and $\gamma$ describe the logarithmic slope of the profile at intermediate ($x \sim x_c$), large ($x \gg x_c$), and small ($x \ll x_c$) radii, respectively, and $P_{200,c}$ is the self-similar amplitude for pressure at $r_{200,c}$ given by~\cite{Kaiser1986,Voit2005}:
\be
\label{eq.P200c}
P_{200,c} = \frac{200 \, G M_{200,c} \rho_{cr}(z) \Omega_b}{2\, \Omega_m r_{200,c}} \,.
\ee
In~\cite{Battagliaetal2012} this parametrization is fit to the stacked pressure profiles of clusters extracted from the simulations described above.  Note that due to degeneracies the parameters $\alpha$ and $\gamma$ are not varied in the fit; they are fixed to $\alpha=1.0$ and $\gamma=-0.3$, which agree with many other studies~(e.g., \cite{Nagaietal2007,Arnaudetal2010,Plaggeetal2012,Plancketal2013}.  In addition to constraining the amplitude of the remaining parameters, \cite{Battagliaetal2012} also fit power-law mass and redshift dependences, with the following results:
\beqn
P_0(M_{200,c},z) & = & 18.1 \left( \frac{M_{200,c}}{10^{14} \,\, M_{\odot}} \right)^{0.154} \left( 1+z \right)^{-0.758} \label{eq.P0batt} \\
x_c(M_{200,c},z) & = & 0.497 \left( \frac{M_{200,c}}{10^{14} \,\, M_{\odot}} \right)^{-0.00865} \left( 1+z \right)^{0.731} \label{eq.xcbatt} \\
\beta(M_{200,c},z) & = & 4.35 \left( \frac{M_{200,c}}{10^{14} \,\, M_{\odot}} \right)^{0.0393} \left( 1+z \right)^{0.415} \label{eq.betabatt} \,.
\eeqn
Note that the denominator of the mass-dependent factor has units of $M_{\odot}$ rather than $M_{\odot}/h$ as used elsewhere in this paper.  The mass and redshift dependence of these parameters captures deviations from simple self-similar cluster pressure profiles.  These deviations arise from non-gravitational energy injections due to AGN and supernova feedback, star formation in the ICM, and non-thermal processes such as turbulence and bulk motions~\cite{Battagliaetal2012,Battagliaetal2012b}.  Eqs.~(\ref{eq.GNFW})--(\ref{eq.betabatt}) completely specify the ICM electron pressure profile as a function of mass and redshift, and provide the remaining ingredient needed for the halo model calculations of the tSZ power spectrum described in Section~\ref{sec:tSZPShalomodel}, in addition to the halo mass function and halo bias.  We will refer to this model of the ICM pressure profile as the Battaglia model.

Although it is derived solely from numerical simulations, we note that the Battaglia pressure profile is in good agreement with a number of observations of cluster pressure profiles, including those based on the REXCESS X-ray sample of massive, $z<0.3$ clusters~\cite{Arnaudetal2010}, independent studies of low-mass groups at $z<0.12$ with Chandra~\cite{Sunetal2011}, and early Planck measurements of the stacked pressure profile of $z<0.5$ clusters~\cite{Plancketal2013}.

We allow for a realistic degree of uncertainty in the ICM pressure profile by freeing the amplitude of the parameters that describe the overall normalization ($P_0$) and the outer logarithmic slope ($\beta$).  To be clear, we do not free the mass and redshift dependences for these parameters given in Eqs.~(\ref{eq.P0batt}) and~(\ref{eq.betabatt}), only the overall amplitudes in those expressions.  The outer slope $\beta$ is known to be highly degenerate with the scale radius $x_c$~(e.g., \cite{Battagliaetal2012,Plaggeetal2012}), and thus it is only feasible to free one of these parameters.  The other slope parameters in Eq.~(\ref{eq.GNFW}) are fixed to their Battaglia values, which match the standard values in the literature.  We parametrize the freedom in $P_0$ and $\beta$ by introducing new parameters $C_{P_0}$ and $C_{\beta}$ defined by:
\beqn
P_0(M_{200,c},z) & = & C_{P_0} \, \times \, 18.1 \left( \frac{M_{200,c}}{10^{14} \,\, M_{\odot}} \right)^{0.154} \left( 1+z \right)^{-0.758} \label{eq.CP0def} \\
\beta(M_{200,c},z) & = & C_{\beta} \, \times \, 4.35 \left( \frac{M_{200,c}}{10^{14} \,\, M_{\odot}} \right)^{0.0393} \left( 1+z \right)^{0.415} \label{eq.Cbetadef} \,.
\eeqn
These parameters thus describe multiplicative overall changes to the amplitudes of the $P_0$ and $\beta$ parameters.  The fiducial Battaglia profile corresponds to $\left\{ C_{P_0}, C_{\beta} \right\} = \left\{1,1\right\}$.  We discuss our priors for these parameters in Section~\ref{sec:forecast}.


\subsection{Parameter Dependences}
\label{sec:params}

Including both cosmological and astrophysical parameters, our model is specified by the following quantities:
\beq
\label{eq.paramslist}
\left\{ \Omega_b h^2, \Omega_c h^2, \Omega_{\Lambda}, \sigma_8, n_s, C_{P_0}, C_{\beta}, (f_{\mathrm{NL}}, M_{\nu}) \right\} \,,
\eeq
which take the following values in our (WMAP9+BAO+$H_0$~\cite{Hinshawetal2012}) fiducial model:
\beq
\label{eq.paramsfid}
\left\{ 0.02240, 0.1146, 0.7181, 0.817, 0.9646, 1.0, 1.0, (0.0, 0.0) \right\} \,.
\eeq
As a reminder, the $\Lambda$CDM parameters are (in order of their appearance in Eq.~(\ref{eq.paramslist})) the physical baryon density, the physical cold dark matter density, the vacuum energy density, the rms matter density fluctuation on comoving scales of $8 \,\, \mathrm{Mpc}/h$ at $z=0$, and the scalar spectral index.  The ICM physics parameters $C_{P_0}$ and $C_{\beta}$ are defined in Eqs.~(\ref{eq.CP0def}) and (\ref{eq.Cbetadef}), respectively, \fnl is defined by Eq.~(\ref{eq.fNLlocmodel}), and \Mnu is the sum of the neutrino masses in units of eV.  We have placed \fnl and \Mnu in parentheses in Eq.~(\ref{eq.paramslist}) in order to make it clear that we only consider scenarios in which these parameters are varied separately: for all cosmologies that we consider with \fnl$\neq 0$, we set \Mnu$=0$, and for all cosmologies that we consider with \Mnu$>0$, we set \fnl$=0$.  In other words, we only investigate one-parameter extensions of the $\Lambda$CDM concordance model.

For the primary $\Lambda$CDM cosmological parameters, we use the parametrization adopted by the WMAP team (e.g., \cite{Hinshawetal2012}), as the primordial CMB data best constrain this set.  The only exception to this convention is our use of $\sigma_8$, which stands in place of the primordial amplitude of scalar perturbations, $A_s$.  We use $\sigma_8$ both because it is conventional in the tSZ power spectrum literature and because it is a direct measure of the low-redshift amplitude of matter density perturbations, which is physically related more closely to the tSZ signal than $A_s$.  However, this choice leads to slightly counterintuitive results when considering cosmologies with \Mnu$>0$, because in order to keep $\sigma_8$ fixed for such scenarios we must increase $A_s$ (to compensate for the suppression induced by \Mnu in the matter power spectrum).

For the fiducial model specified by the values in Eq.~(\ref{eq.paramsfid}), we find that the tSZ power spectrum amplitude at $\ell=3000$ is $\ell (\ell+1) C_{\ell}^y /2\pi = 7.21 \,\, \mu\mathrm{K}^2$ at $\nu = 150$ GHz.  This corresponds to $7.59 \,\, \mu\mathrm{K}^2$ at $\nu = 148$ GHz (the relevant ACT frequency) and $6.66 \,\, \mu\mathrm{K}^2$ at $\nu = 152.9$ GHz (the relevant SPT frequency).  The most recent measurements from ACT and SPT find corresponding constraints at these frequencies of $3.4 \pm 1.4 \,\, \mu\mathrm{K}^2$~\cite{Sieversetal2013} and $3.09 \pm 0.60 \,\, \mu\mathrm{K}^2$~\cite{Crawfordetal2013} (using their more conservative error estimate).  Note that the SPT constraint includes information from the tSZ bispectrum, which reduces the error by a factor of $\sim$2.  Although it appears that our fiducial model predicts a level of tSZ power too high to be consistent with these observations, the results are highly dependent on the true value of $\sigma_8$, due to the steep dependence of the tSZ power spectrum on this parameter.  For example, recomputing our model predictions for $\sigma_8 = 0.79$ gives $5.52 \,\, \mu\mathrm{K}^2$ at $\nu = 148$ GHz and $4.84 \,\, \mu\mathrm{K}^2$ at $\nu = 152.9$ GHz, which are consistent at $3\sigma$ with the corresponding ACT and SPT constraints.  Given that $\sigma_8 = 0.79$ is within the $2\sigma$ error bar for WMAP9~\cite{Hinshawetal2012}, it is difficult to assess the extent to which our fiducial model may be discrepant with the ACT and SPT results.  The difference can easily be explained by small changes in $\sigma_8$ and is also sensitive to variations in the ICM physics, which we have kept fixed in these calculations.  We conclude that our model is not in significant tension with current tSZ measurements (or other cosmological parameter constraints), and is thus a reasonable fiducial case around which to consider variations.

Figs.~\ref{fig.fNL1h2h} and~\ref{fig.Mnu1h2h} show the tSZ power spectra for our fiducial model and several variations around it, including the individual contributions of the one- and two-halo terms.  In the fiducial case, the two-halo term is essentially negligible for $\ell \gtrsim 300$, as found by earlier studies~\cite{Komatsu-Kitayama1999}, and it only overtakes the one-halo term at very low-$\ell$ ($\ell \lesssim 4$).  However, for \fnl$\neq 0$, the influence of the two-halo term is greatly enhanced due to the scale-dependent bias described in Section~\ref{sec:bias}, which leads to a characteristic upturn in the tSZ power spectrum at low-$\ell$.  In addition, \fnl induces an overall amplitude change in both the one- and two-halo terms due to its effect on the halo mass function described in Section~\ref{sec:HMFNG}.  While this amplitude change is degenerate with the effects of other parameters on the tSZ power spectrum (e.g., $\sigma_8$), the low-$\ell$ upturn caused by the scale-dependent bias is a unique signature of primordial non-Gaussianity, which motivates our assessment of forecasts on \fnl using this observable later in the paper.

Fig.~\ref{fig.Mnu1h2h} shows the results of similar calculations for \Mnu$>0$.  In this case, the effect is simply an overall amplitude shift in the one- and two-halo terms, and hence the total tSZ power spectrum.  The amplitude shift is caused by the change in the halo mass function described in Section~\ref{sec:HMFMnu}.  Note that the sign of the amplitude change is somewhat counterintuitive, but arises due to our choice of $\sigma_8$ as a fundamental parameter instead of $A_s$, as mentioned above.  In order to keep $\sigma_8$ fixed while increasing \Mnu, we must increase $A_s$, which leads to an increase in the tSZ power spectrum amplitude.  Although this effect is degenerate with that of $\sigma_8$ and other parameters, the change in the tSZ power spectrum amplitude is rather large even for small neutrino masses ($\approx 12$\% for \Mnu$=0.1$ eV, which is larger than the amplitude change caused by \fnl$=100$).  This sensitivity suggests that the tSZ power spectrum may be a useful observable for constraints on the neutrino mass sum.

\begin{figure}
\begin{minipage}[b]{0.495\linewidth}
\centering
\includegraphics[width=\textwidth]{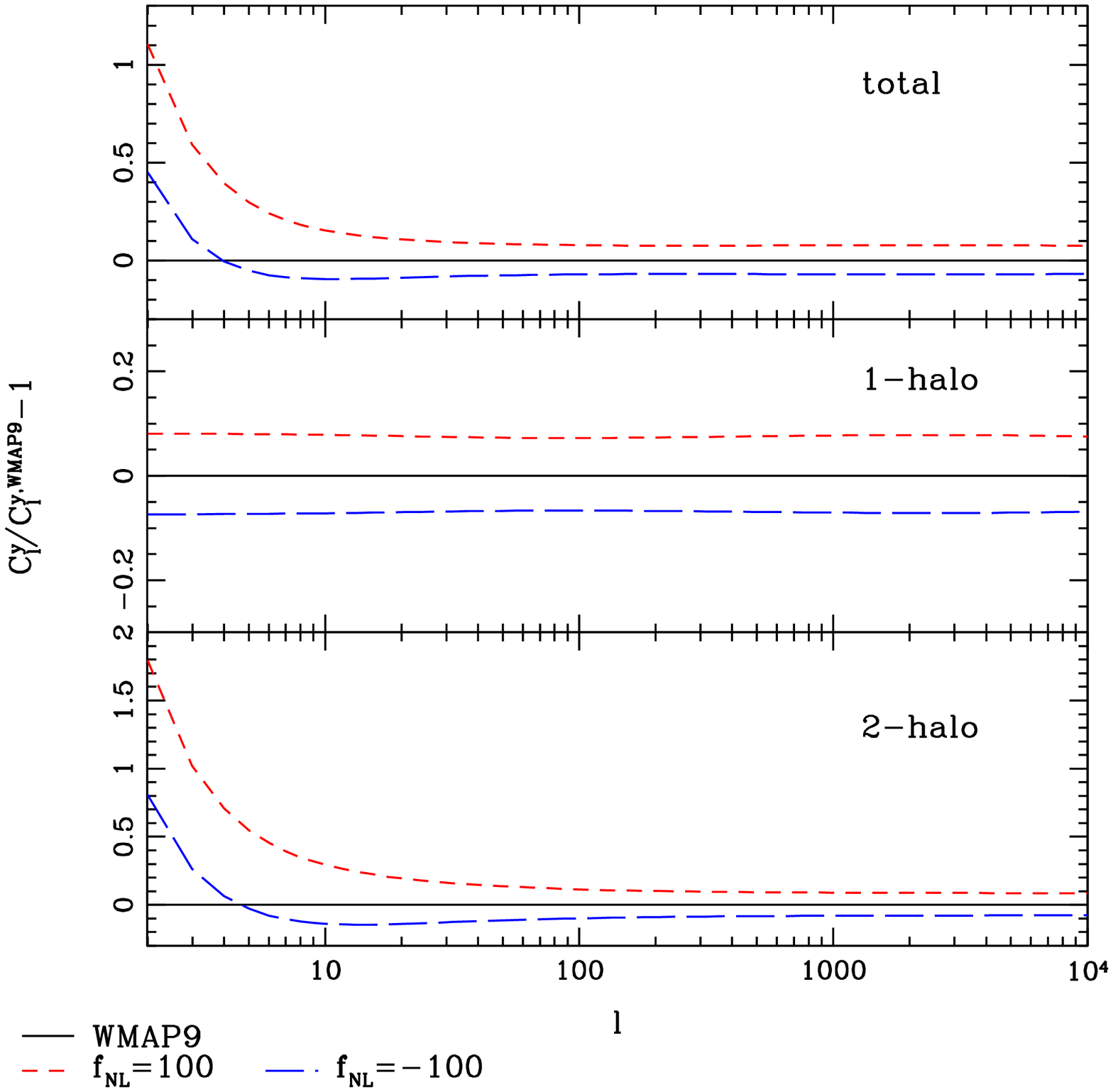}
\caption{The fractional difference between the tSZ power spectrum computed using our fiducial model and power spectra computed for \fnl$=\pm 100$.\label{fig.fNLparamdep}}
\end{minipage}
\begin{minipage}[b]{0.495\linewidth}
\centering
\includegraphics[width=\textwidth]{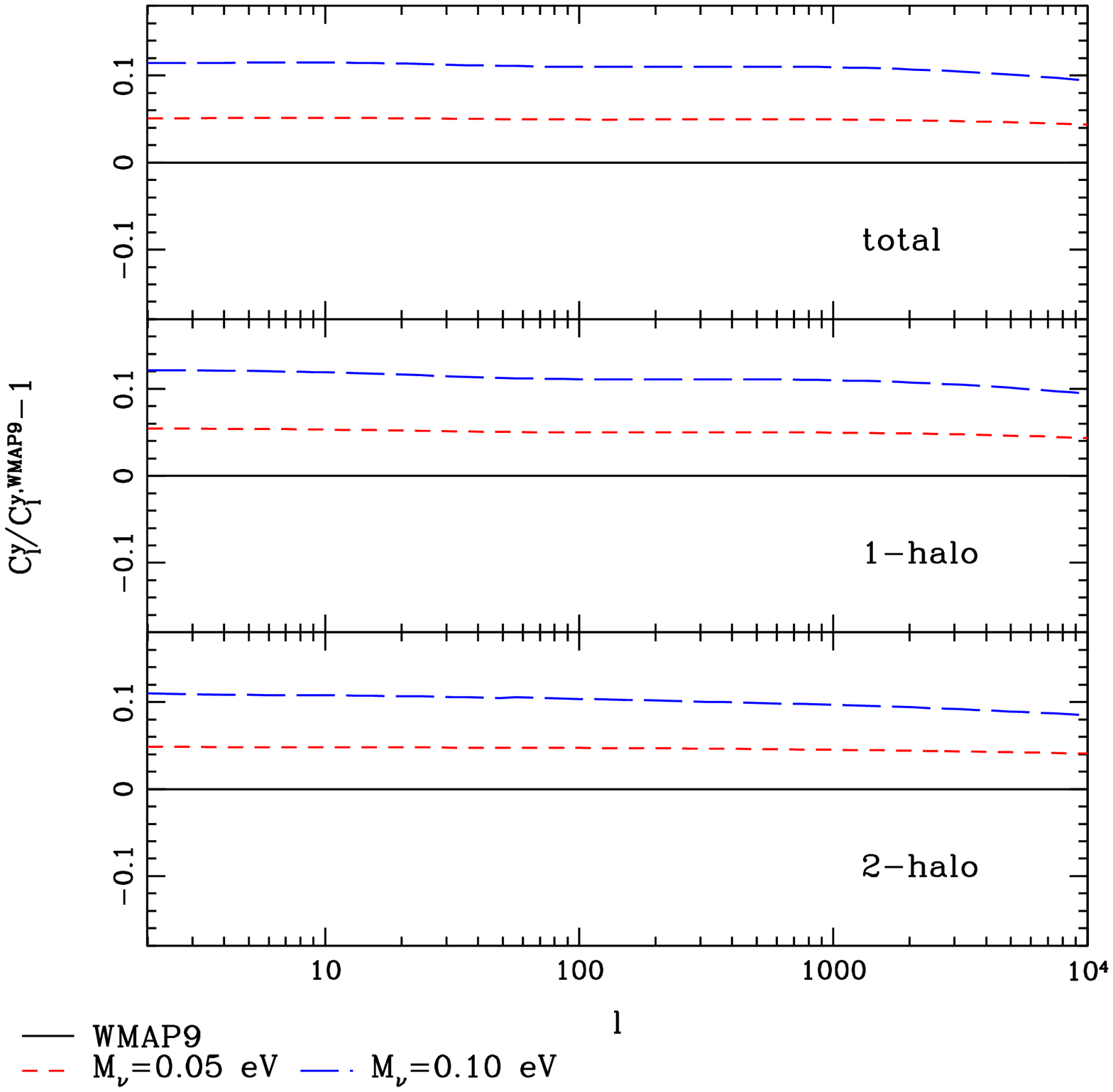}
\caption{The fractional difference between the tSZ power spectrum computed using our fiducial model and power spectra computed for \Mnu$=0.05$ eV and 0.10 eV. \label{fig.Mnuparamdep}}
\end{minipage}
\end{figure}

\begin{figure}
\begin{minipage}[b]{0.495\linewidth}
\centering
\includegraphics[width=\textwidth]{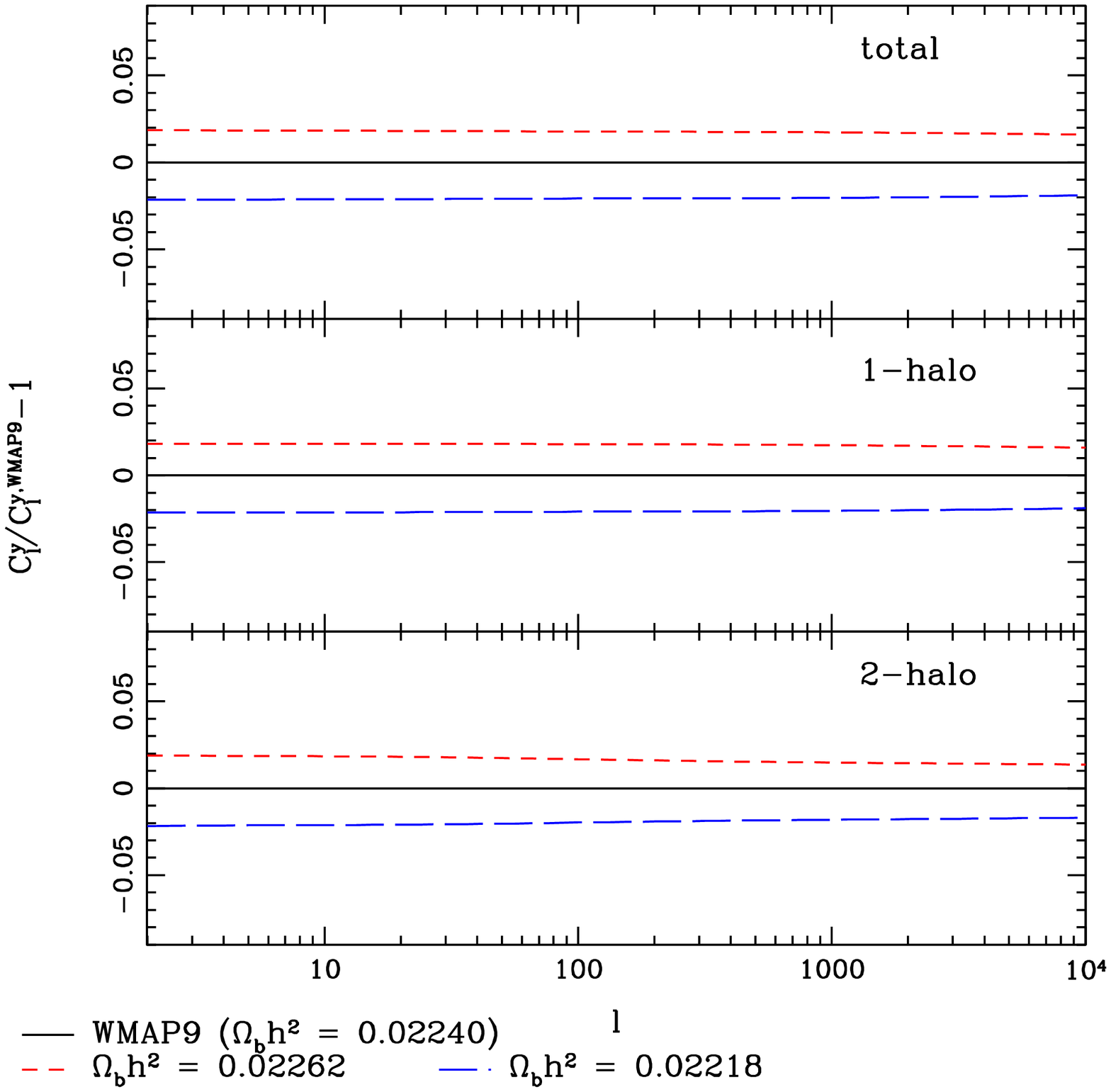}
\caption{The fractional difference between the tSZ power spectrum computed using our fiducial model and power spectra computed for $\Omega_b h^2 = 0.02262$ and 0.02218. \label{fig.ombh2paramdep}}
\end{minipage}
\begin{minipage}[b]{0.495\linewidth}
\centering
\includegraphics[width=\textwidth]{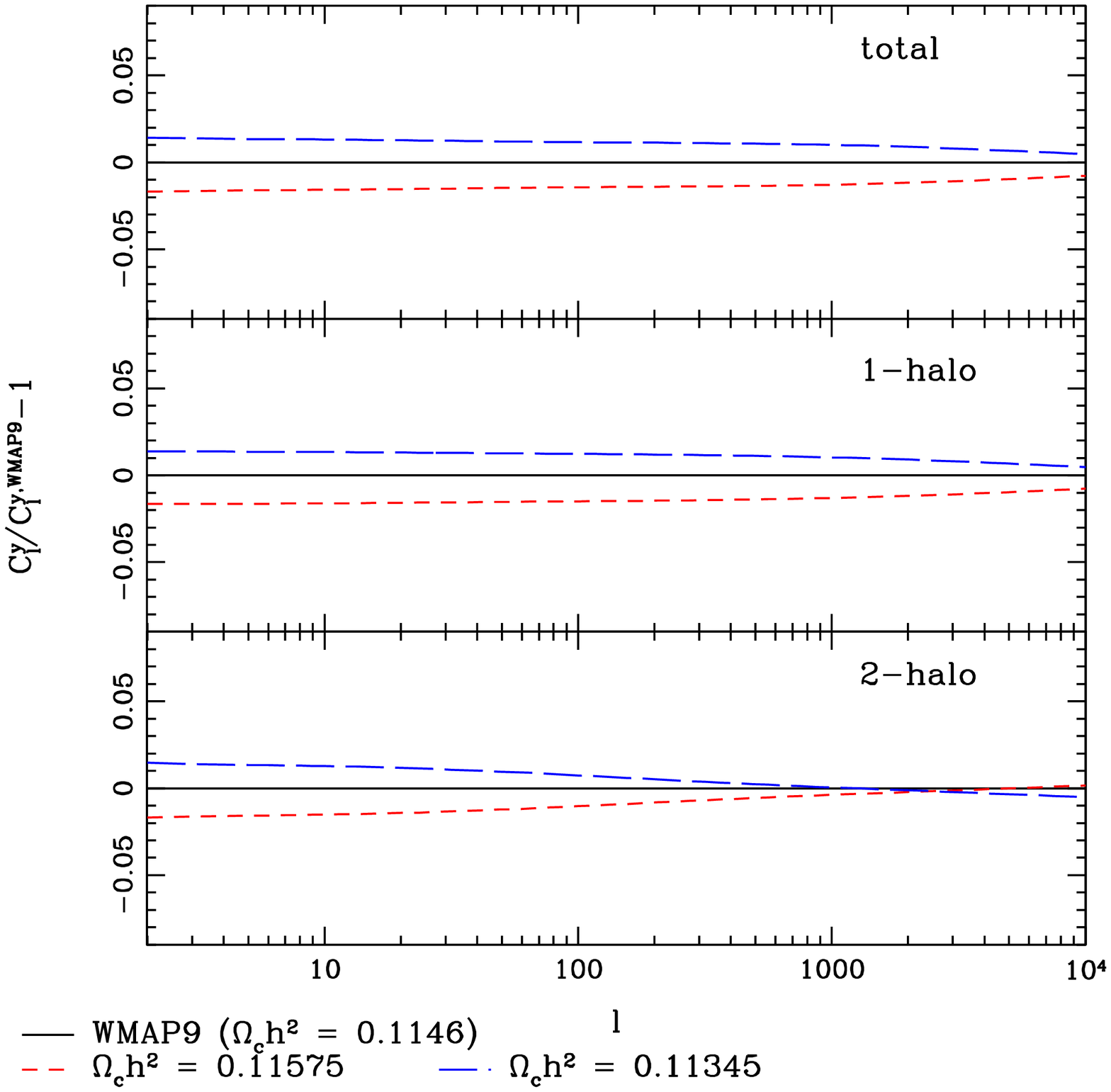}
\caption{The fractional difference between the tSZ power spectrum computed using our fiducial model and power spectra computed for $\Omega_c h^2 = 0.11575$ and 0.11345. \label{fig.omch2paramdep}}
\end{minipage}
\end{figure}

\begin{figure}
\begin{minipage}[b]{0.495\linewidth}
\centering
\includegraphics[width=\textwidth]{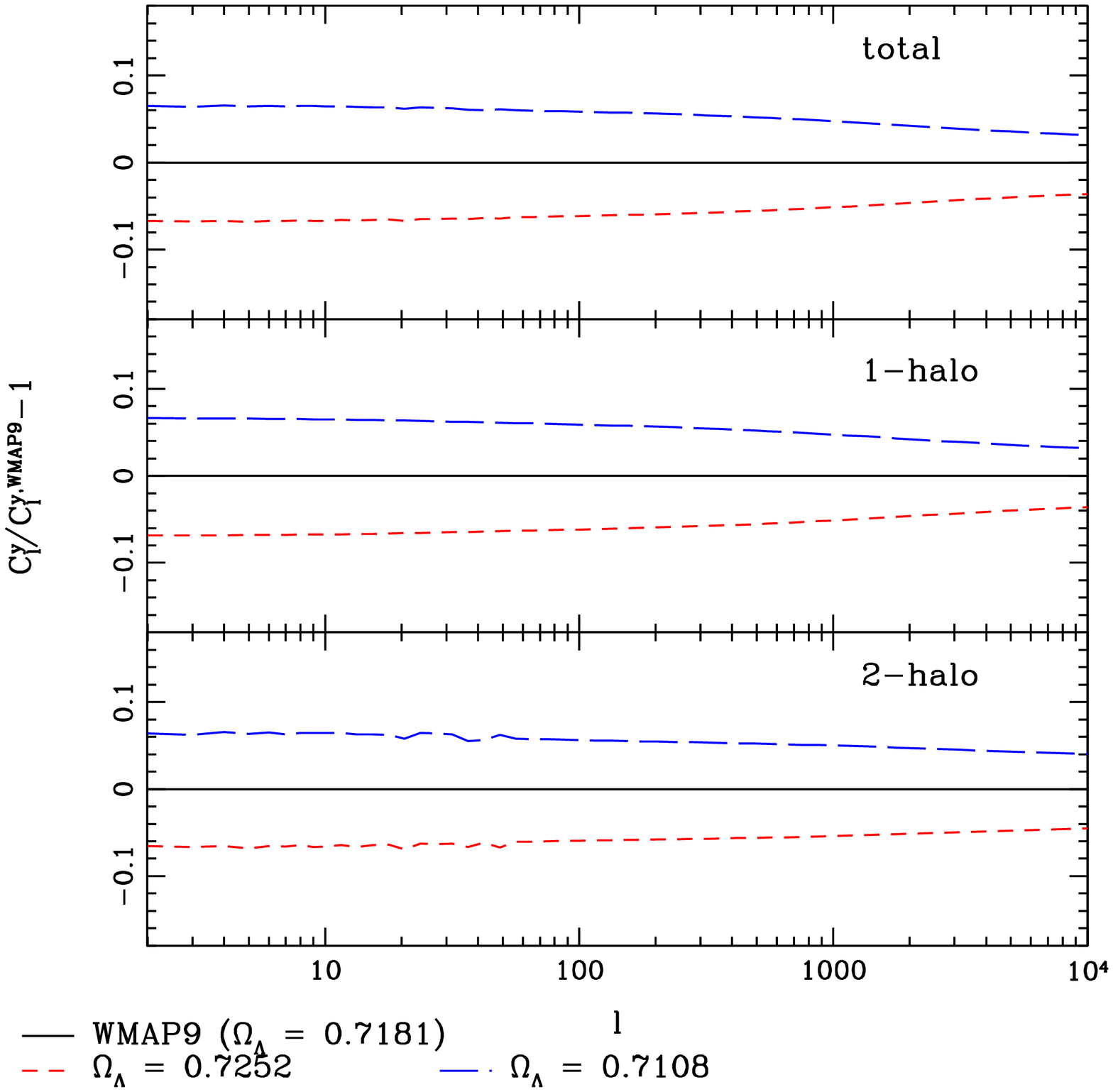}
\caption{The fractional difference between the tSZ power spectrum computed using our fiducial model and power spectra computed for $\Omega_{\Lambda} = 0.7252$ and 0.7108. \label{fig.omLparamdep}}
\end{minipage}
\begin{minipage}[b]{0.495\linewidth}
\centering
\includegraphics[width=\textwidth]{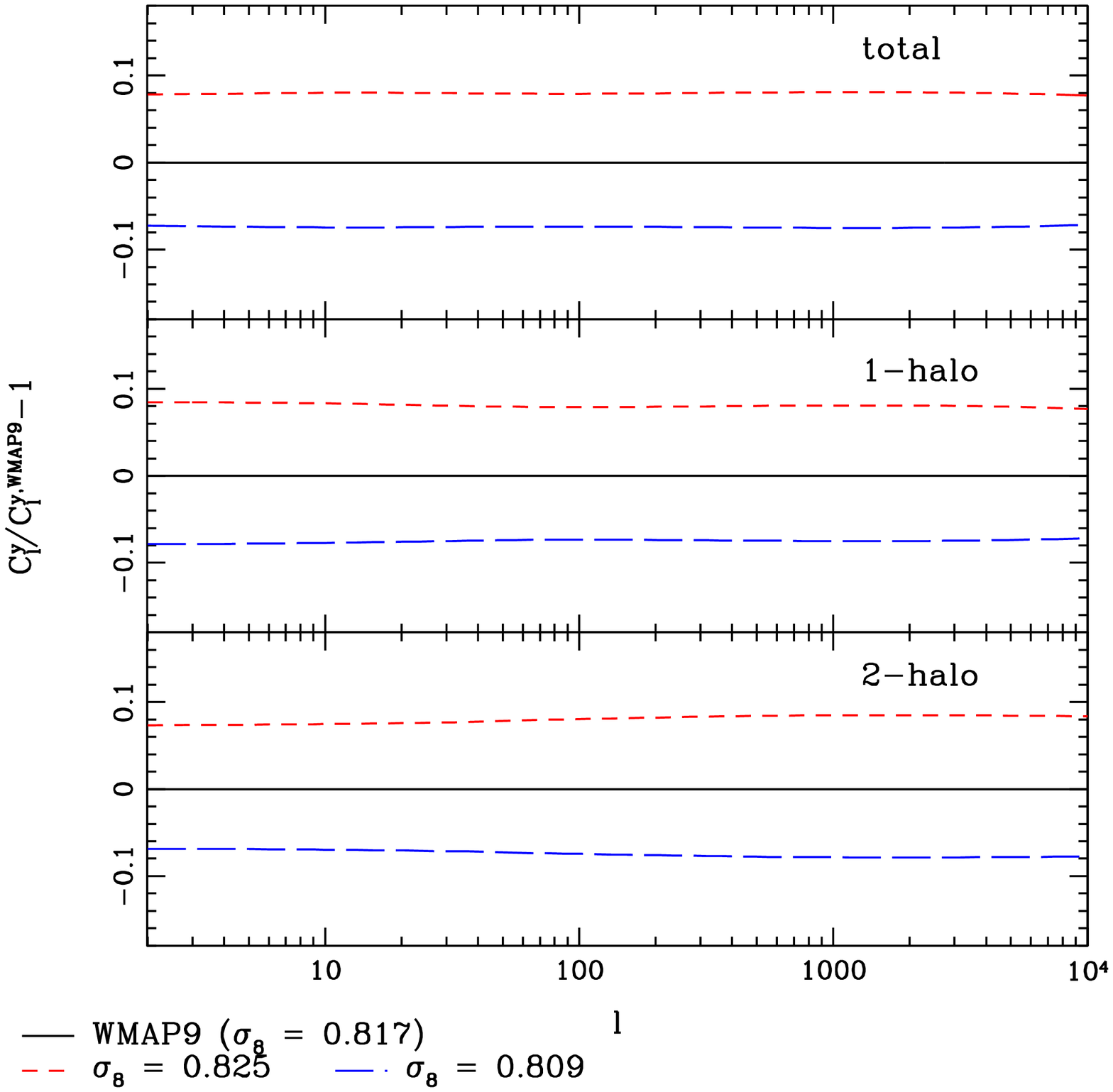}
\caption{The fractional difference between the tSZ power spectrum computed using our fiducial model and power spectra computed for $\sigma_8 = 0.825$ and 0.809. \label{fig.sig8paramdep}}
\end{minipage}
\end{figure}

\begin{figure}
\begin{minipage}[b]{0.495\linewidth}
\centering
\includegraphics[width=\textwidth]{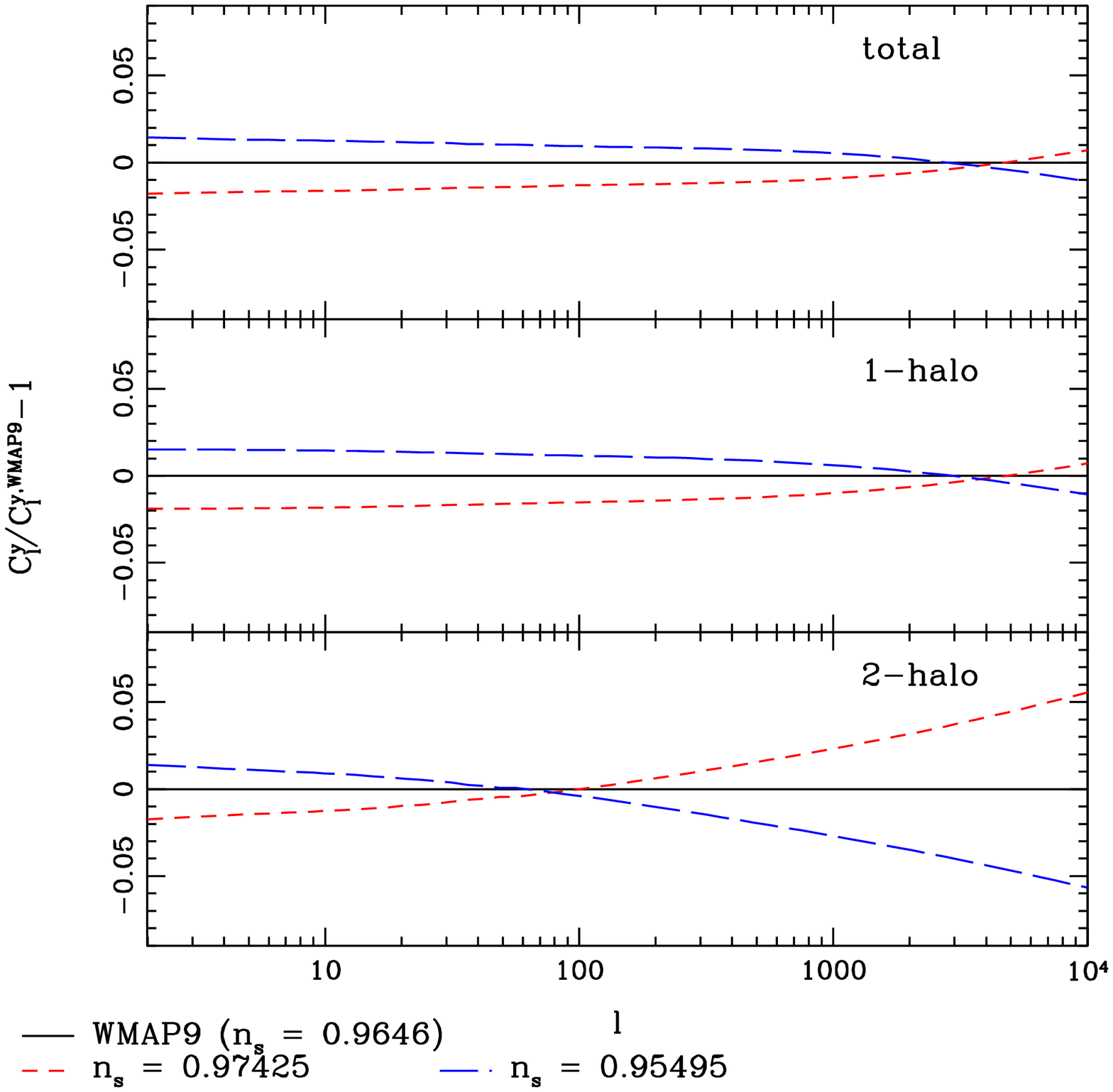}
\caption{The fractional difference between the tSZ power spectrum computed using our fiducial model and power spectra computed for $n_s = 0.97425$ and 0.95495. \label{fig.nsparamdep}}
\end{minipage}
\begin{minipage}[b]{0.495\linewidth}
\centering
\includegraphics[width=\textwidth]{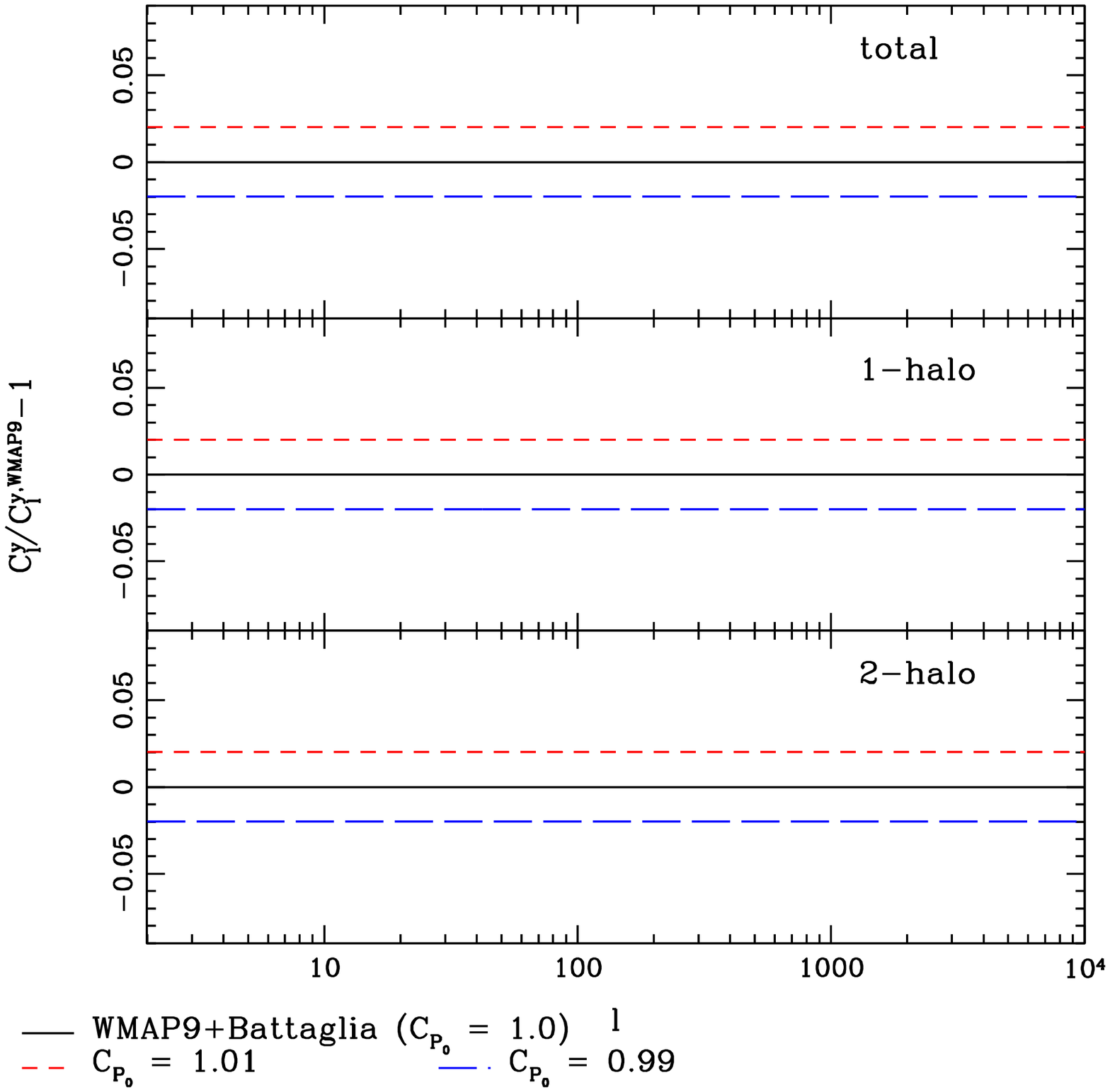}
\caption{The fractional difference between the tSZ power spectrum computed using our fiducial model and power spectra computed for $C_{P_0} = 1.01$ and 0.99. \label{fig.P0paramdep}}
\end{minipage}
\end{figure}

\begin{figure}
\begin{minipage}[b]{0.495\linewidth}
\centering
\includegraphics[width=\textwidth]{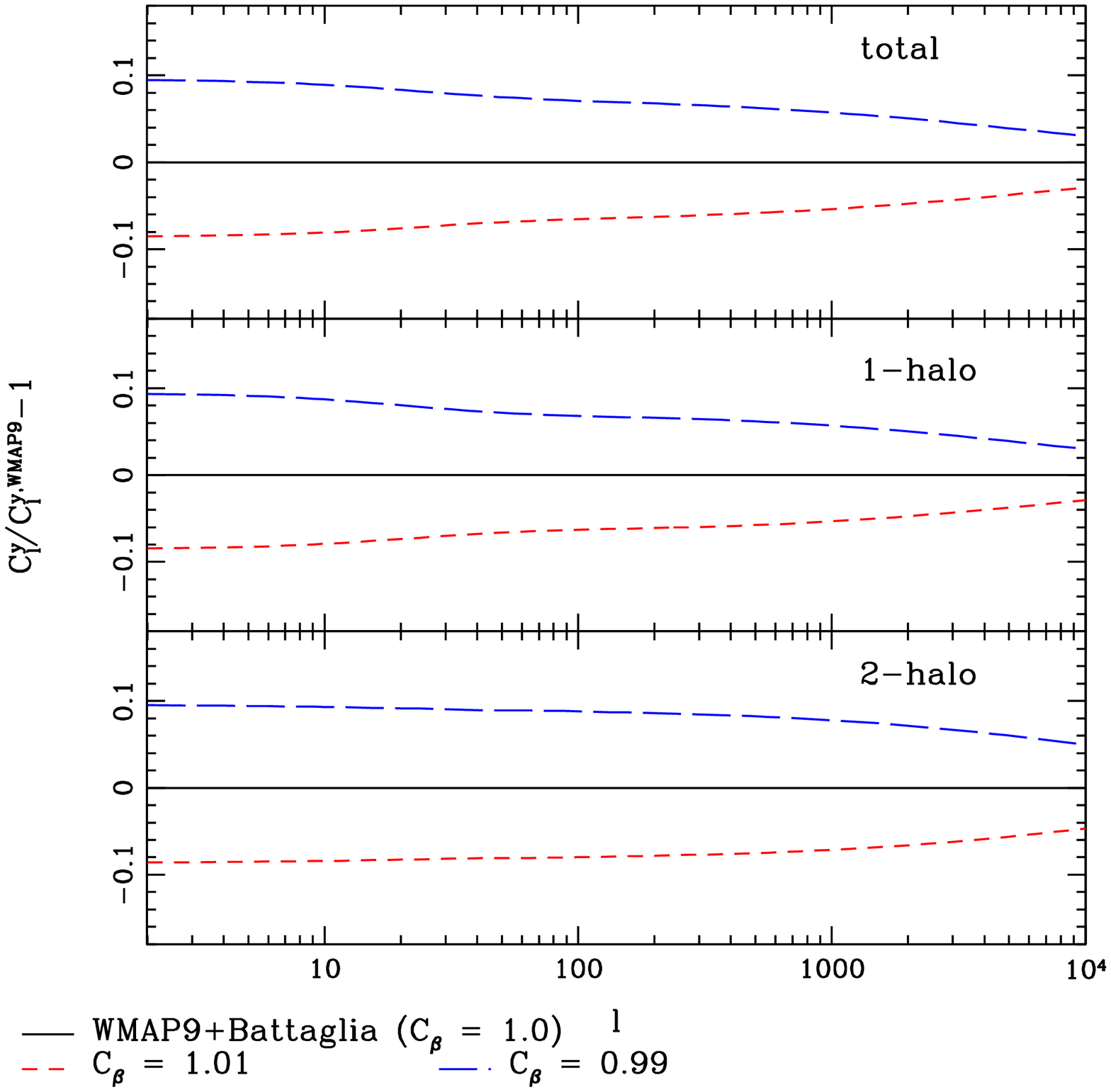}
\caption{The fractional difference between the tSZ power spectrum computed using our fiducial model and power spectra computed for $C_{\beta} = 1.01$ and 0.99. \label{fig.betaparamdep}}
\end{minipage}
\end{figure}

We demonstrate the physical effects of each parameter in our model on the tSZ power spectrum in Figs.~\ref{fig.fNLparamdep}--\ref{fig.betaparamdep}, including the effects on both the one- and two-halo terms individually.  Note that the limits on the vertical axis in each plot differ, so care must be taken in assessing the amplitude of the change caused by each parameter.  Except for \fnl and \Mnu, the figures show $\pm 1$\% variations in each of the parameters, which facilitates easier comparisons between their relative influences on the tSZ power spectrum.  On large angular scales ($\ell \lesssim 100$), the most important parameters (neglecting \fnl and \Mnu) are $\sigma_8$, $\Omega_{\Lambda}$, and $C_{\beta}$.  On very large angular scales ($\ell < 10$), the effect of \fnl is highly significant, but its relative importance is difficult to assess, since the true value of \fnl may be unmeasurably small.  Note, however, that \Mnu is important over the entire $\ell$ range we consider, even if its true value is as small as $0.1$ eV.  Comparison of Figs.~\ref{fig.Mnuparamdep} and~\ref{fig.sig8paramdep} indicates that the amplitude change induced by \Mnu$=0.1$ eV (for fixed $\sigma_8$) is actually slightly larger than that caused by a $1$\% change in $\sigma_8$ around its fiducial value.

We now provide physical interpretations of the effects shown in Figs.~\ref{fig.fNLparamdep}--\ref{fig.betaparamdep}:
\begin{itemize}
\item $f_{\mathrm{NL}}$ (\ref{fig.fNLparamdep}): The change to the halo mass function discussed in Section~\ref{sec:HMFNG} leads to an overall increase (decrease) in the amplitude of the one-halo term for \fnl$>0$ ($<0$).  This increase or decrease is essentially $\ell$-independent, is also seen at $\ell>100$ in the two-halo term, and is $\simeq \pm 10$\% for \fnl$=\pm 100$.  More significantly, the influence of the scale-dependent halo bias induced by \fnl$\neq 0$ is clearly seen in the dramatic increase of the two-halo term at low-$\ell$.  This increase is significant enough to be seen in the total power spectrum despite the typical smallness of the two-halo term relative to the one-halo term for a Gaussian cosmology.
\item $M_{\nu}$ (\ref{fig.Mnuparamdep}): The presence of massive neutrinos leads to a decrease in the number of galaxy clusters at late times, as discussed in Section~\ref{sec:HMFMnu}.  This decrease would lead one to expect a corresponding decrease in the amplitude of the tSZ signal, but Fig.~\ref{fig.Mnuparamdep} shows an increase.  This increase is a result of our choice of parameters --- we hold $\sigma_8$ constant while increasing \Mnu, which means that we must simultaneously increase $A_s$, the initial amplitude of scalar fluctuations.  This increase in $A_s$ (for fixed $\sigma_8$) leads to the increase in the tSZ power spectrum amplitude seen in Fig.~\ref{fig.Mnuparamdep}.  The effect appears to be essentially $\ell$-independent, although it tapers off slightly at very high-$\ell$.
\item $\Omega_b h^2$ (\ref{fig.ombh2paramdep}): Increasing (decreasing) the amount of baryons in the universe leads to a corresponding increase (decrease) in the amount of gas in galaxy clusters, and thus a straightforward overall amplitude shift in the tSZ power spectrum (which goes like $f_{gas}^2$).
\item $\Omega_c h^2$ (\ref{fig.omch2paramdep}): In principle, one would expect that changing $\Omega_c h^2$ should change the tSZ power spectrum, but it turns out to have very little effect, as pointed out in~\cite{Komatsu-Seljak2002}, who argue that the effect of increasing (decreasing) $\Omega_c h^2$ on the halo mass function is cancelled in the tSZ power spectrum by the associated decrease (increase) in the comoving volume to a given redshift.   We suspect that the small increase (decrease) seen in Fig.~(\ref{fig.omch2paramdep}) when decreasing (increasing) $\Omega_c h^2$ is due to the fact that we hold $\Omega_{\Lambda}$ constant when varying $\Omega_c h^2$.  Thus, $\Omega_m \equiv 1 - \Omega_{\Lambda}$ is also held constant, and thus $\Omega_b = \Omega_m - \Omega_c$ is decreased (increased) when $\Omega_c$ is increased (decreased).  This decrease (increase) in the baryon fraction leads to a corresponding decrease (increase) in the tSZ power spectrum amplitude, as discussed in the previous item.  The slight $\ell$-dependence of the $\Omega_c h^2$ variations may be due to the associated change in $h$ required to keep $\Omega_m$ constant, which leads to a change in the angular diameter distance to each cluster, and hence a change in the angular scale associated with a given physical scale.  Increasing (decreasing) $\Omega_c h^2$ requires increasing (decreasing) $h$ in order to leave $\Omega_m$ unchanged, which decreases (increases) the distance to each cluster, shifting a given physical scale in the spectrum to lower (higher) multipoles.  However, it is hard to completely disentangle all of the effects described here, and in any case the overall influence of $\Omega_c h^2$ is quite small.
\item $\Omega_{\Lambda}$ (\ref{fig.omLparamdep}): An increase (decrease) in $\Omega_{\Lambda}$ has several effects which all tend to decrease (increase) the amplitude of the tSZ power spectrum.  First, $\Omega_m$ is decreased (increased), which leads to fewer (more) halos, although this effect is compensated by the change in the comoving volume as described above.  Second, for fixed $\Omega_b/\Omega_c$, this decrease (increase) in $\Omega_m$ leads to fewer (more) baryons in clusters, and thus less (more) tSZ power.  Third, more (less) vacuum energy leads to more (less) suppression of late-time structure formation due to the decaying of gravitational potentials, and thus less (more) tSZ power.  All of these effects combine coherently to produce the fairly large changes caused by $\Omega_{\Lambda}$ seen in Fig.~\ref{fig.omLparamdep}.  The slight $\ell$-dependence may be due to the associated change in $h$ required to keep $\Omega_c h^2$ and $\Omega_b h^2$ constant, similar (though in the opposite direction) to that discussed in the $\Omega_c h^2$ case above.  Regardless, this effect is clearly subdominant to the amplitude shift caused by $\Omega_{\Lambda}$, which is only slightly smaller on large angular scales than that caused by $\sigma_8$ (for a $1$\% change in either parameter).	
\item $\sigma_8$ (\ref{fig.sig8paramdep}): Increasing (decreasing) $\sigma_8$ leads to a significant overall increase (decrease) in the amplitude of the tSZ power spectrum, as has been known for many years~(e.g.~\cite{Komatsu-Seljak2002}).  The effect is essentially $\ell$-independent and appears in both the one- and two-halo terms.
\item $n_s$ (\ref{fig.nsparamdep}): An increase (decrease) in $n_s$ leads to more (less) power in the primordial spectrum at wavenumbers above (below) the pivot, which we set at the WMAP value $k_{piv} = 0.002 \,\, \mathrm{Mpc}^{-1}$ (no $h$).  Since the halo mass function on cluster scales probes much smaller scales than the pivot (i.e., much higher wavenumbers $k \sim 0.1-1 \,\, h/\mathrm{Mpc}$), an increase (decrease) in $n_s$ should lead to more (fewer) halos at late times.  However, since we require $\sigma_8$ to remain constant while increasing (decreasing) $n_s$, we must decrease (increase) $A_s$ in order to compensate for the change in power on small scales.  This is similar to the situation for \Mnu described above.  Thus, an increase (decrease) in $n_s$ actually leads to a small decrease (increase) in the tSZ power spectrum on most scales, at least for the one-halo term.  The cross-over in the two-halo term is likely related to the pivot scale after it is weighted by the kernel in Eq.~(\ref{eq.yCl2hLimberquote}), but this is somewhat non-trivial to estimate.  Regardless, the overall effect of $n_s$ on the tSZ power spectrum is quite small.
\item $C_{P_0}$ (\ref{fig.P0paramdep}): Since $C_{P_0}$ sets the overall normalization of the ICM pressure profile (or, equivalently, the zero-point of the $Y-M$ relation), the tSZ power spectrum simply goes like $C_{P_0}^2$.
\item $C_{\beta}$ (\ref{fig.betaparamdep}): Since $C_{\beta}$ sets the logarithmic slope of the ICM pressure profile at large radii (see Eq.~(\ref{eq.GNFW})), it significantly influences the total integrated thermal energy of each cluster, and thus the large-angular-scale behavior of the tSZ power spectrum.  An increase (decrease) in $C_{\beta}$ leads to a decrease (increase) in the pressure profile at large radii, and therefore a corresponding decrease (increase) in the tSZ power spectrum on angular scales corresponding to the cluster outskirts and beyond.  On smaller angular scales, the effect should eventually vanish, since the pressure profile on small scales is determined by the other slope parameters in the pressure profile.  This trend is indeed seen at high-$\ell$ in Fig.~\ref{fig.betaparamdep}.  Note that a $1$\% change in $C_{\beta}$ leads to a much larger change in the tSZ power spectrum at nearly all angular scales than a $1$\% change in $C_{P_0}$, suggesting that simply determining the zero-point of the $Y-M$ relation may not provide sufficient knowledge of the ICM physics to break the long-standing ICM-cosmology degeneracy in tSZ power spectrum measurements.  It appears that constraints on the shape of the pressure profile itself will be necessary.
\end{itemize}

\section{Experimental Considerations}
\label{sec:exp}
\begin{figure}
\includegraphics[width=0.95\textwidth]{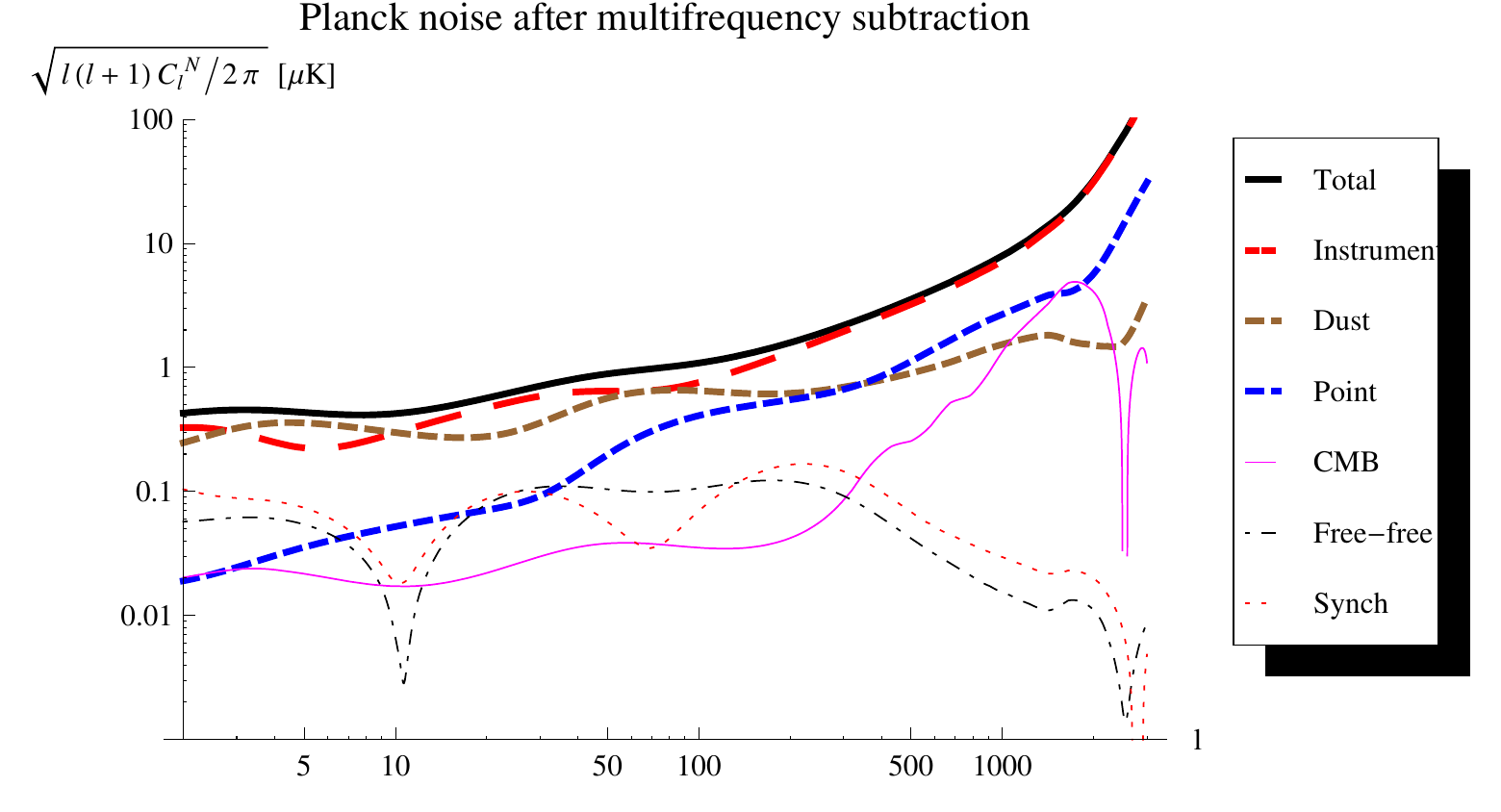}
\includegraphics[width=0.95\textwidth]{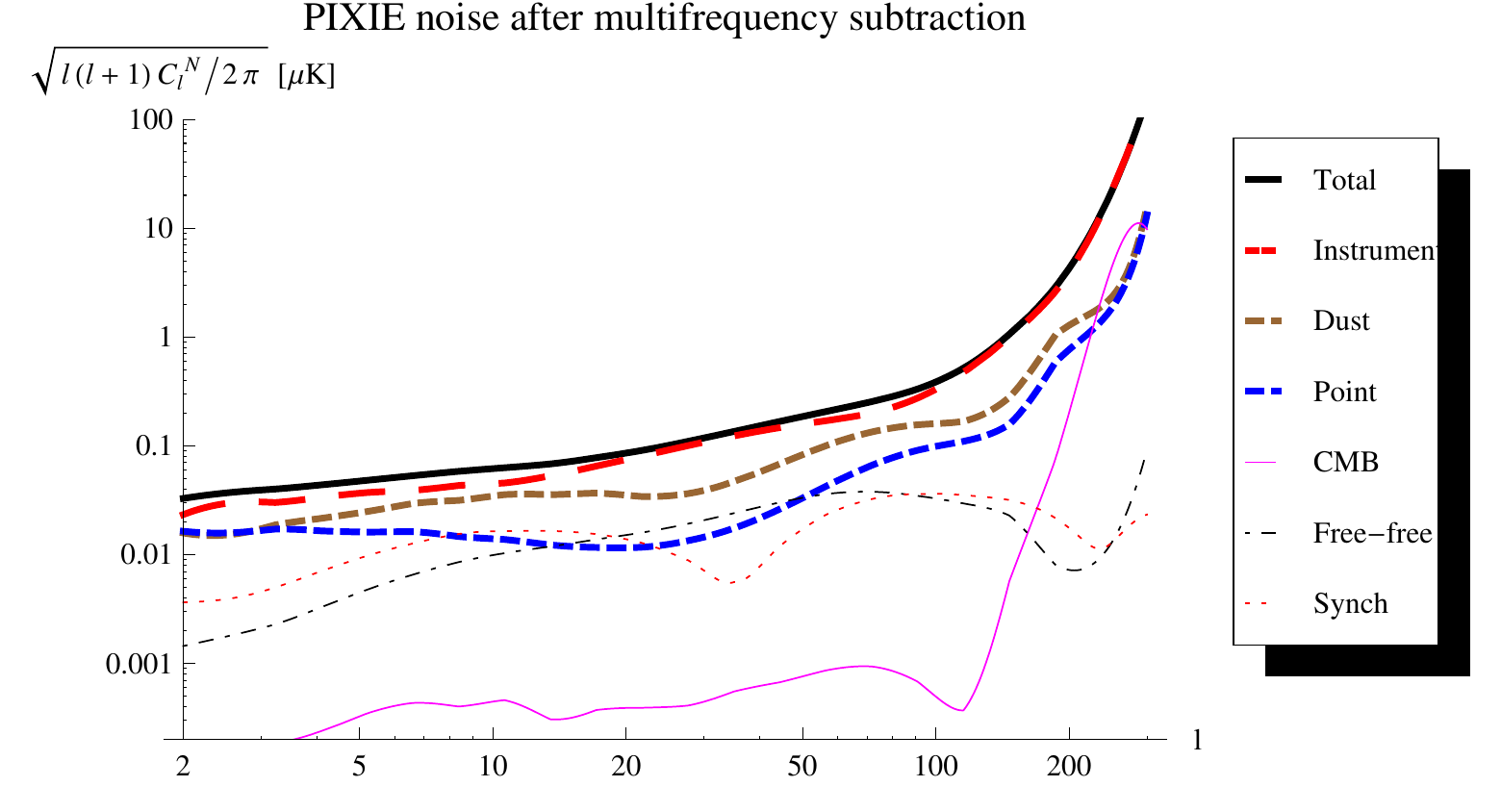}
\caption{The two plots show the various contributions to the total noise per $\ell$ and $m$ after multifrequency subtraction for Planck (top panel) and PIXIE (bottom panel). Because of the many frequency channels both experiments can subtract the various foregrounds and the total noise is not significantly different from the instrumental noise alone.  \label{f:p}}
\end{figure}

In this section we estimate the noise in the measurement of the tSZ power spectrum. The first ingredient is instrumental noise. We describe it for the Planck experiment and for an experiment with the same specifications as the proposed PIXIE satellite~\cite{Kogutetal2011}. The second ingredient is foregrounds\footnote{To be precise we will consider both \textit{foregrounds}, e.g.~from our galaxy, and \textit{backgrounds}, e.g.~the CMB. On the other hand, in order to avoid repeating the cumbersome expression ``foregrounds and backgrounds'' we will collectively refer to all these contributions as foregrounds, sacrificing some semantic precision for the sake of an easier read.}. We try to give a rather complete account of all these signals and study how they can be handled using multifrequency subtraction. Our final results are in Fig.~\ref{f:p}. Because of the several frequency channels, Planck and to a much larger extent PIXIE can remove all foregrounds and have a sensitivity to the tSZ power spectrum mostly determined by instrumental noise.

\subsection{Multifrequency Subtraction}
\label{sec:multifreq}
We discuss and implement multifrequency subtraction\footnote{We are thankful to K.~Smith for pointing us in this direction.} along the lines of \cite{THEO,CHT}. The main idea is to find a particular combination of frequency channels that minimize the variance of some desired signal, in our case the tSZ power spectrum. We hence start from 
\be\label{w}
\aSZ_{\ell m}=\sum_{\nu_{i}} \frac{w_{i} a_{\ell m}(\nu_{i})}{g_{\nu_{i}}}\,,
\ee
where $a^{SZ}$ refers to our estimator for the tSZ signal at 150 GHz (the conversion to a different frequency is straightforward), $\nu_{i}$ are the different frequency channels relevant for a given experiment, $w_{i}$ are the weights for each channel, $a_{\ell m}(\nu_{i})$ are spherical harmonic coefficients of the total measured temperature anisotropies at each frequency and finally $g_{\nu_{i}}$ is the tSZ spectral function defined in Section~\ref{sec:tSZPS}, allowing us to convert from Compton-$y$ to $\Delta T$.  We can decompose the total signal according to $a_{\ell m}=a^{SZ}_{\ell m}+\sum_{f} a_{\ell m}^{f}$ with $f$ enumerating all other contributions. We will assume that $\ex{a_{\ell m}^{f}a_{\ell m}^{f'}}\propto\delta_{ff'}$, i.e.~different contributions are uncorrelated with each other. Dropping for the moment the $\ell$ and $m$ indices, the variance of $\hat a^{SZ}$ is then found to be
\be\label{asq}
\ex{  \aSZ \aSZ}=C^{SZ} \left(\sum_{\nu_{i}} w_{i}\right)^{2}+\sum_{\nu_{i}\nu_{j}} w_{i} w_{j}\sum_{f} \frac{C^{f}(\nu_{i},\nu_{j})}{g_{\nu_{i}} g_{\nu_{j}}}\,,
\ee
where $C^{SZ}=C_{\ell}^{SZ}$ is the tSZ power spectrum at 150 GHz as given in Eq. (\ref{eq.Cell}) and $C^{f}(\nu_{i},\nu_{j})$ (again the $\ell$ index is implicit) is the cross-correlation at different frequencies of the $a_{\ell m}^{f}$ of each foreground component (we will enumerate and describe these contributions shortly). To simplify the notation, in the following we will use
\be
\label{eq.Cij}
C(\nu_{i},\nu_{j})=C_{ij}\equiv\sum_{f} \frac{C^{f}(\nu_{i},\nu_{j})}{g_{\nu_{i}} g_{\nu_{j}}}\,.
\ee
We now want to minimize $\ex{(\aSZ)^{2}}$ with the constraint that the weights describe a unit response to a tSZ signal, i.e., $\sum_{i} w_{i}=1$. This can be done using a Langrange multiplier $\lambda$ and solving the system 
\be
\partial_i \left[\ex{(\aSZ)^{2}}+\lambda \left(\sum_{i} w_{i}-1\right)\right]=\partial_{\lambda} \left[\ex{(\aSZ)^{2}}+\lambda \left(\sum_{i} w_{i}-1\right)\right]=0\,.
\ee 
Because of the constraint $\sum_{i} w_{i}=1$, the $C^{SZ}$ term in $\ex{(\aSZ)^{2}}$ is independent of $w_{i}$ (alternatively one can keep this term and see that it drops out at the end of the computation). Then the solution of the first equation can be written as
\be
w_{i}=-\lambda (C^{-1})_{ij} e_{j}=0\,,
\ee
where $e_{j}=1$ is just a vector with all ones and $(C^{-1})_{ij}$ is the inverse of $C_{ij}$ in Eq.~(\ref{eq.Cij}). This solution can then be plugged back into the constraint $\sum_{i} w_{i}=1$ to give
\be
w_{i}=\frac{(C^{-1})_{ij}e_{j}}{e_{k}(C^{-1})_{kl}e_{l}}\,,
\ee
which is our final solution for the minimum-variance weights. From Eq.~(\ref{asq}) we see that the total noise in each $\aSZ_{\ell m}$ after multifrequency subtraction is
\be
N_{\ell}=w_{i}C_{ij} w_{j}\,,\label{Nf}
\ee
and the partial contributions to $N_{\ell}$ from each foreground can be obtained by substituting $C$ with $C^{f}$ (recall that there is an implicit $\ell$ index on $C_{ij}$). Notice that averaging over all $m$'s for each $\ell$ and assuming that a given experiment covers only a fraction $f_{sky}$ of the sky, the final noise in each $\ell$ is $N_{\ell}/(f_{sky} (2\ell+1))$.

 
\subsection{Foregrounds}
\label{sec:foregrounds}
We will consider the following sources of noise: instrumental noise ($N$), CMB ($CMB$), synchrotron ($Synch$), free-free ($ff$), radio and IR point sources ($Radio$ and $IR$) and thermal dust ($Dust$).  We now discuss each of them in turn.

\begin{table}
\centering
\begin{tabular}{|c|ccccccc cc|} 
\hline	
  $\nu$ [GHz] & 30 & 44 & 70 & 100 & 143 & 217 & 353 & 545 & 857 \tabularnewline
   \hline
FWHM [arcmin]  & 33  & 24 & 14 & 10 & 7.1 & 5.0 & 5.0 & 5.0 & 5.0                       \tabularnewline
 \hline 
$10^{6}\Delta T / T_{CMB}  $   &    2.0 & 2.7 & 4.7 & 2.5 & 2.2 & 4.8 & 14.7 & 147 & 6700 \tabularnewline
 \hline 
 \end{tabular} 
 \caption{For the nine frequency bands for Planck we report the central frequency (in GHz), the Full Width at Half Maximum (FWHM, in arcminutes, to be converted into radians in the noise computation) of each pixel, and the 1$\sigma$ sensitivity to temperature per square pixel~\cite{PlanckBB}.\label{t:Planck} }
 \end{table}
 
We assume that the noise is Gaussian with a covariance matrix diagonal in $l$-space and uncorrelated between different frequencies. Then \cite{Knox}
\beqn
C_{\ell}^{N}(\nu,\nu')=\delta_{\nu\nu'} \Delta T(\nu)^{2} e^{\ell(\ell+1) \theta(\nu)^{2}} \left(8\ln2\right) \theta(\nu)^{2}\,,
\eeqn
where the beam size in radians at each frequency is $\theta(\nu)={\rm FWHM}(\nu) (8\ln 2)^{-1/2} \times \pi / (180\times 60)$. The frequency channels $\nu$, the FWHM$(\nu)$ (Full Width at Half Maximum) and $\Delta T(\nu)$ depend on the experiment. In the following we consider the Planck satellite with specifications given in Table~\ref{t:Planck} and the proposed PIXIE satellite~\cite{Kogutetal2011}. The latter is a fourth generation CMB satellite targeting primordial tensor modes through the polarization of the CMB. PIXIE will cover frequencies between $30$ GHz and 6 THz with an angular resolution of $1^{\circ}.6$ Gaussian FWHM corresponding to $\ell_{max}\equiv \theta^{-1}\simeq 84$. The frequency coverage will be divided into 400 frequency channels each with a typical sensitivity of $\Delta I=4\times 10^{-24}\,\textrm{W}\, \textrm{m}^{-2} \textrm{sr}^{-1} \textrm{Hz}^{-1}$ in each of 49152 sky pixels. In order to get $\Delta T$ we can use Planck's law with respect to CMB temperature
\be\label{I}
I(\nu,T_{CMB})=\frac{2h}{c^{2}}\frac{\nu^{3}}{e^{\nu/ (56.8 GHz)}-1}\quad \Rightarrow \quad \Delta T(\nu)=\left[\frac{\partial I(\nu,T)}{\partial T}\right]^{-1}_{T_{CMB}} \Delta I\,,
\ee
where we used the numerical value of fundamental constants and $T_{CMB}=2.725 \, \mathrm{K}$ to write $h\nu/(k_{B}T_{CMB})=\nu/(56.8 \textrm{GHz})$. For example one finds $\Delta T(150 \textrm{GHz})\simeq 1.00\, \mu $K. 

For all foregrounds except point sources we use the models and parameters discussed in~\cite{THEO}. We assume that different components are uncorrelated and for each component $f$ we define
\be
C^{f}(\nu_{i},\nu_{j})=\frac{\Theta^{f}(\nu_{i})\Theta^{f}(\nu_{j})}{\Theta^{f}(\nu_{0})^{2}}\,R(\nu_{i},\nu_{j})\, C^{f}_{\ell}\,,
\ee
where $\Theta^{f}(\nu)$ encodes the frequency dependence, $C^{f}_{\ell}$ provides the $\ell$-dependence and normalization at some fiducial frequency $\nu_{0}$ (which will be different for different components) and finally $R(\nu_{i},\nu_{j})$ accounts for the frequency coherence. The latter ingredient was used in \cite{THEO, CHT} and discussed in \cite{T98}. The general picture is that the auto-correlation of some contribution $f$ at two different frequencies might not be perfect. Instrumental noise is an extreme case of this in which two different frequency channels have completely uncorrelated noise, i.e. $R(\nu_{i},\nu_{j})=\delta_{ij}$. The CMB sits at the opposite extreme in that it follows a blackbody spectrum to very high accuracy, hence being perfectly coherent between any two frequencies: $R(\nu_{i},\nu_{j})=1$ for any $i,j$. All other foregrounds lie in between these two extrema, having an $R(\nu_{i},\nu_{j})$ that starts at unity for $i=j$ and goes to zero as the frequencies are taken apart from each other. To model this Tegmark \cite{T98} proposed using
\be\label{fdeco}
R(\nu_{i},\nu_{j})=\exp \left\{- \frac{1}{2} \left[\frac{\log \left(\nu_{i}/\nu_{j}\right)}{\xi^{f}}\right]^{2}\right\}\,,
\ee
where $\xi^{f}$ depends on the foreground and can be estimated as $\xi^{f}\sim \left(\sqrt{2} \Delta \alpha\right)^{-1}$ with $\Delta \alpha$ being the variance across the sky of the spectral index of that particular component $f$. In the following we will write the frequency covariance as $R(\nu_{i},\nu_{j},\Delta \alpha)$, e.g. for the we CMB we will have $R(\nu_{i},\nu_{j},0)$ while for instrumental noise $R(\nu_{i},\nu_{j},\infty)$.

We will parameterize the frequency dependences of the various components as
\be
\Theta^{CMB}(\nu)&=& 1 \,,\\
\Theta^{ff}(\nu)&=& \nu^{-2.15} c(\nu) \,,\\
\Theta^{Dust}(\nu)&=& \frac{c(\nu) \tilde c(\nu) \nu^{3+1.7}}{e^{\frac{\nu}{56.8 \textrm{GHz}} \frac{2.725 K}{18 K}}-1} \,,\\
\Theta^{Synch}(\nu)&=& \nu^{-2.8} c(\nu)  \,,\\
\Theta^{Radio}(\nu)&=& \nu^{-0.5}\left[\frac{\partial I(\nu,T)}{\partial T}\right]^{-1}_{T_{CMB}} \,,\\
\Theta^{IR}(\nu)&=& \nu^{2.1} I(\nu,9.7 K) \left[\frac{\partial I(\nu,T)}{\partial T}\right]^{-1}_{T_{CMB}} \,.
\ee
where we used Eq.~(\ref{I}) and 
\be
c(\nu)\equiv  \left[\frac{2 \sinh \left(x/2\right)}{x}\right]^{2} \,,\quad \tilde c (\nu) \propto \nu^{-2}\,.
\ee
Adding the information about the angular scale dependence we get
\be
C^{CMB}_{\ell}(\nu_{i},\nu_{j})&=& C^{CMB}_{\ell}\,,\\
C^{ff}_{\ell}(\nu_{i},\nu_{j})&=& \frac{\Theta^{ff}(\nu_{i})\Theta^{ff}(\nu_{j})}{\Theta^{ff}(31 \textrm{GHz})^{2}} (70 \mu \textrm{K})^{2} \ell^{-3} R(\nu_{i},\nu_{j},0.02)\,,\\
C^{Dust}_{\ell}(\nu_{i},\nu_{j})&=&\frac{\Theta^{Dust}(\nu_{i})\Theta^{Dust}(\nu_{j})}{\Theta^{Dust}(90 \textrm{GHz})^{2}} (24 \mu \textrm{K})^{2} \ell^{-3} R(\nu_{i},\nu_{j},0.3) \,,\\
C^{Synch}_{\ell}(\nu_{i},\nu_{j})&=& \frac{\Theta^{Synch}(\nu_{i})\Theta^{Synch}(\nu_{j})}{\Theta^{Synch}(19 \textrm{GHz})^{2}} (101 \mu \textrm{K})^{2} \ell^{-2.4} R(\nu_{i},\nu_{j},0.15)\,,\\
C^{Radio}_{\ell}(\nu_{i},\nu_{j})&=&\frac{\Theta^{Radio}(\nu_{i})\Theta^{Radio}(\nu_{j})}{\Theta^{Radio}(31 \textrm{GHz})^{2}} (\sqrt{3} \mu \textrm{K})^{2}\frac{2\pi}{\ell(\ell+1)}\left(\frac{\ell}{3000}\right)^{2} R(\nu_{i},\nu_{j},0.5) \,,\\
C^{IR}_{l}(\nu_{i},\nu_{j})&=&\frac{\Theta^{IR}(\nu_{i})\Theta^{IR}(\nu_{j})}{\Theta^{IR}(31 \textrm{GHz})^{2}} \frac{2\pi}{\ell(\ell+1)} \left[\left(\frac{\ell}{3000} \right)^{2} 7 \mu \textrm{K}^{2}+ \left(\frac{\ell}{3000}\right)^{2-1.2} 5.7 \mu \textrm{K}^{2} \right] R(\nu_{i},\nu_{j},0.3) \,.
\ee
For the CMB, $C_{\ell}^{CMB}$ is obtained using CAMB with the parameters of our fiducial cosmology. The parameters in the free-free, synchrotron and thermal dust components have been taken from the Middle Of the Road values in \cite{THEO} (their Table 2 and text). The parameterization of the IR and Radio point sources follows \cite{ACT}.

 
\subsection{Noise After Multifrequency Subtraction}
\label{sec:finalnoise}
Using the formulae in the last two sections we can estimate what the total variance in $\aSZ$ will be after multifrequency subtraction. We denote the final result by $N_{\ell}$ for the total noise and by $N^{f}_{\ell}$ for each foreground component (see around Eq.~(\ref{Nf})). Then we plot $ \left[\ell(\ell+1) N_{\ell} /2\pi\right]^{1/2}$ and $ \left[\ell(\ell+1) N_{\ell}^{f} /2\pi\right]^{1/2}$ in units of $\mu$K for Planck and PIXIE in Fig.~\ref{f:p}. With this choice we can compare directly with the results of \cite{CHT} and see that they agree for Planck once one accounts for the fact that we are constraining tSZ at 150 GHz while there the tSZ in the Rayleigh-Jeans tail is considered, which brings a factor of about 4 difference in $N_{\ell}$.

The results for PIXIE are new. The proposed PIXIE design features 400 logarithmically-spaced frequency channels. This would require one to work with a very large multifrequency matrix which quickly becomes computationally expensive. Also, since $C$ is very close to a singular matrix, the numerical inversion introduces some unavoidable error that becomes too large for matrices larger than about $35\times 35$. For this reason, we decide to perform the computation binning the initial 400 channel into a smaller more manageable number. As we decrease the number of bins (i.e. bin more and more channels together) we expect two main effects to influence the final result. First, when the number of channels become comparable with the number of foregrounds that we want to subtract, the multifrequency subtraction will become very inefficient. Since we stay well away from this limit of very heavy binning, this is not an issue for us. Second, as we decrease the number of bins, the separation in frequency between adjacent bins grows larger. Because of the frequency decoherence (see discussion around Eq.~(\ref{fdeco})), when the bins are very far apart, they are contaminated by uncorrelated foregrounds and again the subtraction becomes inefficient. For a rough estimate of when this happens we take
\be\label{ine}
 \left[\frac{b}{400}\log \left(\frac{6000 {\rm GHz}}{30 {\rm GHz}}\right) \sqrt{2}\Delta \alpha \right]^{2}\frac{1}{2}\leq 1\,,
\ee 
where $b$ is the number of channels that we put in a bin, $(6000/30)^{1/400}$ is the logarithmic spacing and for $\Delta \alpha$ we take the largest one appearing in the foregrounds, i.e.~$\Delta \alpha\sim .5$ for radio sources (dust and IR point sources have a comparable value). Then one finds that Eq.~(\ref{ine}) starts being violated around $b\sim8$, which is what we take in our analysis. The last point is that if we want to cover the frequency most relevant for tSZ with 35 bins each containing 8 channels, we cannot start from the lowest frequency covered by PIXIE, namely $30$ GHz. We decide instead to start at 45 GHz, since the signal at lower frequencies is swamped anyhow by synchrotron and free-free radiation. Summarizing, we take 35 logarithmically spaced frequency channels between 45 GHZ and 1836 GHz and use them for the multifrequency subtraction. Given the arguments above we do not expect that using more channels will improve the final noise appreciably. The final noise for PIXIE after multifrequency subtraction is shown in the bottom panel of Fig.~\ref{f:p}. 

\begin{figure}
\includegraphics[width=0.95\textwidth]{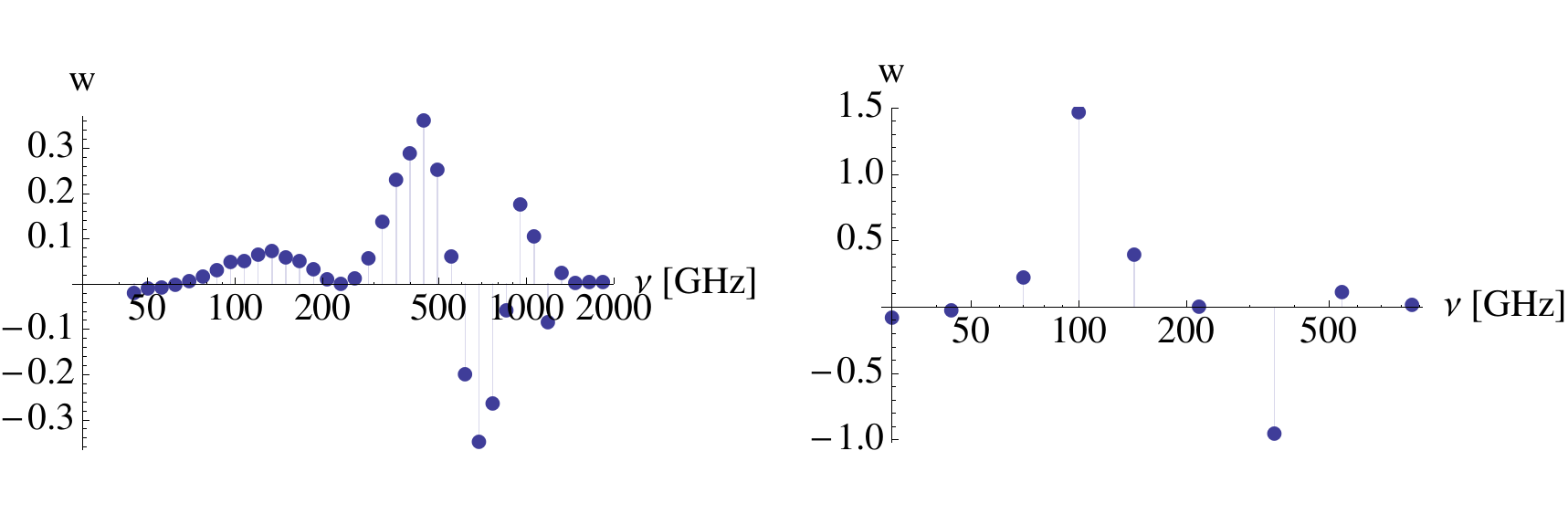}
\caption{The plots show the weights appearing in Eq.~(\ref{w}) for PIXIE (left) and Planck (right) as a function of $\nu$, for $\ell = 30$. \label{f:w}}
\end{figure}
\begin{figure}
\includegraphics[width=.7 \textwidth]{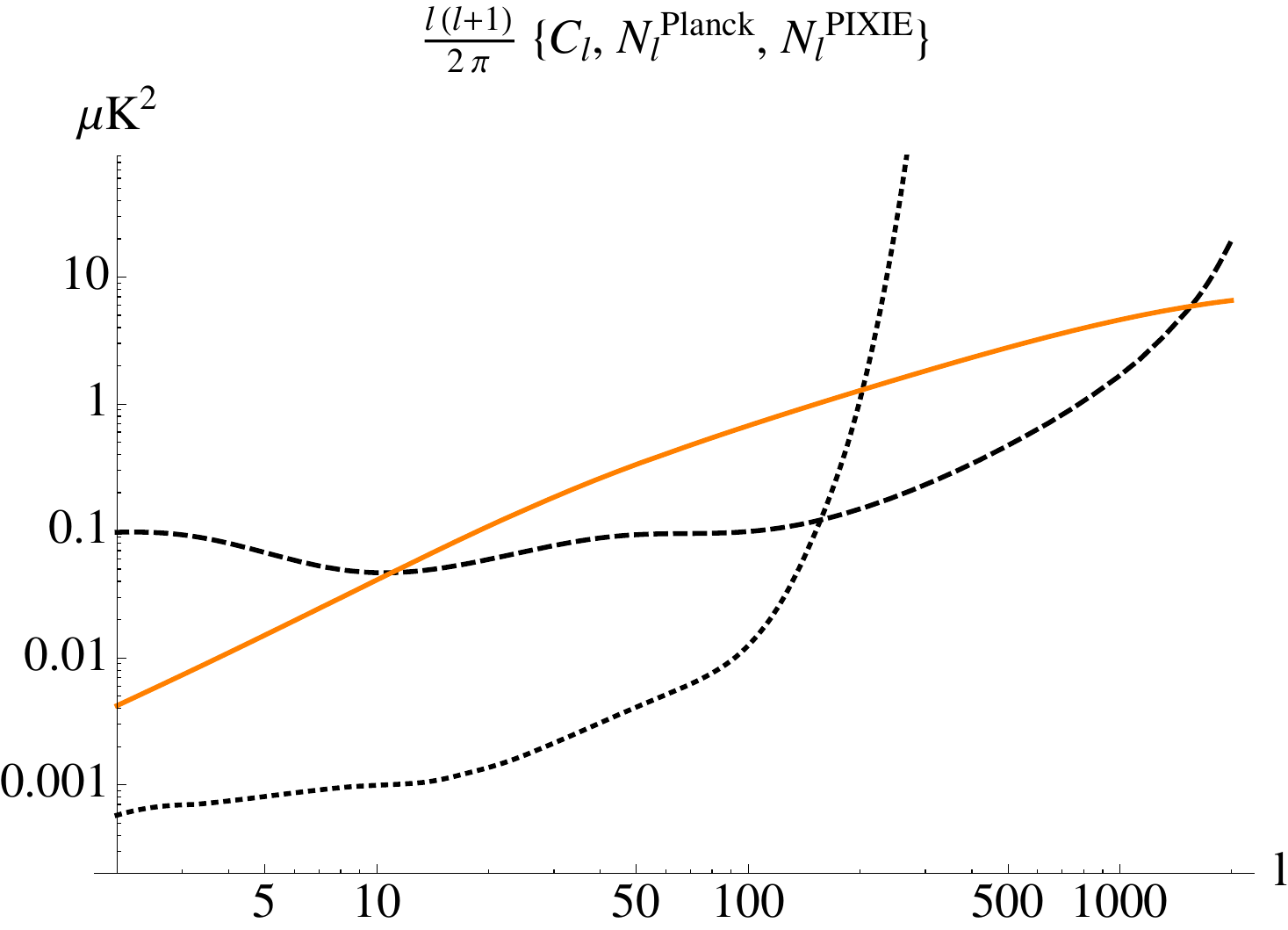}
\caption{The plot compares the noise $N_{\ell} [f_{sky} (2\ell+1)]^{-1/2}$ for Planck (black dashed line) and PIXIE (black dotted line) with the expected tSZ power spectrum $C^{SZ}_{\ell} \equiv C^y_{\ell}$ (continuous orange line) at 150 GHz with $f_{sky}=0.7$. \label{f:n}}
\end{figure}

We show the weights $w_{i}$ at $\ell=30$ (as an example) for Planck and PIXIE in Fig.~\ref{f:w}. As expected in both cases the weights are close to zero at $217$ GHz, which is the null of the tSZ signal. Also in both cases, very low and very high frequencies have very small weights. Finally in Fig.~\ref{f:n} we compare the total noises for Planck and PIXIE after summing over $m$'s and for a partial sky coverage $f_{sky}=0.7$, i.e.~$N_{\ell}[f_{sky} (2\ell+1)]^{-1/2}$, with the expected tSZ signal. PIXIE leads to an improvement of more than two orders of magnitude at low $\ell$'s, which is where the effect of the scale- dependent bias due to primordial non-Gaussianity arises. Since the PIXIE beam corresponds to $\ell_{max}\sim 84$, the PIXIE noise becomes very large beyond $\ell \sim $ few hundred where Planck is still expected to have signal-to-noise greater than one.


\section{Covariance Matrix of the tSZ Power Spectrum}
\label{sec:cov}

In order to forecast parameter constraints and the detection SNR of the tSZ power spectrum, we must compute its covariance matrix.  The covariance matrix contains a Gaussian contribution from the total (signal+noise) tSZ power spectrum observed in a given experiment, as well as a non-Gaussian cosmic variance contribution from the tSZ angular trispectrum.  We compute the covariance matrix for three different experiments: Planck, PIXIE, and a future cosmic variance (CV)-limited experiment.  The experimental noise after foreground subtraction is computed for Planck and PIXIE using the methods described in Section~\ref{sec:exp}.  For PIXIE, we assume a maximum multipole $\ell_{max}=300$, while for Planck and the CV-limited experiment we assume a maximum multipole $\ell_{max}=3000$.  In the PIXIE and Planck cases, these values are well into the noise-dominated regime, so there is no reason to go to higher multipoles.  For the CV-limited experiment, one can clearly compute up to as high a multipole as desired; however, it is unrealistic to imagine a satellite experiment being launched in the foreseeable future with noise levels better than PIXIE and angular resolution better than Planck, so we choose to adopt the semi-realistic value of $\ell_{max}=3000$ for the CV-limited experiment.  In all cases, we assume that the total available sky fraction used in the analysis is $f_{sky} = 0.7$, i.e., $30$\% of the sky is masked due to unavoidable contamination from foregrounds in our Galaxy.

In the remainder of this section, we outline the halo model-based calculations used to compute the tSZ power spectrum covariance matrix and then discuss in detail the different masking scenarios that we consider to reduce the level of cosmic variance error in the results.


\subsubsection{Halo Model Formalism}
\label{sec:covhalomodel}

We compute the tSZ power spectrum covariance matrix using the halo model approach, as was used for the power spectrum itself in Section~\ref{sec:tSZPShalomodel}.  We provide additional background on these calculations in Appendix~\ref{appendix}.  The total tSZ power spectrum covariance matrix, $M^{y}_{\ell\ell'}$, is given by Eq.~(\ref{eq.yClcov}):
\be
\label{eq.Mellellpr}
M^y_{\ell\ell'} =  \frac{1}{4 \pi f_{\mathrm{sky}}} \left( \frac{4 \pi (C^y_{\ell} + N_{\ell})^2}{\ell+1/2} \delta_{\ell\ell'} + T^y_{\ell\ell'} \right) \,,
\ee
where $C^y_{\ell}$ is the tSZ power spectrum given by Eq.~(\ref{eq.Cell}), $N_{\ell}$ is the noise power spectrum after foreground removal given by Eq.~(\ref{Nf}), and $T^y_{\ell\ell'}$ is the tSZ angular trispectrum.  Note that we have neglected an additional term in the covariance matrix that arises from the so-called ``halo sample variance'' (HSV) effect (e.g., see Eq.~(18) in~\cite{Satoetal2009} --- although that result is for the weak lensing power spectrum, the tSZ result is directly analogous).  The HSV term becomes negligible in the limit of a full-sky survey, which is all we consider in this paper; thus, we do not expect this approximation to affect our results.  Furthermore, we approximate the trispectrum contribution in Eq.~(\ref{eq.Mellellpr}) by the one-halo term only, which has been shown to dominate the trispectrum on nearly all angular scales~\cite{Cooray2001}.  We also restrict ourselves to the flat-sky limit, although the exact result is given in Appendix~\ref{appendix}.  The tSZ trispectrum is thus given by Eq.~(\ref{eq.yTl1hflatsky}):
\be
\label{eq.Tellellprquote}
T_{\ell\ell'}^{y,1h} & \approx & \int dz \frac{d^2V}{dz d\Omega} \int dM \frac{dn}{dM} \left| \tilde{y}_{\ell}(M,z) \right|^2 \left| \tilde{y}_{\ell'}(M,z) \right|^2 \,.
\ee
Justifications for our approximations are given in Appendix~\ref{appendix}.  We compute Eq.~(\ref{eq.Mellellpr}) for our fiducial WMAP9 cosmology using each of the masking scenarios discussed in the following section.  These results are then combined with the parameter variations discussed in Section~\ref{sec:params} in order to compute Fisher matrix forecasts in Section~\ref{sec:forecast}.


\subsubsection{Masking}
\label{sec:covmasking}

\begin{figure}
  \begin{center}
   \includegraphics[trim=0cm 0cm 0cm 0cm, clip=true, totalheight=0.5\textheight]{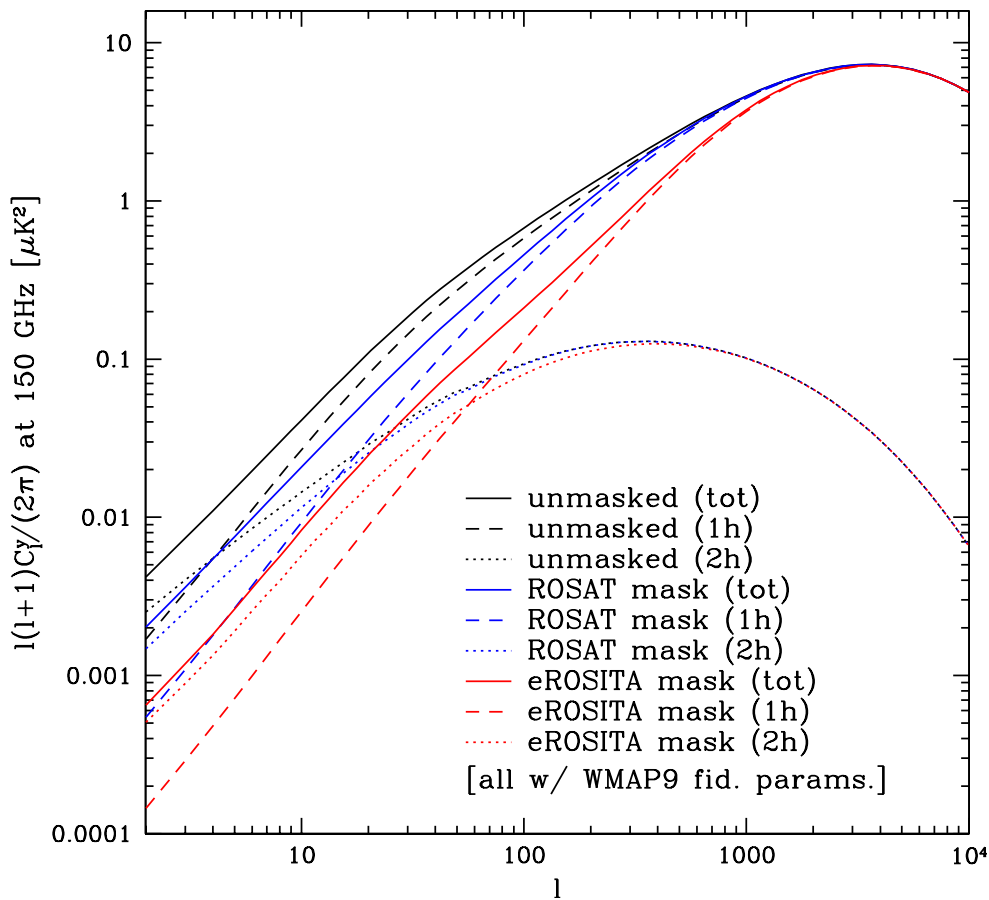}
    \caption{This plot shows the effect of different masking scenarios on the tSZ power spectrum calculated for our fiducial model.  The amplitude of the one-halo term is suppressed relative to the two-halo term, leading to an increase in the multipole where their contributions are equal.  In addition there is an overall decrease in the total signal at $\ell \lesssim 200$, as one would expect after masking the massive, nearby clusters that dominate the one-halo term in this regime. \label{fig.fidmaskingcomp}}
  \end{center}
\end{figure}

For all of the experiments that we consider, the tSZ power spectrum covariance matrix in Eq.~(\ref{eq.Mellellpr}) is dominated by the trispectrum contribution over at least part of the multipole range relevant to that experiment.  This issue is worse for PIXIE than for Planck, due to its lower noise levels, but even the Planck tSZ covariance is dominated by the trispectrum at some multipoles ($10 \lesssim \ell \lesssim 100$).  Fortunately, the trispectrum contribution can be significantly reduced by simply masking the massive, nearby galaxy clusters that dominate the signal, especially at low-$\ell$~\cite{Komatsu-Kitayama1999}.

However, there are some complications in this procedure.  It is not obvious \emph{a priori} that one should mask as many clusters as possible, since at some point the power spectrum signal itself will begin to decrease enough that the overall SNR decreases, despite the decrease in the noise.  In addition, one must take care to use a very pure and complete cluster sample with a well-known selection function to do the masking; otherwise, it will be extremely difficult to properly account for the masking in the corresponding theoretical computations of the tSZ power spectrum.  Finally, there is the unavoidable problem of scatter in the cluster mass-observable relation (e.g., $L_X$--$M$ or $Y$--$M$), which will introduce additional uncertainty in the theoretical calculation of the masked tSZ power spectrum.  Moreover, this uncertainty may be hard to precisely quantify.

In this paper, we consider a set of masking scenarios that approximately correspond to catalogs from existing and future all-sky X-ray surveys.  At present, X-ray cluster surveys likely possess the most well-understood selection functions, as compared to those derived from optical, SZ, or weak lensing data~(e.g., \cite{Melinetal2005,Bohringeretal2001}).  Furthermore, the scatter in the mass-observable relation for the X-ray quantity $Y_X$ (a measure of the integrated gas pressure) is believed to be quite small, possibly $<10$\%~\cite{Vikhlininetal2009b,Vikhlininetal2009}.  Although existing X-ray catalogs are only complete for fairly high masses and low redshifts (due to the steep decrease in X-ray surface brightness with redshift), these clusters are exactly the ones that need to be masked to suppress the tSZ trispectrum.  Furthermore, the future catalogs from eROSITA, an upcoming X-ray satellite designed for an all-sky survey, will be complete to much lower masses and very high redshifts.  Overall, it seems that X-ray-based masking is the most robust option at present.

In order to simplify our calculations and avoid the need to specify the details of any individual X-ray survey, we compute the effects of masking by removing all clusters in the tSZ calculations that lie above a mass threshold $M_{mask}$ \emph{and} below a redshift cutoff $z_{mask}$.  This method also circumvents the issue of modeling scatter in the mass-observable relation.  Clearly a more sophisticated approach will be needed for the analysis of real data, but these choices allow us to explore several possibilities fairly quickly.

\begin{figure}
  \begin{center}
    \includegraphics[trim=0cm 0cm 0cm 0cm, clip=true, totalheight=0.5\textheight]{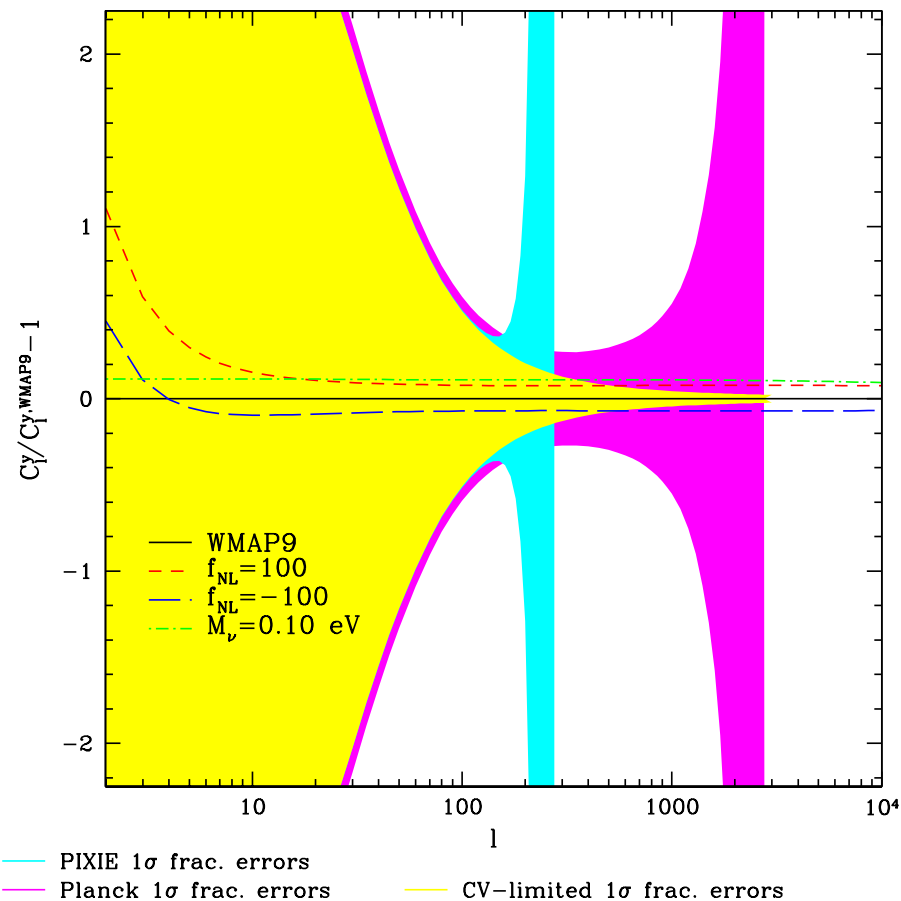}
    \caption{This plot shows the fractional difference between the tSZ power spectrum of our fiducial model and three different parameter variations, as labeled in the figure.  In addition, we show the $1\sigma$ fractional errors on the tSZ power spectrum for each of the three experiments we consider, computed from the square root of the diagonal of the covariance matrix.  In this unmasked calculation, the overwhelming influence of cosmic variance due to the tSZ trispectrum is clearly seen at low-$\ell$. \label{fig.fNLMnuwerrs}}
  \end{center}
\end{figure}

\begin{figure}
  \begin{center}
    \includegraphics[trim=0cm 0cm 0cm 0cm, clip=true, totalheight=0.5\textheight]{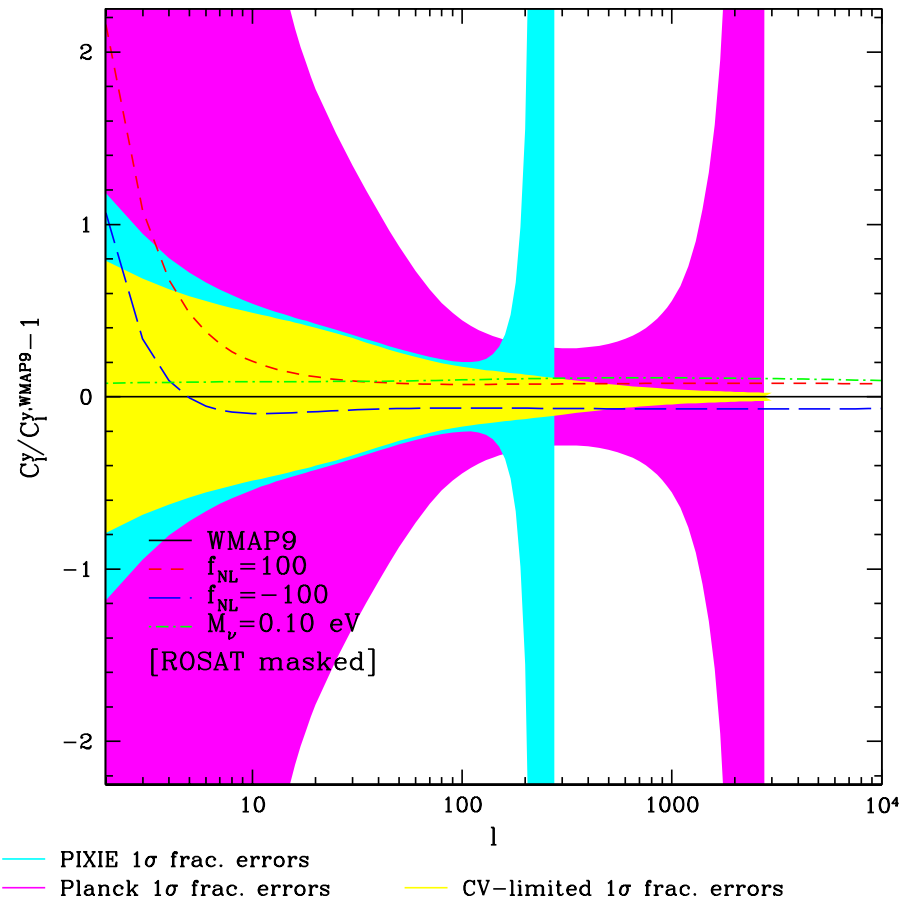}
    \caption{This plot is identical to Fig.~\ref{fig.fNLMnuwerrs} except that it has been computed for the ROSAT-masked scenario (see Section~\ref{sec:covmasking}).  Compared to Fig.~\ref{fig.fNLMnuwerrs}, it is clear that the errors at low-$\ell$ for PIXIE and the CV-limited case have been dramatically reduced by the masking procedure.  The errors from Planck are less affected because it is dominated by instrumental noise over most of its multipole range.   \label{fig.fNLMnuwerrsmasked}}
  \end{center}
\end{figure}

\begin{figure}
  \begin{center}
    \includegraphics[trim=0cm 0cm 0cm 0cm, clip=true, totalheight=0.5\textheight]{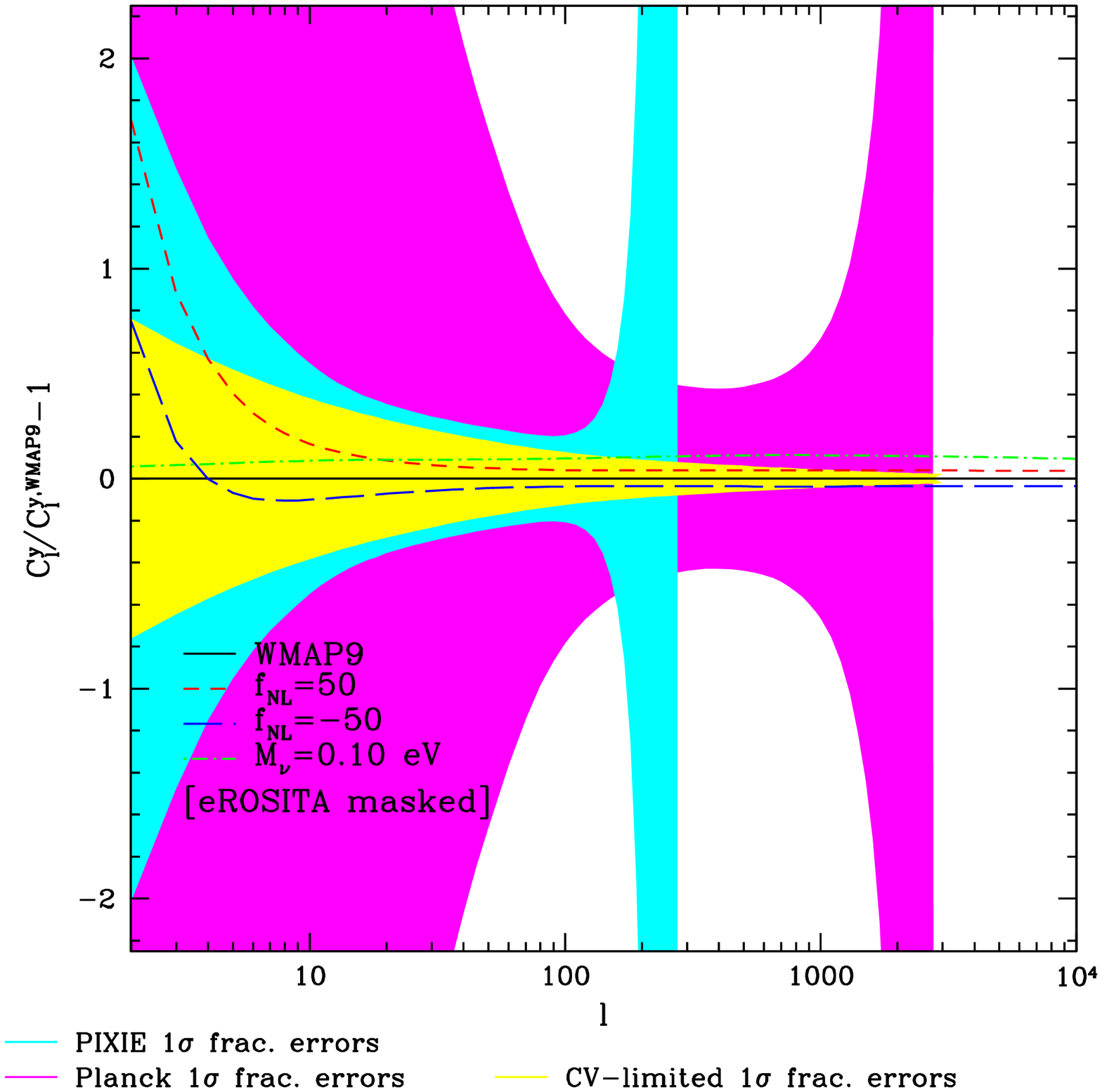}
    \caption{This plot is identical to Fig.~\ref{fig.fNLMnuwerrs} except that it has been computed for the eROSITA-masked scenario (see Section~\ref{sec:covmasking}).  Compared to Fig.~\ref{fig.fNLMnuwerrs}, it is clear that the errors have been significantly reduced at low-$\ell$, though again the effect is minimal for Planck.  The comparison to the ROSAT-masked case in Fig.~\ref{fig.fNLMnuwerrsmasked} is more complicated: it is clear that the errors are further reduced for the CV-limited case, but the PIXIE fractional errors actually increase at low-$\ell$ as a result of the reduction in the amplitude of the signal there due to the masking.  However, the PIXIE fractional errors at higher multipoles are in fact slightly reduced compared to the ROSAT-masked case, although it is hard to see by eye in the plot. \label{fig.fNLMnuwerrssupermasked}}
  \end{center}
\end{figure}

First, we consider a masking criterion based on the results of the ROSAT All-Sky Survey (RASS)~\cite{Vogesetal1999}.  The catalogs derived from RASS (e.g., BCS/eBCS~\cite{Ebelingetal1998,Ebelingetal2000}, REFLEX~\cite{Bohringeretal2001}) are nominally complete to a flux limit $f_X > 3 \times 10^{-12}$ erg/cm$^2$/s (0.5--2.0 keV band).  At $z=0.05$, this flux corresponds to a luminosity $L_X \approx 9 \times 10^{42} \, \, h^2$ erg/s.  Using the scaling relations from~\cite{Popessoetal2005} (for either optical or X-ray masses) to convert from $L_X$ to mass, this luminosity corresponds to $M \approx 10^{14} \,\, M_{\odot}/h$, where $M$ is the virial mass defined in Section~\ref{sec:HMF} (we have also used the NFW profile~\cite{NFW1997} and the concentration-mass relation from~\cite{Duffyetal2008} in this conversion).  Thus, the catalogs derived from RASS should be $\sim 100$\% complete for clusters with $M \gtrsim \mathrm{few} \times 10^{14} \,\, M_{\odot}/h$ at $z<0.05$.  To be conservative, we set $M_{mask} = 5 \times 10^{14} \,\, M_{\odot}/h$, which corresponds to a luminosity and flux of $L_X \approx 7 \times 10^{44} \, \, h^2$ erg/s and $f_X \approx 2 \times 10^{-11}$ erg/cm$^2$/s at $z=0.05$, well above the flux limit given above.  We will refer to this masking choice ($M_{mask} = 5 \times 10^{14} \,\, M_{\odot}/h$, $z_{mask} = 0.05$) as ``ROSAT-masked" in the remainder of the paper.

Second, we consider a masking criterion based on the upcoming all-sky survey conducted by eROSITA.  The eROSITA cluster catalogs are expected to be nominally complete to a flux limit $f_X > 4 \times 10^{-14}$ erg/cm$^2$/s (0.5--2.0 keV band)~\cite{Merlonietal2012}.  This limit is low enough that it will be possible to essentially mask clusters arbitrarily at low redshifts.  At $z=0.05$, this flux corresponds to $L_X \approx 10^{41} \, \, h^2$ erg/s, far below the emission from even a $\sim 10^{13} \,\, M_{\odot}/h$ group, for which $L_X \sim 10^{42} \, \, h^2$ erg/s (extrapolating the scaling relations from~\cite{Popessoetal2005}).  Thus, there is a wide range of possible masking choices based on the eROSITA catalogs.  In principle, it would be best to fully explore the ($M_{mask}$, $z_{mask}$) parameter space and find the values that optimize the SNR for the tSZ power spectrum, or perhaps optimize the constraints on some particular parameter, such as \fnl.  This optimization is beyond the scope of this paper.  We anticipate that masking heavily above $z \sim 0.1$--0.2 will eventually begin to decrease the tSZ power spectrum signal too significantly, so we make a reasonably conservative cut at $z_{mask} = 0.15$.  As mentioned, at these redshifts eROSITA will detect essentially all clusters (at $z=0.15$, the flux limit corresponds to $L_X \approx 1.2 \times 10^{42} \, \, h^2$ erg/s, which roughly scales to $M \approx 2 \times 10^{13} \,\, M_{\odot}/h$ using the same scaling relations as above).  Thus, we can choose the mass threshold at essentially any value.  Our final values are ($M_{mask} = 2 \times 10^{14} \,\, M_{\odot}/h$, $z_{mask} = 0.15$); we will refer to this masking choice as ``eROSITA-masked'' in the remainder of the paper.

Finally, we also consider a completely unmasked calculation, both due to its theoretical simplicity and as a means to assess how much the choice of masking can improve the SNR on the tSZ power spectrum, as well as how the forecasted parameter constraints change.  Throughout the paper, unless figures or tables are labelled with a masking choice, they have been calculated in the unmasked case.

\begin{figure}
  \begin{center}
    \includegraphics[trim=0cm 0cm 0cm 0cm, clip=true, totalheight=0.5\textheight]{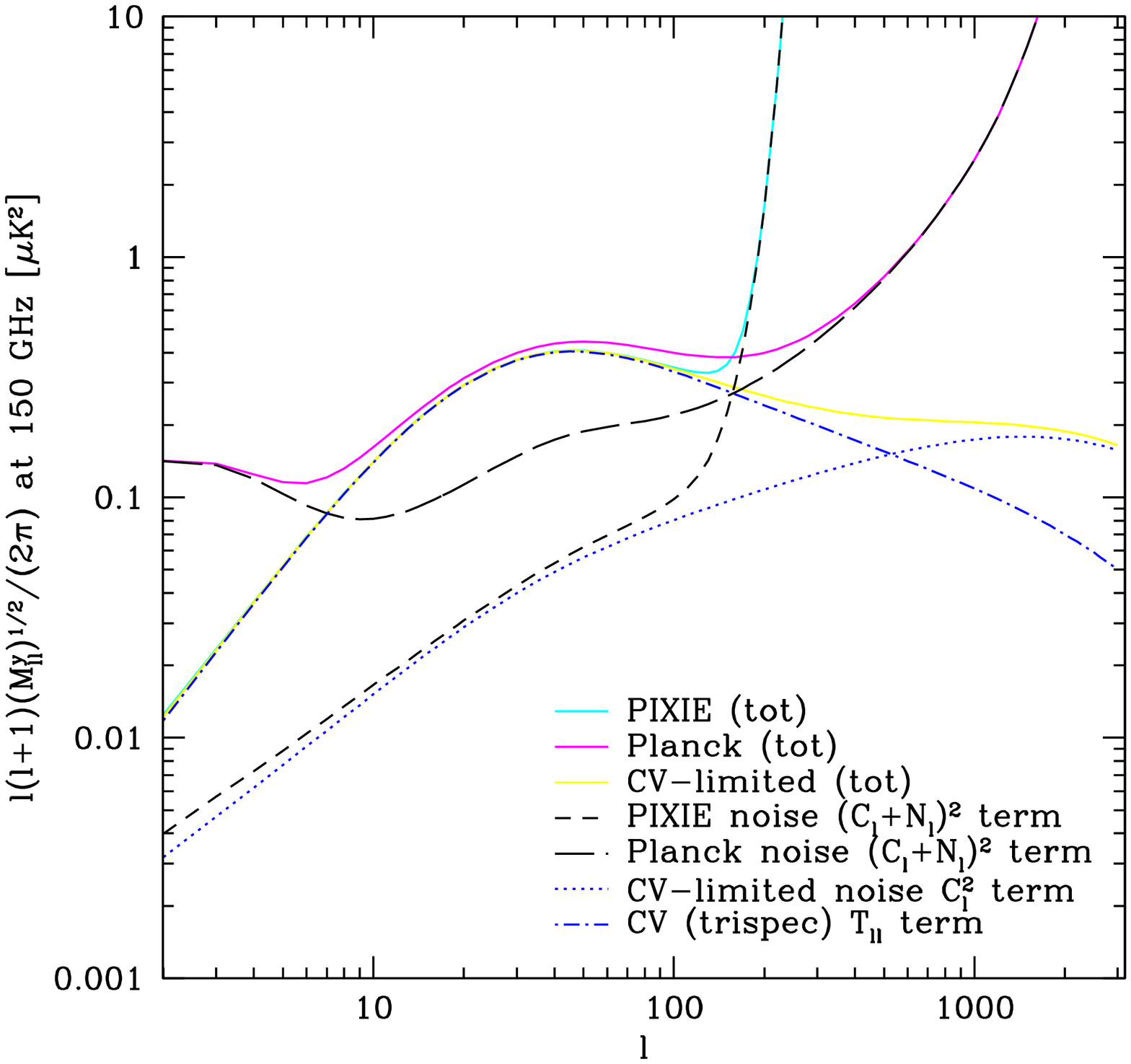}
    \caption{This plot shows the square root of the diagonal elements of the covariance matrix of the tSZ power spectrum, as well as the contributions from the Gaussian and non-Gaussian terms in Eq.~(\ref{eq.Mellellpr}).  Results are shown for the total in each experiment, as well as the different Gaussian terms in each experiment and the trispectrum contribution that is identical for all three.  See the text for discussion. \label{fig.covcomps}}
  \end{center}
\end{figure}

\begin{figure}
  \begin{center}
    \includegraphics[trim=0cm 0cm 0cm 0cm, clip=true, totalheight=0.5\textheight]{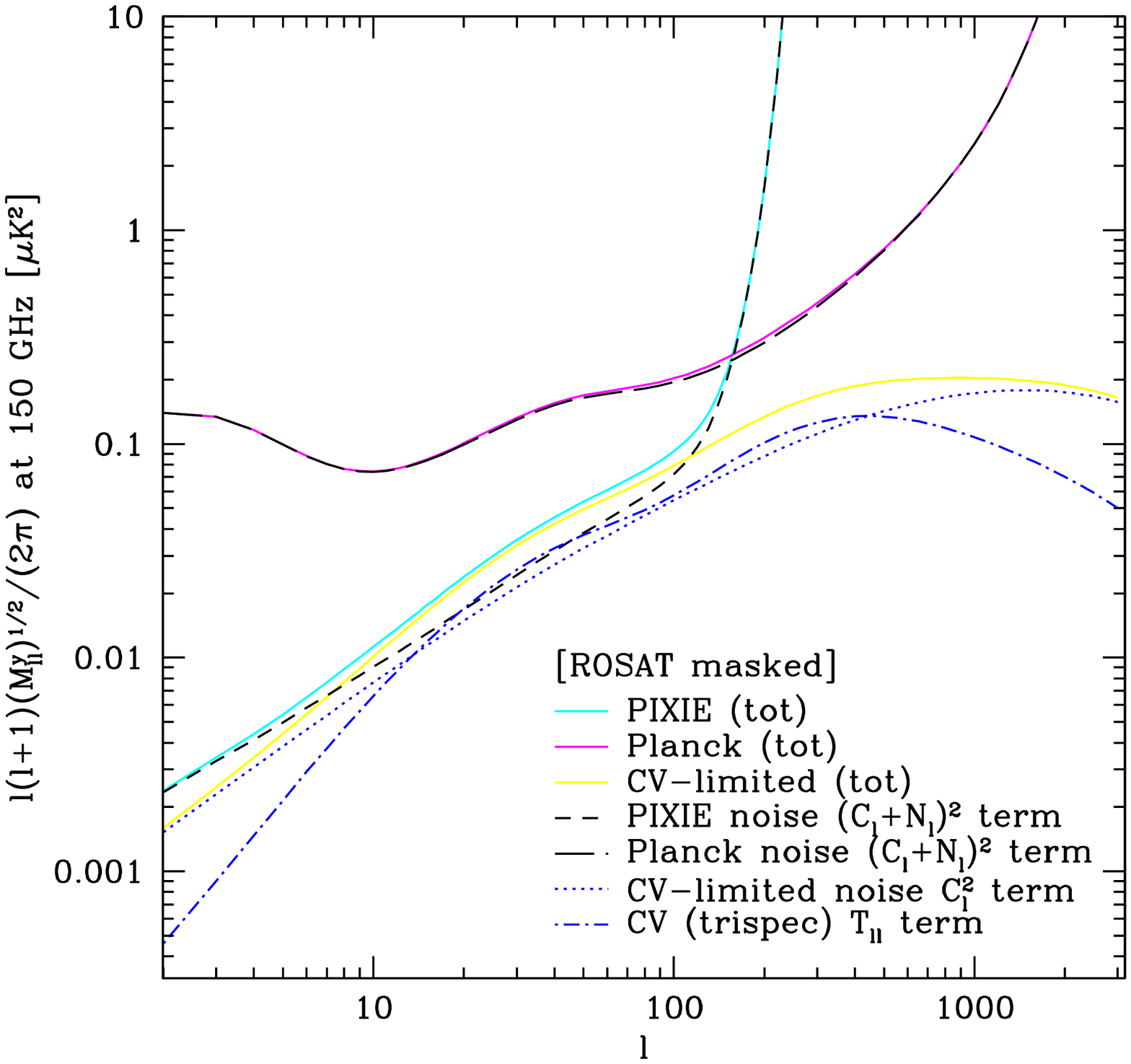}
    \caption{This plot is identical to Fig.~\ref{fig.covcomps} except that it has been computed for the ROSAT-masked scenario (see Section~\ref{sec:covmasking}). Results are shown for the total in each experiment, as well as the different Gaussian terms in each experiment and the trispectrum contribution that is identical for all three.  See the text for discussion. \label{fig.covcompsmasked}}
  \end{center}
\end{figure}

\begin{figure}
  \begin{center}
    \includegraphics[trim=0cm 0cm 0cm 0cm, clip=true, totalheight=0.5\textheight]{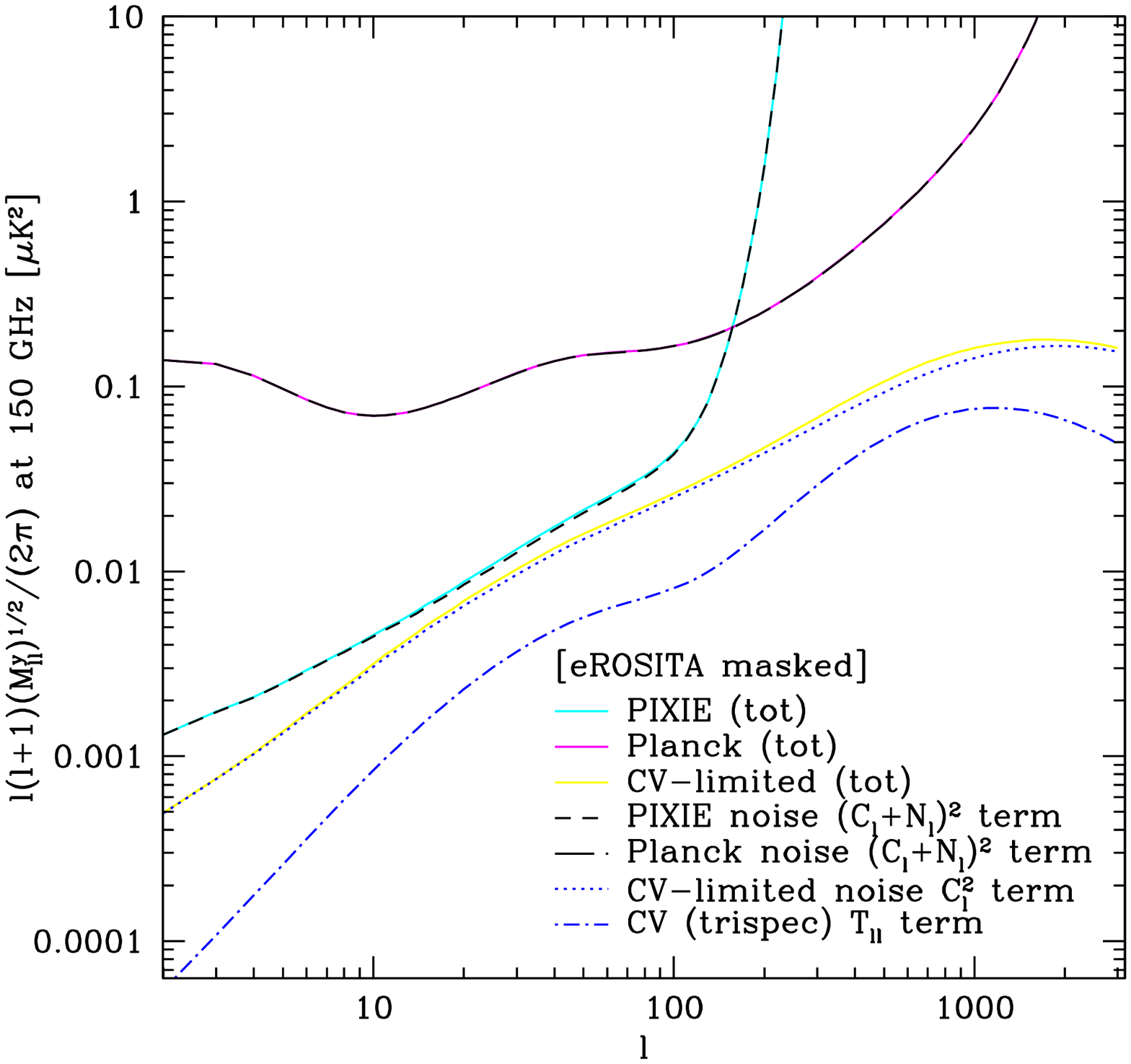}
    \caption{This plot is identical to Fig.~\ref{fig.covcomps} except that it has been computed for the eROSITA-masked scenario (see Section~\ref{sec:covmasking}). Results are shown for the total in each experiment, as well as the different Gaussian terms in each experiment and the trispectrum contribution that is identical for all three.  See the text for discussion. \label{fig.covcompssupermasked}}
  \end{center}
\end{figure}

In Fig.~\ref{fig.fidmaskingcomp} we demonstrate the effects of masking on the tSZ power spectrum computed using our fiducial parameters.  The main effects are a decrease in the amplitude of the one-halo term relative to the two-halo term at low-$\ell$ and an overall decrease in the amplitude of the power spectrum at low-$\ell$.  In the unmasked case, the one-halo term and two-halo term are roughly equal at $\ell \simeq 4$, while this cross-over multipole increases to $\simeq 14$ for the ROSAT-masked case and $\simeq 56$ for the eROSITA-masked case.  Note that the enhancement of the two-halo term relative to the one-halo term increases the sensitivity of the power spectrum to \fnl through the effect of the scale-dependent bias shown in Fig.~\ref{fig.fNL1h2h}.  Although the amplitude of the total signal is decreased by masking, the suppression of the cosmic variance error due to the trispectrum is much larger, as seen in Figs.~\ref{fig.fNLMnuwerrs}--\ref{fig.covcompssupermasked}.

Figs.~\ref{fig.fNLMnuwerrs}--\ref{fig.fNLMnuwerrssupermasked} demonstrate the significant reduction in error due to masking.  These figures show the fractional difference with respect to our fiducial WMAP9 results for the tSZ power spectrum computed in three different cosmologies (\fnl$=100$ (or 50), \fnl$=-100$ (or $-50$), and \Mnu$=0.10$ eV).  Note that these power spectra are computed using the same masking criterion as used in the covariance matrix.  The figures also show the $1\sigma$ fractional errors on the tSZ power spectrum for each of the three experiments we consider.  These fractional errors have been computed from the diagonal of the covariance matrix, although we emphasize that the entire covariance matrix is used in all calculations presented in the paper.  The overall implication of Figs.~\ref{fig.fNLMnuwerrs}--\ref{fig.fNLMnuwerrssupermasked} is that masking is in general highly beneficial for tSZ power spectrum measurements.  Although the improvements for Planck are marginal due to the fact that it is instrumental noise-dominated (see the subsequent figures), the improvements for PIXIE and the CV-limited case are dramatic.  We note, however, that the reduction in the signal for the eROSITA-masked case is large enough at low-$\ell$ that the fractional errors for PIXIE are actually larger than in the ROSAT-masked case, because the signal at low-$\ell$ is becoming smaller than the PIXIE instrumental noise.  Nonetheless, at higher multipoles the PIXIE fractional errors are in fact slightly smaller for the eROSITA-masked case than ROSAT-masked.  This trend suggests that the optimal choice of masking for a given experiment depends on the noise levels of the experiment and the multipole range that one would like to measure with the highest SNR.  For the CV-limited case, it is perhaps unsurprising that the fractional errors continue to decrease with heavier masking, as the trispectrum term is further and further suppressed relative to the Gaussian term in Eq.~(\ref{eq.Mellellpr}).  It is unclear how far one can push this masking before the SNR starts to decrease due to the reduction in the signal, but Fig.~\ref{fig.covcompssupermasked} (discussed in the next paragraph) suggests that our eROSITA masking scenario is not too far from this limit.

Figs.~\ref{fig.covcomps}--\ref{fig.covcompssupermasked} show the contributions to (and total of) the diagonal of the covariance matrix from the Gaussian and non-Gaussian terms in Eq.~(\ref{eq.Mellellpr}) for each of the experimental and masking scenarios.  From Fig.~\ref{fig.covcomps}, one can see that Planck is dominated by instrumental noise over most of its multipole range, the exception being a window from $10 \lesssim \ell \lesssim 100$ where the trispectrum contribution dominates.  Fig.~\ref{fig.covcomps} also demonstrates that PIXIE is near the CV-limited case over essentially its entire multipole range.  This explains the large decrease in the fractional errors for PIXIE after applying the ROSAT-masking in Fig.~\ref{fig.fNLMnuwerrsmasked}.  Fig.~\ref{fig.covcompsmasked} shows the covariance matrix contributions in the ROSAT-masked case, where it is clear that Planck is now completely dominated by instrumental noise at all multipoles.  Thus, further masking is not beneficial for Planck.  PIXIE is still near the CV-limited case for the ROSAT-masked scenario, except at low-$\ell$ where its instrumental noise starts to become significant.  This suggests that further masking will decrease the SNR at low multipoles for PIXIE, as indeed can be seen in the eROSITA-masked scenario of Fig.~\ref{fig.fNLMnuwerrssupermasked}.  Finally, Fig.~\ref{fig.covcompssupermasked} shows the covariance matrix contributions in the eROSITA-masked case, where one can see that the trispectrum term has now been suppressed below the Gaussian term at all multipoles, even for the CV-limited experiment.  This trend likely suggests that further masking will begin to lead to a reduction in the SNR even for the CV-limited case, although we have not computed this precisely, as we have no masking cases beyond eROSITA.  Both PIXIE and Planck are clearly dominated by instrumental noise for the eROSITA-masked case, suggesting that further masking is unlikely to lead to significant improvements in SNR for these experiments.



\section{Forecasted Constraints}
\label{sec:forecast}
Having described our model for the signal and its covariance matrix, we now use the Fisher matrix formalism~\cite{Fisher1935,Knox,Tegmarketal1997,Jungmanetal1996} to forecast the expected constraints on the parameters listed in Eq.~(\ref{eq.paramslist}) for a total of nine different experimental specifications (Planck, PIXIE and CV-limited) and masking choices (unmasked, ROSAT-masked, eROSITA-masked).  We compute derivatives of the tSZ power spectrum signal with respect to each parameter around the fiducial values given in Eq.~(\ref{eq.paramsfid}).  The Fisher matrix is given by
\beq
\label{eq.Fisher}
F_{ij} = \frac{\partial C_{\ell}^y}{\partial p_i} \left( M^{-1} \right)_{\ell \ell'} \frac{\partial C_{\ell'}^y}{\partial p_j} \,,
\eeq
where $p_i$ is the $i^{\mathrm{th}}$ parameter in Eq.~(\ref{eq.paramslist}), $C_{\ell}^y$ is the tSZ power spectrum given by Eq.~(\ref{eq.Cell}), and $\left( M^{-1} \right)_{\ell \ell'}$ is the inverse of the covariance matrix given by Eq.~(\ref{eq.Mellellpr}).  Note that we only consider \fnl$\neq 0$ and \Mnu$>0$ cosmologies separately; in the interest of simplicity, we only seek to constrain minimal one-parameter extensions of the $\Lambda$CDM standard model.  Thus, our Fisher matrices are eight-by-eight, containing the five relevant $\Lambda$CDM parameters, the two ICM physics parameters ($C_{P_0}$ and $C_{\beta}$) and either \fnl or \Mnu.  The unmarginalized $1\sigma$ error on parameter $p_i$ is given by $1/\sqrt{F_{ii}}$, and it describes the best possible error one can obtain when all other parameters are known exactly.  The marginalized $1\sigma$ error is given by $\sqrt{(F^{-1})_{ii}}$, and it describes the case in which all other parameters are constrained from the same set of data.  In the following sections, we quote the unmarginalized $1\sigma$ errors on each of the parameters from the tSZ power spectrum alone, as well as marginalized $1\sigma$ errors computed by the following methods.  The results are summarized in Fig.~\ref{tab:params}.  In addition, we provide the complete Fisher matrices for these calculations online\footnote{{\tt http://www.astro.princeton.edu/\textasciitilde jch/tSZFisher/ }}.

In order to compute marginalized constraints, we find that it is necessary to include external data (in addition to basic priors, which are described below), because the degeneracies between the various parameters are too strong for the tSZ signal alone to provide meaningful constraints.  For this purpose, we include a Fisher matrix computed for the imminent results from the Planck measurements of the primordial CMB temperature fluctuation power spectrum\footnote{We are thankful to M.~Takada for providing a computation of the Planck CMB Fisher matrix.}.  This Fisher matrix includes our five primary cosmological parameters $\left\{\Omega_b h^2, \Omega_c h^2, \Omega_{\Lambda}, \sigma_8, n_s\right\}$, as well as the optical depth to reionization, $\tau$ (note that it does not include any non-$\Lambda$CDM parameters, i.e., \fnl and \Mnu are not included).  We marginalize over $\tau$, since the tSZ signal is insensitive to this parameter.  In addition, we include a prior on $H_0$ from the results of~\cite{Riessetal2011}, who find a $\approx 3.3$\% constraint on this parameter from cosmological distance ladder measurements.  Although their central value is slightly discrepant from our fiducial value, we simply include the statistical power of this measurement as a representative measure of current constraints on $h$.  To include this prior, we transform to a parameter set in which $h$ lies on the diagonal, add a Gaussian prior of width $\Delta h = 0.033 \times 0.697 = 0.023$, where $h=0.697$ is the value for our fiducial cosmology, and then transform back to our original parameters.  We add the resulting ``Planck CMB + $H_0$'' Fisher matrix to the appropriate five-dimensional sub-matrix of our tSZ Fisher matrices in order to break degeneracies between the various parameters, and investigate to what extent the tSZ signal can improve on the upcoming Planck CMB measurements of the primary cosmological parameters.

In addition to the Planck CMB prior matrix for the primary cosmological parameters, we also place simple priors on the ICM physics parameters $C_{P_0}$ and $C_{\beta}$.  We adopt $1\sigma$ Gaussian priors of $\Delta C_{P_0} = 0.2 = \Delta C_{\beta}$ (recall that these parameters are normalized versions of the GNFW pressure profile amplitude and outer logarithmic slope).  These values correspond roughly to the variances determined for these parameters due to the scatter between cluster pressure profiles in the simulations from which the fiducial model was obtained\footnote{N.~Battaglia, priv.~comm.}.  They also encompass values obtained from various X-ray and SZ observations~\cite{Arnaudetal2010,Plaggeetal2012,Plancketal2013}, although we note that our use of $P_{200,c}$ in Eq.~(\ref{eq.GNFW}) makes direct comparisons of the $P_0$ parameter somewhat nontrivial between our model and others in the literature.  Finally, these priors guarantee that the unphysical values $C_{P_0} = 0$ and $C_{\beta} = 0$ are highly disfavored ($>5\sigma)$.  We find that the Fisher forecasts computed below are often strong enough to provide constraints on the ICM parameters well below the width of these priors (at least for $C_{\beta}$, which is currently the less well-constrained of the two), and thus we conclude that the resulting errors are robust.

We place no priors on \fnl or \Mnu as our goal is to assess the detectability of these parameters using information in the tSZ power spectrum, with no external constraints (apart from those necessary to break degeneracies amongst the primary cosmological parameters, as implemented in our Planck CMB+$H_0$ prior matrix).

In the following subsections we describe our main parameter forecast results, summarized in Fig.~\ref{tab:params}, for the two separate cases in which the standard cosmological and astrophysical parameters (see Eq.~(\ref{eq.paramslist})) have been supplemented by either \fnl or \Mnu.


\subsection{\fnl}
\label{sec:fNLconstraints}

\begin{figure}
\begin{center}
\includegraphics[width=\textwidth]{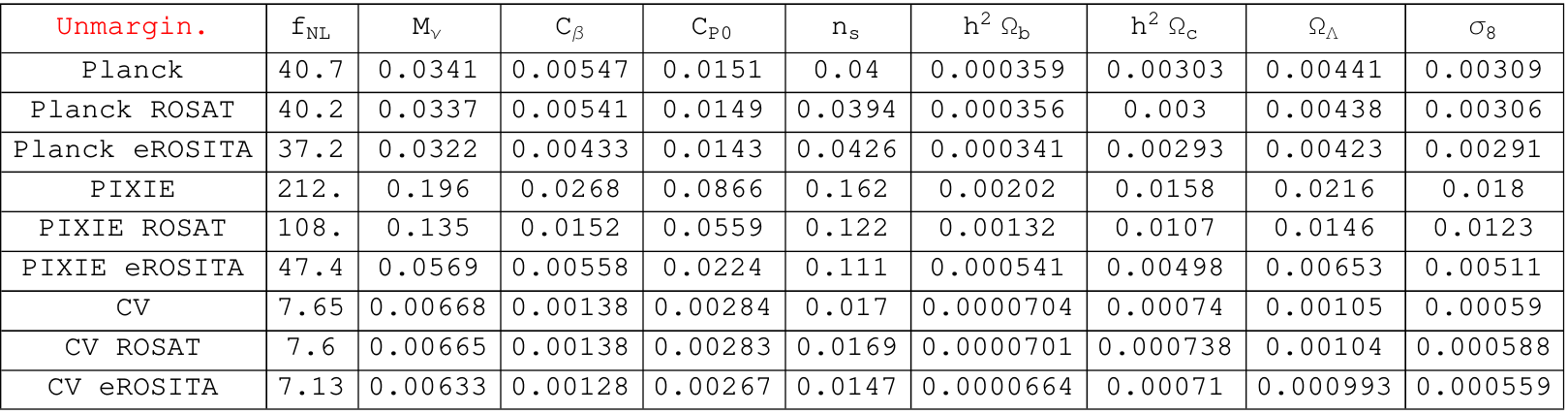}\\[.5cm]
\includegraphics[width=.47\textwidth]{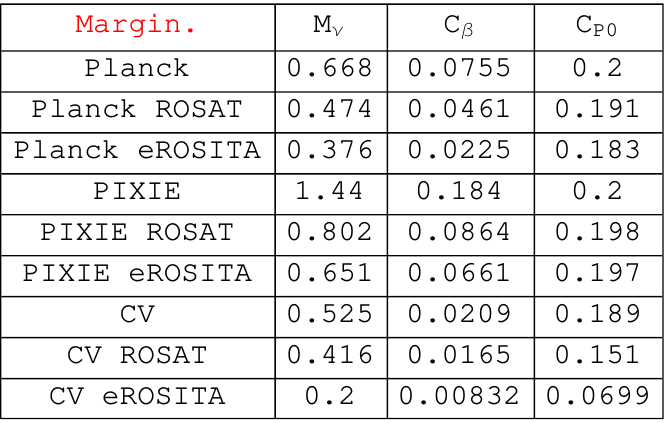}\qquad
\includegraphics[width=.45\textwidth]{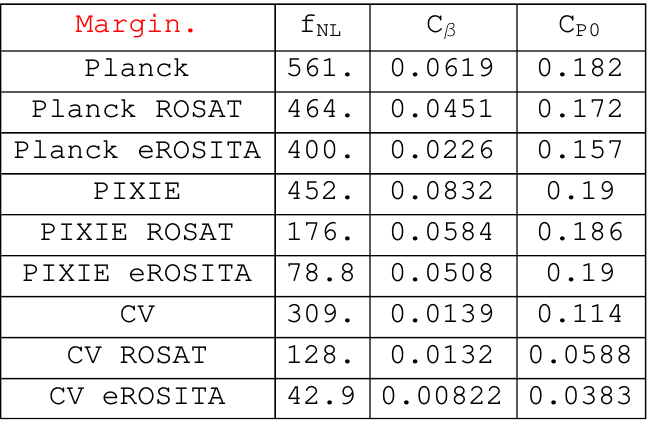}\\[.5cm]
\includegraphics[width=\textwidth]{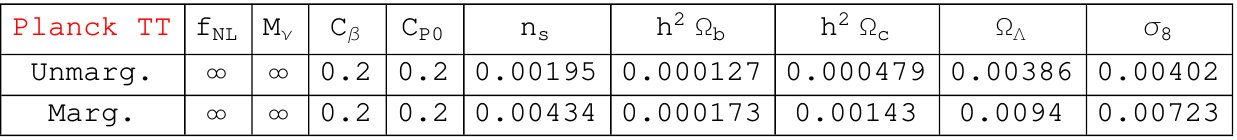}
\caption{The tables show the estimated unmarginalized and marginalized $1\sigma$ error bars for the indicated parameters and for a total of nine different experimental specifications (Planck, PIXIE, and CV-limited) and cluster masking choices (no masking, ROSAT masking, and eROSITA masking). The unmarginalized errors (top table) are derived using only the tSZ power spectrum. The marginalized errors (two central tables) are derived by adding as an external prior the forecasted Planck constraints from the CMB $TT$ power spectrum plus a prior on $H_{0}$ from~\cite{Riessetal2011}. Since adding the tSZ power spectrum leads to effectively no improvement in the errors on the $\Lambda$CDM parameters with respect to the Planck $TT$ priors, we do not show those numbers in the two central tables (marginalized tSZ plus Planck CMB+$H_0$ priors).  They are equal to the numbers in the bottom table, which show the marginalized Planck CMB+$H_0$ priors, to $\approx 1\%$ precision.  Note that all constraints on \Mnu are in units of eV, while the other parameters are dimensionless. \label{tab:params}}
\end{center}
\end{figure}

We discuss local primordial non-Gaussianity and its effects on the tSZ power spectrum in Sections~\ref{sec:LSS} and~\ref{sec:params}. The current strongest bounds come from WMAP9 and correspond to $-3 < f_{\mathrm{NL}} < 77$ at $95\%$ CL~\cite{Bennettetal2012}, but Planck will likely improve on these limits by a factor of $\approx 3$.  From the top table in Fig.~\ref{tab:params}, one can see that at the unmarginalized level the tSZ power spectrum is quite sensitive to \fnl, with an unmarginalized CV limit of $\Delta f_{\mathrm{NL}} \approx 7$.  On the other hand, \fnl is very degenerate with all the other parameters and the marginalized bounds are far weaker than the current bounds (see the central right table in Fig.~\ref{tab:params}).  Adding the tSZ power spectrum constraint to the current bounds in quadrature improves the overall constraint beyond the current bounds by at most a few percent, even in the most optimistic scenario (CV-limited experiment with eROSITA masking).  We have checked that no single parameter is driving the marginalized error on \fnl away from the unmarginalized value, which would be comparable with constraints from the Planck primordial temperature bispectrum measurements.  On the contrary, in order to get closer to the unmarginalized bound, all other parameters, both cosmological and astrophysical, would need to be constrained much better.  We point out that if one ever hopes to constrain \fnl using the tSZ power spectrum, it is crucial to obtain high SNR on the lowest possible multipoles, where the signature of the scale-dependent bias breaks the degeneracy between \fnl and other parameters that influence the overall amplitude of the power spectrum.  This fact explains why PIXIE would achieve much tighter marginalized constraints on \fnl using the tSZ power spectrum than Planck would (as seen in the central right table in Fig.~\ref{tab:params}), despite Planck's larger multipole range.  Nevertheless, even the PIXIE constraints are unlikely to be competitive with those from other probes of primordial non-Gaussianity.

\begin{figure}
\begin{center}
\includegraphics[width=.6\textwidth]{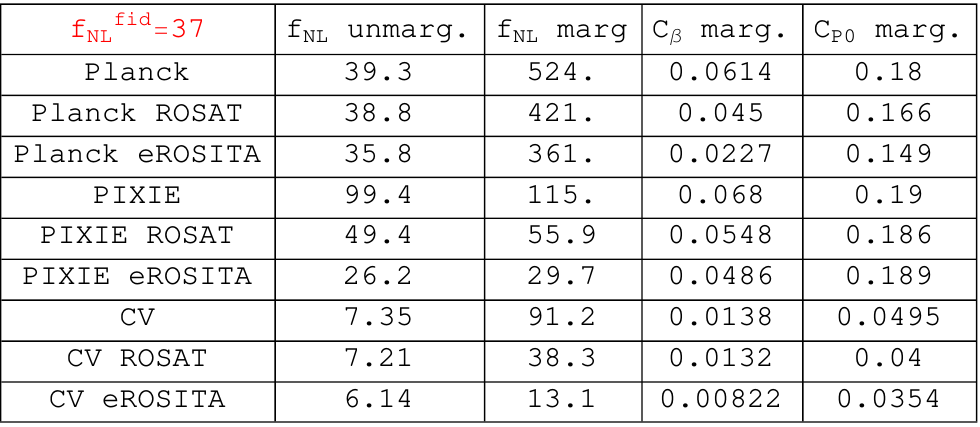}
\caption{The same as in the right central table of Fig.~\ref{tab:params}, but for a fiducial \fnl$=37$, the central value of WMAP9~\cite{Bennettetal2012}.  A CV-limited experiment could lead to a $3\sigma$ detection. \label{tab:fNL37}}
\end{center}
\end{figure}

Thus far we have only considered constraints around our fiducial cosmology with \fnl$=0$.  If instead we assume a fiducial \fnl$=37$, corresponding to the central value of WMAP9~\cite{Bennettetal2012}, we find as before that Planck and PIXIE would have less than a $2\sigma$ detection.  However, for a CV-limited experiment using eROSITA masking, we find a marginalized error $\Delta f_{\mathrm{NL}} \simeq 13$.  Thus, if \fnl turns out to be of this magnitude, one could obtain a $3\sigma$ detection using methods completely independent of the primordial CMB temperature bispectrum.  The full results of these calculations are given in Fig.~\ref{tab:fNL37}.


\subsection{\Mnu}
\label{sec:Mnuconstraints}

We discuss massive neutrinos and their effects on the tSZ power spectrum in Sections~\ref{sec:LSS} and~\ref{sec:params}.  The current strongest bounds are in the range \Mnu$\lesssim 0.3$ eV~\cite{Sieversetal2013,Vikhlininetal2009,Mantzetal2010b,dePutteretal2012}.  However, a $3\sigma$ detection near this mass scale was recently claimed in~\cite{Houetal2012}.  From the top table in Fig.~\ref{tab:params}, it is clear that the unmarginalized constraints on $M_{\nu}$ from the tSZ power spectrum alone are quite strong.  The Planck tSZ power spectrum unmarginalized error $\Delta M_{\nu} \simeq 0.03$ eV is slightly smaller than the lower bounds from neutrino oscillations, and the unmarginalized error for a CV-limited experiment would lead to a very robust detection of $M_{\nu}$. The bounds from PIXIE are much weaker due to its lower angular resolution; the change induced by \Mnu in the tSZ power spectrum is effectively an overall amplitude shift, as seen in Fig.~\ref{fig.Mnuparamdep}, and thus one can gain much more leverage on this parameter by going to higher multipoles.  For PIXIE, one must mask heavily in order to reduce the CV errors to a sufficient level at the low multipoles where it observes in order to measure the effect of small neutrino masses.  Masking makes much less of a difference in the Planck and CV-limited constraints on \Mnu because most of their constraining power comes from higher multipoles where the masking procedure does not significantly reduce the error.

After adding the external Planck CMB+$H_0$ priors on the $\Lambda$CDM parameters (which are summarized in the bottom table of Fig.~\ref{tab:params}), the marginalized error for Planck with eROSITA masking is $\Delta M_{\nu} \simeq 0.37$ eV, comparable with current bounds. This could be useful to strengthen current bounds and confirm or reject a detection at this level.  For a CV-limited experiment, the error is $\Delta M_{\nu} \simeq 0.20$ eV, a factor of two better than Planck.  Note that we have derived these bounds after fully marginalizing over the primary $\Lambda$CDM parameters and the ICM gas physics parameters, so they are unlikely to be overly optimistic.  In addition, we have only used the information on \Mnu contained the tSZ power spectrum; we have not used any information from the Planck CMB temperature power spectrum regarding \Mnu.  Including primordial CMB constraints could significantly tighten the forecasted Planck errors to a level well below the current constraints.  Given the imminent release of the Planck sky maps, we leave this as an avenue to be pursued with data.

Thus far we have only considered constraints around our fiducial cosmology with \Mnu$=0$.  If instead we assume a fiducial \Mnu$=0.1$ eV, roughly corresponding to the minimum value in the inverted neutrino hierarchy~\cite{McKeown-Vogel2004}, we find that the marginalized Planck bound after eROSITA masking becomes slightly tighter than the current upper limits in the literature $\simeq 0.3$ eV.  Note that the forecasted errors depend on the fiducial neutrino mass assumed; if the actual neutrino mass is near the current upper limits, the eROSITA-masked Planck $1\sigma$ error would lie below the actual mass, suggesting a possible marginal detection.  However, we have focused on forecasts for \Mnu near the minimum allowed values in order to be conservative.

\begin{figure}
\begin{center}
\includegraphics[width=.6\textwidth]{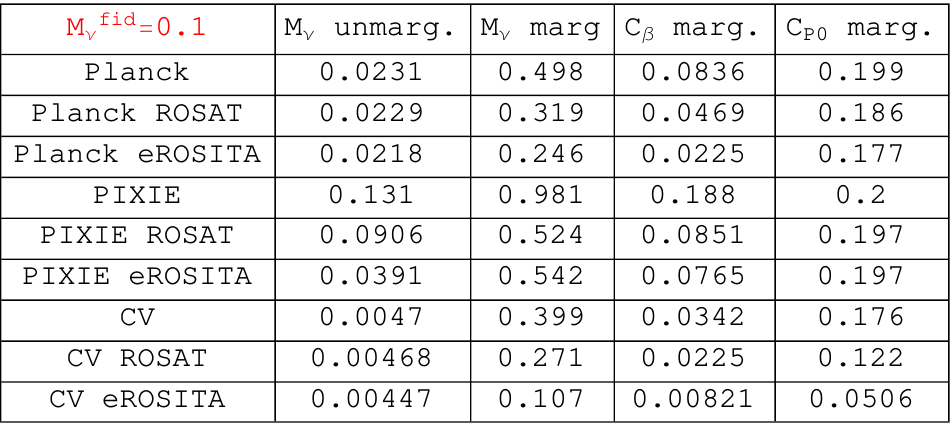}
\caption{The same as in the left central table of Fig.~\ref{tab:params}, but for a fiducial \Mnu$=0.1$ eV, similar to the minimum allowed value in the inverted neutrino hierarchy~\cite{McKeown-Vogel2004}. Note that all constraints on \Mnu are in units of eV. \label{tab:Mnu0p1}}
\end{center}
\end{figure}

\begin{figure}
\begin{center}
\includegraphics[width=\textwidth]{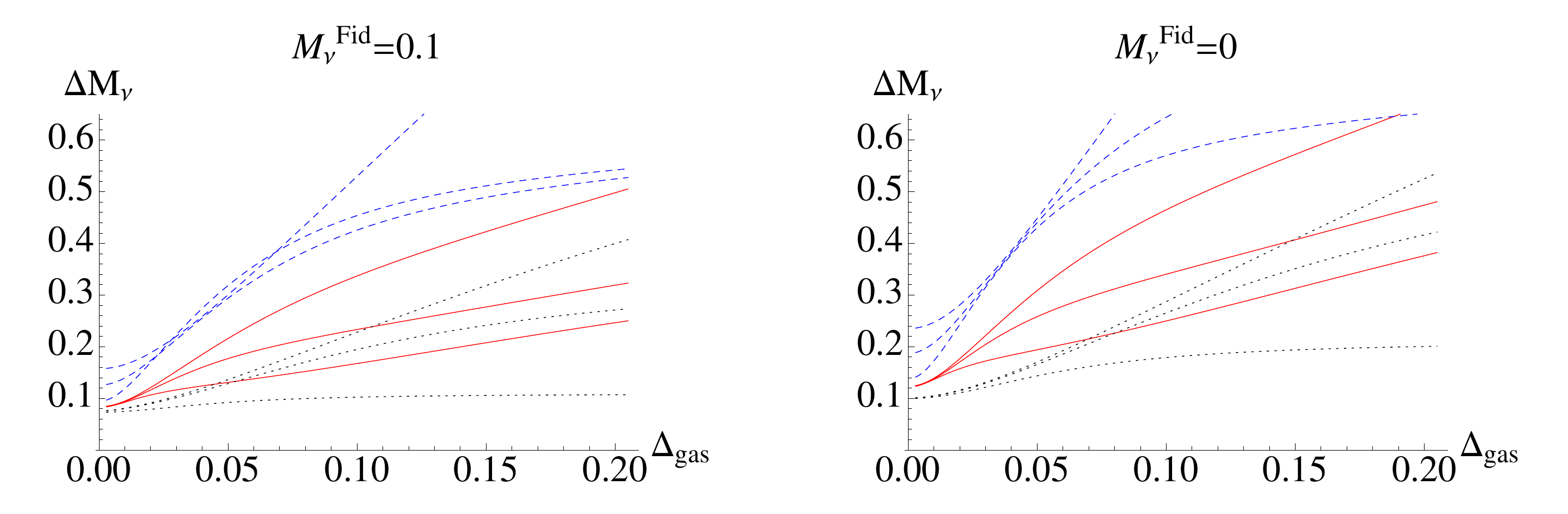}
\caption{These plots show the effect of strengthening the prior on the ICM gas physics parameters $C_{P_0}$ and $C_{\beta}$ on our forecasted $1\sigma$ error on the sum of the neutrino masses (vertical axis, in units of eV).  The left panel assumes a fiducial neutrino mass of \Mnu$=0.1$ eV, while the right panel assumes a fiducial value of \Mnu$=0$ eV.  The solid red curves show the results for Planck, assuming (from top to bottom) no masking, ROSAT masking, or eROSITA masking.  The blue dashed curves show the results for PIXIE, with the masking scenarios in the same order. The dotted black curves show the results for a CV-limited experiment, again with the masking scenarios in the same order.  The horizontal axis in both plots is the value of the $1\sigma$ Gaussian prior $\Delta C_{P_0} = \Delta C_{\beta} \equiv \Delta_{gas}$ placed on the gas physics parameters (our fiducial value is 0.2).  Note that we vary the priors simultaneously, keeping them fixed to the same value.  The plots demonstrate that even modest improvements in the external priors on $C_{P_0}$ and $C_{\beta}$ could lead to significant decreases in the expected error on \Mnu from tSZ power spectrum measurements. \label{fig.MnuCbetaprior}}
\end{center}
\end{figure}

More interestingly, we find that the primary degeneracy of \Mnu is with the ICM physics parameters $C_{P_0}$ and $C_{\beta}$.  These degeneracies are comparable to but much stronger than that of \Mnu with any of the $\Lambda$CDM parameters (after imposing the Planck CMB+$H_0$ prior), which suggests that strong external constraints on the ICM pressure profile would be of great use in constraining \Mnu using tSZ power spectrum measurements.  We investigate this issue in detail in Fig.~\ref{fig.MnuCbetaprior}, which shows the forecasted $1\sigma$ error on \Mnu as a function of the width of the Gaussian prior placed on $C_{P_0}$ and $C_{\beta}$, which is 0.2 in our standard case.  We provide plots for both the fiducial \Mnu$=0$ case and the case where the fiducial \Mnu$=0.1$ eV.  The latter plot indicates that strengthening the prior on $C_{P_0}$ and $C_{\beta}$ by a factor of two would lead to a $\simeq 40$\% decrease in the forecasted error on \Mnu for the eROSITA-masked Planck experiment.  Although the improvement eventually saturates as the degeneracies with other parameters become important for very tight priors on $C_{P_0}$ and $C_{\beta}$, these results demonstrate the importance of strong external constraints on the ICM physics in tightening the bounds on cosmological parameters from tSZ measurements.  Fortunately, it will be possible to derive such external constraints from detailed studies of the ICM pressure profile with Planck~(e.g., \cite{Plancketal2013}) and eROSITA themselves, which portends a bright future for tSZ-based cosmological constraints.

We conclude that the tSZ power spectrum is a promising probe for the neutrino masses and could lead to interesting results already using Planck data.  We find $\Delta M_{\nu} \simeq 0.3$ eV around a fiducial \Mnu$=0$ using the tSZ power spectrum alone to probe \Mnu (including other data in the \Mnu constraint would tighten this bound).  Considering larger values of the fiducial \Mnu and/or imposing stronger priors on the ICM gas physics parameters $C_{P_0}$ and $C_{\beta}$ can shrink this bound considerably, perhaps to the level of a $2\sigma$ measurement using Planck data alone.


\subsection{Other Parameters}
\label{sec:otherparamconstraints}

Although our focus in this paper has been on constraining currently unknown extensions to the $\Lambda$CDM standard model, we also compute the forecasted constraints on all of the standard parameters in our model.  Unfortunately, we find that adding the tSZ power spectrum data to the priors from the Planck CMB$+H_0$ Fisher matrix yields essentially no improvement in the forecasted errors, which are already very small using Planck$+H_0$ alone.  Even for a CV-limited experiment with eROSITA masking, the marginalized constraints on the five $\Lambda$CDM parameters in our model only improve by $\approx 1-2\%$ beyond the marginalized priors from Planck$+H_0$ (which are given in the bottom table in Fig.~\ref{tab:params}).  The primary cosmological utility of tSZ measurements is in providing a low-redshift probe of the amplitude of fluctuations, which permits a constraint on \Mnu when combined with a high-redshift probe of this amplitude from the CMB.

In contrast to the results for the cosmological parameters, the forecasted constraints on the ICM physics parameters $C_{P_0}$ and $C_{\beta}$ are more encouraging, as seen in the central tables in Fig.~\ref{tab:params}, as well as Figs.~\ref{tab:fNL37} and~\ref{tab:Mnu0p1}.  This is especially true of the outer slope parametrized by $C_{\beta}$, for which we forecast a $\approx 6-8\%$ constraint using the unmasked Planck data, which decreases to nearly the percent level after masking with eROSITA.  Most current observational measurements of this parameter do not have reported error bars, but the analysis of SPT stacked SZ profiles in~\cite{Plaggeetal2012} found a $\sim 40$\% uncertainty in this parameter after marginalization, which is probably a representative value.  We thus expect this error to decrease dramatically very shortly.  However, it will be difficult for Planck or PIXIE to constrain the overall normalization of the ICM pressure profile using the tSZ power spectrum alone due to its strong degeneracy with the cosmological parameters (e.g., $\sigma_8$).  This is reflected in the fact that the marginalized constraints on $C_{P_0}$ in Figs.~\ref{tab:params}--\ref{tab:Mnu0p1} are often near the bound imposed by our prior, the exceptions being either of the masked Planck measurements or any of the CV-limited measurements.  The best approach to obtaining constraints on $C_{P_0}$ is likely using cross-correlations between the tSZ signal and lensing maps of the dark matter distribution.  However, it is also worth noting that we have not considered the minimal $\Lambda$CDM case in these calculations, in which both \fnl and \Mnu are fixed to zero.  In such a scenario, the bounds on the ICM physics parameters would be stronger than those quoted here.  However, given that we know \Mnu$>0$ in our universe, it is perhaps most reasonable to look at the marginalized constraints in that case as an example of future constraints on $C_{P_0}$ and $C_{\beta}$.


\subsection{Forecasted SNR}
\label{sec:SNR}

The cumulative SNR on the tSZ power spectrum using multipoles $\ell < \ell_{max}$ is given by
\be
\label{eq.CumSNR}
\mathrm{SNR}_{cumul}(\ell<\ell_{max}) = \sqrt{\sum_{{\ell,\ell'}=2}^{\ell_{max}} C_{\ell}^y (M_{\ell\ell'}^y)^{-1} C_{\ell'}^y} \,,
\ee
where $(M_{\ell\ell'}^y)^{-1}$ refers to the inverse of the $\ell_{max}$-by-$\ell_{max}$ submatrix of the full covariance matrix.  The cumulative SNR provides a simple way to assess the constraining power of a given experimental and masking choice on the tSZ power spectrum, without regard to constraints on particular parameters.  We compute Eq.~(\ref{eq.CumSNR}) for our fiducial cosmology using each of the experiment (Planck, PIXIE, CV-limited) and masking options (unmasked, ROSAT-masked, eROSITA-masked) considered in the previous sections.  Note that the covariance matrix in Eq.~(\ref{eq.CumSNR}) includes all contributions from cosmic variance (Gaussian and non-Gaussian) and experimental noise after foreground removal, as discussed in Section~\ref{sec:cov}.  Note that our approach includes the trispectrum (or sample variance) contribution to the covariance matrix in calculating the SNR, which is perhaps more conservative than an approach in which only the Gaussian errors are considered in assessing the SNR.

\begin{figure}
  \begin{center}
    \includegraphics[trim=0cm 0cm 0cm 0cm, clip=true, totalheight=0.5\textheight]{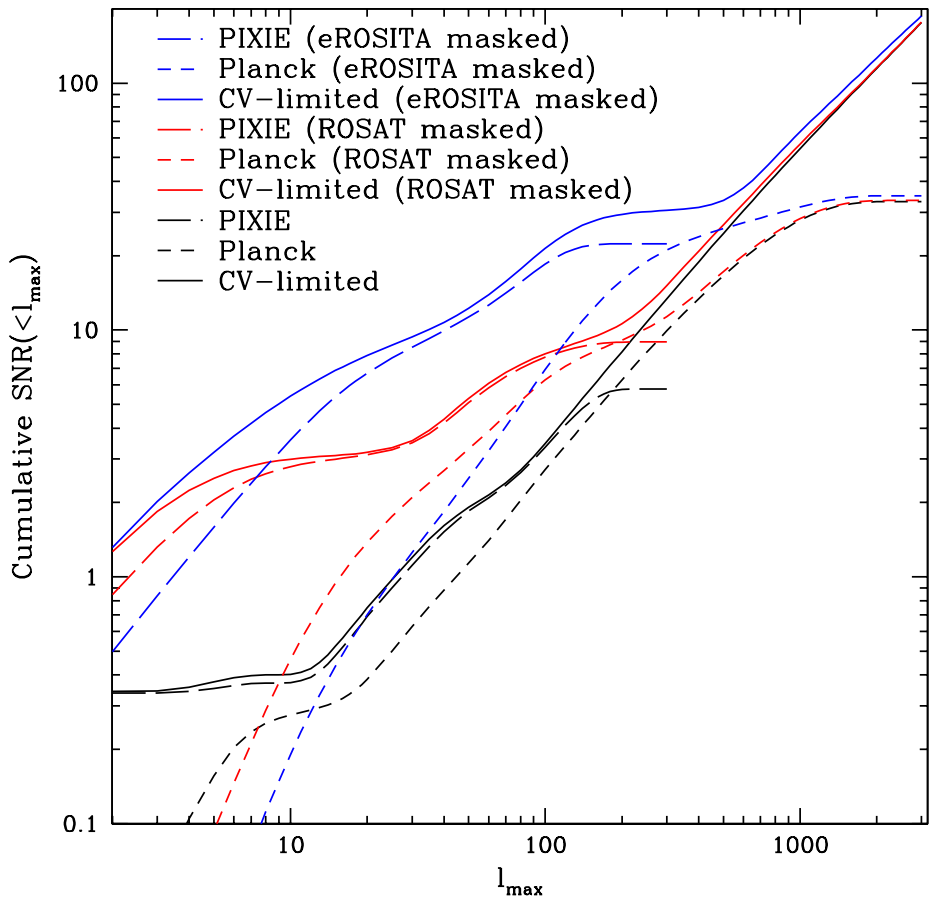}
    \caption{This plot shows the cumulative SNR achievable on the tSZ power spectrum for each of nine different experimental and making scenarios.  The solid curves display results for a CV-limited experiment, the short dashed curves show results for Planck, and the long dashed curves show results for PIXIE.  The different colors correspond to different masking options, as noted in the figure.  Note that PIXIE is close to the CV limit over its signal-dominated multipole range.  The total SNR using the imminent Planck data is $\approx 35$, essentially independent of the masking option used (note that masking can increase the cumulative SNR up to lower multipoles, however, as compared to the unmasked case).\label{fig.SNRplot}}
  \end{center}
\end{figure}

The results of these calculations are shown in Fig.~\ref{fig.SNRplot}.  We find that Planck can detect the tSZ power spectrum with a cumulative SNR $\approx 35$ using multipoles $\ell < 3000$ (see Appendix~\ref{appendixPlanck} for a comparison with the recently-released initial tSZ results from the Planck collaboration~\cite{Plancketal2013b}).  This result is essentially independent of the masking scenario, although masking more heavily can lead to greater cumulative SNR using lower values of $\ell_{max}$, as compared to the unmasked case.  However, masking leads to significant improvement in the PIXIE results: the unmasked PIXIE cumulative SNR using $\ell < 300$ is $\approx 5.8$, while the ROSAT- and eROSITA-masked results are $\approx 8.9$ and $22$, respectively.  These results follow from the fact that PIXIE is nearly CV-limited for $\ell < 100$, as seen in the unmasked curves shown in Fig.~\ref{fig.SNRplot} (and discussed earlier in Section~\ref{sec:cov}).  For the masked cases, the CV errors are reduced sufficiently that the PIXIE noise starts to become important at $\ell \lesssim 10$.

There is one subtlety of the masking procedure that can be understood by considering Fig.~\ref{fig.SNRplot} in combination with Figs.~\ref{fig.covcomps}--\ref{fig.covcompssupermasked}.  Looking at the ROSAT-masked case for Planck in Fig.~\ref{fig.covcompsmasked}, it appears that the Planck errors are dominated by the Gaussian instrumental noise term at all $\ell$, and hence that masking further for Planck should be harmful rather than helpful; this appears to be confirmed by the fact that the Planck fractional errors in the eROSITA-masked case in Fig.~\ref{fig.fNLMnuwerrssupermasked} are indeed larger than in the ROSAT-masked case in Fig.~\ref{fig.fNLMnuwerrsmasked}.  However, it is clear in Fig.~\ref{fig.SNRplot} that the eROSITA-masked Planck case has a larger total SNR than the ROSAT-masked case.  The resolution of this apparent discrepancy lies in the fact that the masking continues to suppress the off-diagonal terms in the covariance matrix, which arise solely from the trispectrum term in Eq.~(\ref{eq.Mellellpr}), even as the on-diagonal fractional errors begin to increase.  The plots in Figs.~\ref{fig.fNLMnuwerrs}--\ref{fig.covcompssupermasked} only show the diagonal entries in the covariance matrix, and thus one may not realize the effect of the masking on the off-diagonal terms in the covariance matrix without examining the cumulative SNR (this result is also implied by the improved parameter constraints for the eROSITA-masked Planck case given in the previous sections).  These considerations imply that there is likely an optimal masking choice for a given experimental noise level and survey specifications, but obtaining the precise answer to this question lies beyond the scope of this paper.

Overall, Fig.~\ref{fig.SNRplot} indicates that near-term data promises highly significant detections of the tSZ power spectrum on larger angular scales than have been probed thus far by ACT and SPT.

\section{Discussion and Outlook}
\label{sec:discussion}

In this paper we have performed a comprehensive analysis of the possible constraints on cosmological and astrophysical parameters achievable with measurements of the tSZ power spectrum from upcoming full-sky CMB observations, with a particular focus on extensions to the $\Lambda$CDM standard model parametrized by \fnl and \Mnu.  We have included all of the important physical effects due to these additional parameters, including the change to the halo mass function and the scale-dependent halo bias induced by primordial non-Gaussianity.  Our halo model calculations of the tSZ power spectrum include both the one- and two-halo terms, and we use the exact expressions where necessary to obtain accurate results on large angular scales.  We model the ICM pressure profile using parameters that have been found to agree well with existing constraints, and furthermore we model the uncertainty in the ICM physics by freeing two of these parameters.  We also include a realistic treatment of the instrumental noise for the Planck and PIXIE experiments, accounting for the effects of foregrounds by using a multifrequency subtraction technique.  Our calculations of the covariance matrix of the tSZ power spectrum include both the Gaussian noise terms and the non-Gaussian cosmic variance term due to the tSZ trispectrum.  We investigate two masking scenarios motivated by the ROSAT and eROSITA all-sky surveys, which significantly reduce the large errors that would otherwise be induced by the trispectrum term, especially at low-$\ell$.  Finally, we use these calculations to forecast constraints on \fnl, \Mnu, the primary $\Lambda$CDM parameters, and two parameters describing the ICM pressure profile.

Our primary findings are as follows:
\begin{itemize}
\item The tSZ power spectrum can be detected with a total SNR $>30$ using the imminent Planck data up to $\ell=3000$, regardless of masking (see Appendix~\ref{appendixPlanck} for a comparison with the initial tSZ power spectrum results released by the Planck collaboration while this manuscript was under review~\cite{Plancketal2013b}) ;
\item The tSZ power spectrum can be detected with a total SNR between $\approx 6$ and 22 using the future PIXIE data up to $\ell=300$, with the result being sensitive to the level of masking applied to remove massive, nearby clusters;
\item Adding the tSZ power spectrum information to the forecasted constraints from the Planck CMB temperature power spectrum and existing $H_0$ data is unlikely to significantly improve constraints on the primary cosmological parameters, but may give interesting constraints on the extensions we consider:
\begin{itemize}
\item If the true value of \fnl is near the WMAP9 ML value of $\approx 37$, a future CV-limited experiment combined with eROSITA-masking could provide a $3\sigma$ detection, completely independent of the primordial CMB temperature bispectrum; alternatively, PIXIE could give $1-2\sigma$ evidence for such a value of \fnl with this level of masking;
\item If the true value of \Mnu is near 0.1 eV, the Planck tSZ power spectrum with eROSITA masking can provide upper limits competitive with the current upper bounds on \Mnu; with stronger external constraints on the ICM physics, Planck with eROSITA masking could provide $1-2\sigma$ evidence for massive neutrinos from the tSZ power spectrum, depending on the true neutrino mass;
\end{itemize}
\item Regardless of the cosmological constraints, Planck will allow for a very tight constraint on the logarithmic slope of the ICM pressure profile in the outskirts of galaxy clusters, and may also provide some information on the overall normalization of the pressure profile (which sets the zero point of the $Y-M$ relation).
\end{itemize}

Our results are subject to a few caveats.  We have made the usual Fisher matrix approximation that the likelihood function is nearly Gaussian around our fiducial parameter values, but this should be safe for small variations, which are all that we consider (in particular, $\sigma_8$ is tightly constrained by the external Planck CMB prior, and it would be most likely to have a non-Gaussian likelihood).  We have also neglected any tSZ signal from the intergalactic medium, filaments, or other diffuse structures, but the comparison between simulations and halo model calculations in~\cite{Battagliaetal2012} indicates that this approximation should be quite good.  We have also assumed that the mass function parameters are perfectly well known, while in reality some uncertainties remain, especially in the exponential tail.  Given that our most optimistic results involve masking nearly all of the clusters that live in the exponential tail, we believe that our forecasts should be fairly robust to the mass function uncertainties, in contrast to cluster count calculations which are highly sensitive to small changes in the tail of the mass function.  Finally, we have only included the flat-sky version of the one-halo term in our computations of the tSZ power spectrum covariance matrix.  For the masked calculations, the flat-sky result should suffice, since massive, nearby clusters are removed; however, it is possible that the two-, three-, or four-halo terms could eventually become relevant in the masked calculations.  These would be largest at low-$\ell$, however, where our primary interest is in constraining \fnl, which does not appear very optimistic in any case.  Thus, we neglect these corrections for our purposes.

There are many future extensions of this work involving higher-order tSZ statistics and cross-correlations with other tracers of large-scale structure.  Recent work on the tSZ bispectrum and skewness~\cite{Wilsonetal2012,Hill-Sherwin2013,Bhattacharyaetal2012,Crawfordetal2013} indicates that significantly stronger constraints on both cosmology and the ICM physics can be obtained by using higher-order statistics.  These may also be a better place to look for \fnl constraints, as the additional powers of the halo bias could lead to a larger signal at low-$\ell$ than in the power spectrum (controlling systematics will be of paramount importance, as will masking to reduce the very large cosmic variance due to the tSZ six-point function).  Determining an optimal strategy to extract the neutrino mass through combinations of tSZ statistics and cross-spectra with other tracers is also work in progress.  The key factor remains breaking the degeneracy with the ICM physics, or, more optimistically, simultaneously constraining both the ICM and cosmological parameters using tSZ measurements.


\acknowledgments{We are thankful to Masahiro Takada for providing a computation of the Planck CMB Fisher matrix.  We are also grateful to Kendrick Smith for his suggestion of using the multifrequency subtraction techniques to remove foregrounds, as well as many other insightful conversations.  We thank Nick Battaglia, Bruce Draine, Eiichiro Komatsu, Marilena LoVerde, Blake Sherwin, David Spergel, and Matias Zaldarriaga for a number of helpful exchanges.  JCH is supported by NASA Theory Grant NNX12AG72G.  EP is supported in part by the Department of Energy grant DE-FG02-91ER-40671.}


\appendix
\section{Halo Model Derivation of tSZ Statistics}
\label{appendix}
In the halo model, it is assumed that all matter in the universe is bound in halos.  Each halo of virial mass $M$ is assumed to have a density profile, $\rho(\vec{x};M)$, and (for our purposes) an electron pressure profile, $P_e(\vec{x};M)$.  The mass density field at position $\vec{x}$ is then given by the sum of the contributions from all halos in the universe:
\beq
\label{eq.densityfield}
\rho(\vec{x}) = \sum_{i \, \in \, \mathrm{halos}} \rho(\vec{x}-\vec{x}_i;M_i) \,.
\eeq
Similarly, the electron pressure field at position $\vec{x}$ is given by:
\beq
\label{eq.pressurefield}
P_e(\vec{x}) = \sum_{i \, \in \, \mathrm{halos}} P_e(\vec{x}-\vec{x}_i;M_i) \,.
\eeq

For calculations involving the tSZ effect, it is convenient to define a ``3D Compton-$y$'' field that is simply a re-scaling of the electron pressure field:
\beq
\label{eq.3Dcomptonyfield}
y_{3D}(\vec{x}) = \frac{\sigma_T}{m_e c^2} P_e(\vec{x}) \,.
\eeq
Note that the 3D Compton-$y$ field has dimensions of inverse length (it is thus important to be careful about comoving versus physical units --- in our calculations using the Battaglia pressure profile, the pressure is given in physical units, and thus so is $y_{3D}$).  The usual (2D) Compton-$y$ field is then given by the LOS projection of $y_{3D}(\vec{x})$:
\beqn
y(\hat{n}) & = & \int c \, dt \, y_{3D}(\vec{x}(\chi(t),\hat{n})) \nonumber \\
  & = & \int d\chi \, a(\chi) y_{3D}(\vec{x}(\chi,\hat{n})) \,,
\label{eq.Comptony}
\eeqn
where $t$ is the age of the universe at a given epoch, $\chi(t)$ is the comoving distance to that epoch, $a(\chi)$ is the scale factor at that epoch, and $\hat{n}$ is a unit vector on the sky.  We have used $dt/da = 1/(aH)$ and $d\chi/da = -c/(Ha^2)$ in going from the first line to the second line, where $H(a)$ is the Hubble parameter.  
Defining the projection kernel $W^{y}(\chi)$ for the Compton-$y$ field via $y(\vec{\hat{n}}) = \int d\chi W^{y}(\chi) y_{3D}(\vec{x}(\chi,\hat{n}))$, we thus have: 
\beq
\label{eq.WtSZ}
W^{y}(\chi) = a(\chi) \,.
\eeq

Starting from the 3D Compton-$y$ field defined in Eq.~(\ref{eq.3Dcomptonyfield}), we derive the angular power spectrum of the 2D Compton-$y$ field.  First, we calculate the relevant 3D power spectrum by means of the halo model (N.B. in this expression and many others in the following, the redshift dependence will be suppressed for notational simplicity):
\beq
\label{eq.Pofk}
P_{y_{3D}}(\vec{k}) = P_{y_{3D}}^{1h}(\vec{k}) + P_{y_{3D}}^{2h}(\vec{k}) \,,
\eeq
where the one-halo term is
\beq
\label{eq.y3D1h}
P_{y_{3D}}^{1h}(\vec{k}) = \int dM \frac{dn}{dM} \left| \tilde{y}_{3D}(\vec{k};M) \right|^2
\eeq
and the two-halo term is
\beq
\label{eq.y3D2h}
P_{y_{3D}}^{2h}(\vec{k}) = \int dM_1 \frac{dn}{dM_1} b(M_1) \tilde{y}_{3D}(\vec{k};M_1) \int dM_2 \frac{dn}{dM_2} b(M_2) \tilde{y}_{3D}(\vec{k};M_2 ) P_{\mathrm{lin}}(\vec{k}) \,.
\eeq
In these equations, $\tilde{y}_{3D}(\vec{k};M)$ is the Fourier transform of the 3D Compton-$y$ profile for a halo of virial mass $M$:
\beqn
\tilde{y}_{3D}(\vec{k};M) & = & \int d^3 r \, e^{-i \vec{k} \cdot \vec{r}} y_{3D}(\vec{r};M) \nonumber \\
				      & = & \int dr \, 4 \pi r^2 \frac{\sin(kr)}{kr} y_{3D}(r;M) \,,
\label{eq.y3Dtwid}
\eeqn
where $r=|\vec{r}|$, $k=|\vec{k}|$, and we have assumed that $y_{3D}(\vec{r};M)$ is spherically symmetric to obtain the second expression.  Also, in Eqs.~(\ref{eq.y3D1h}) and (\ref{eq.y3D2h}), $dn(M,z)/dM$ is the comoving number density of halos of mass $M$ at redshift $z$, $b(M,z)$ is the bias of halos of mass $M$ at redshift $z$ (which we will later consider to be scale-dependent) and $P_{\mathrm{lin}}(\vec{k})$ is the linear matter power spectrum, as defined in Section~\ref{sec:HMF}.  At this point, it is worth emphasizing that expressions analogous to Eqs.~(\ref{eq.Pofk})--(\ref{eq.y3D2h}) can be written for any field defined at all points in the universe after its profile for each halo of mass $M$ is specified.  These expressions are generic consequences of the halo model.  The primary assumption made is that the halo-halo power spectrum for halos of mass $M_1$ and $M_2$ is given by the linear matter power spectrum multiplied by the relevant bias parameters:
\beq
\label{eq.Phh}
P_{hh}(\vec{k};M_1,M_2) =  b(M_1) b(M_2) P_{\mathrm{lin}}(\vec{k}) \,.
\eeq

\subsection{The One-Halo Term}
We now compute the contribution of the one-halo term to the tSZ power spectrum.  We do the exact calculation first, and then consider the flat-sky limit.  Note that for the one-halo term, the notion of the ``Limber approximation'' is not well-defined --- there are no LOS cancellations to consider, since one halo is by definition fixed at a single redshift.  Thus, in going from the exact calculation to the small-angle (high-$\ell$) limit, we only need to consider the projection of the electron pressure profile from 3D to 2D.  Note that this generalization will only affect very massive, low-redshift clusters, which subtend a significant solid angle on the sky.

Consider a cluster of virial mass $M$ at comoving separation $\vec{\chi}$ with respect to our location.  The 3D Compton-$y$ field due to this cluster at comoving separation $\vec{r}$ with respect to the cluster center is given by:
\beq
\label{eq.y3Dtwidinv}
y_{3D}(\vec{r};M) = \int \frac{d^3 k}{(2\pi)^3} \tilde{y}_{3D} (\vec{k};M) e^{i\vec{k} \cdot (\vec{r}-\vec{\chi})} \,.
\eeq
This expression is simply the inverse transform of Eq.~(\ref{eq.y3Dtwid}).  Projecting along the LOS as in Eq.~(\ref{eq.Comptony}) and using the Rayleigh plane wave expansion, we obtain:
\beqn
y(\hat{n};M) & = & \int d\chi' W^y(\chi') y_{3D}(\vec{r}(\chi',\hat{n});M) \nonumber \\
 & = & \int d\chi' a(\chi') \int \frac{d^3 k}{(2\pi)^3} \tilde{y}_{3D} (\vec{k};M) \left[ \sum_{\ell m} 4\pi i^{\ell} Y^{*}_{\ell m}(\hat{k}) Y_{\ell m} (\hat{n}) j_{\ell}(k\chi') \right] e^{-i\vec{k} \cdot \vec{\chi}} \,,
\label{eq.Comptonyprojgen}
\eeqn
where $\hat{n}$ is a unit vector on the sky and $\hat{k}$ is the direction of $\vec{k}$.  Defining the expansion coefficients $y_{\ell m}(M)$ via 
\beq
\label{eq.ysphharmcoeff}
y(\hat{n};M) = \sum_{\ell m} y_{\ell m}(M) Y_{\ell m}(\hat{n}) \,,
\eeq
we can read them off from Eq.~(\ref{eq.Comptonyprojgen}):
\beqn
y_{\ell m}(M) & = & \int d\chi' a(\chi') \int \frac{d^3 k}{(2\pi)^3} \tilde{y}_{3D} (\vec{k};M) 4\pi i^{\ell} Y^{*}_{\ell m}(\hat{k}) j_{\ell}(k\chi') \left[ \sum_{\ell' m'} 4\pi (-i)^{\ell'} Y_{\ell' m'}(\hat{k}) Y^{*}_{\ell' m'} (\hat{\chi}) j_{\ell'}(k\chi) \right] \nonumber \\
 & = & \int d\chi' a(\chi') \int \frac{2}{\pi} k^2 \, dk \, \tilde{y}_{3D}(k;M) j_{\ell}(k\chi') j_{\ell}(k\chi) Y^{*}_{\ell m} (\hat{\chi}) \,,
\label{eq.ylm}
\eeqn
where we have again used the Rayleigh expansion and have used the orthonormality of the spherical harmonics to do the integral over $\hat{k}$ in going from the first line to the second.  From this expression, we can read off the exact result for the 2D Fourier transform of the projected $y$-profile due to a cluster of mass $M$ at redshift $z$:
\beqn
\tilde{y}_{2D}(\ell;M,z) & = & \int d\chi' a(\chi') \int \frac{2}{\pi} k^2 \, dk \, j_{\ell}(k\chi') j_{\ell}(k\chi(z)) \, \tilde{y}_{3D}(k;M,z) \nonumber \\
   & = & \frac{1}{\sqrt{\chi(z)}} \int \frac{d\chi'}{\sqrt{\chi'}} a(\chi') \int k \, dk \, J_{\ell+1/2}(k\chi') J_{\ell+1/2}(k\chi) \tilde{y}_{3D}(k;M,z) \,,
\label{eq.yelltwidexact1}
\eeqn
where we have rewritten the spherical Bessel functions in terms of Bessel functions of the first kind using $j_{\nu}(x) = \sqrt{\frac{\pi}{2x}} J_{\nu+1/2}(x)$ and we have explicitly included a possible dependence of the $y_{3D}$ profile on redshift (in addition to mass).

The total one-halo term in the tSZ power spectrum is then given by the sum of the individual contributions from every cluster in the universe:
\beq
\label{eq.yCl1hexact1}
C_{\ell}^{y,1h} = \int dz \frac{d^2V}{dz d\Omega} \int dM \frac{dn(M,z)}{dM} \left| \tilde{y}_{2D}(\ell;M,z) \right|^2 \,,
\eeq
where $d^2V/dz d\Omega = c \chi^2(z)/H(z)$ is the comoving volume element per steradian.  Substituting Eq.~(\ref{eq.yelltwidexact1}) into this expression, converting the $\chi'$ integral to a redshift integral, and rearranging the order of the integrals then yields the final result for the exact one-halo term:
\beq
C_{\ell}^{y,1h} = \int \frac{dz}{\chi(z)} \frac{d^2V}{dz d\Omega} \int dM \frac{dn}{dM} \left| \int k \, dk \, J_{\ell+1/2}(k\chi(z)) \tilde{y}_{3D}(k;M,z) \int \frac{c \, dz'}{H(z')(1+z')\sqrt{\chi(z')}} J_{\ell+1/2}(k\chi(z')) \right|^2 \,.
\label{eq.yCl1hexact2}
\eeq
Note that this expression is exact: no flat-sky approximation (or any other) has been used in deriving Eq.~(\ref{eq.yCl1hexact2}). 

In order to recover the flat-sky (i.e., small-angle) limit of Eq.~(\ref{eq.yCl1hexact2}), we use the following $\ell \rightarrow \infty$ limit for the spherical Bessel functions:
\beq
j_{\ell}(x) \rightarrow \sqrt{\frac{\pi}{2\ell+1}} \delta_D(\ell+1/2-x) \,.
\label{eq.sphBessiden}
\eeq
Applying this limit to the first line of Eq.~(\ref{eq.yelltwidexact1}) yields
\beq
\tilde{y}_{2D}(\ell \gg 1;M,z) \approx \frac{a(z)}{\chi^2(z)} \tilde{y}_{3D} \left( \frac{\ell+1/2}{\chi(z)};M,z \right) \,.
\eeq
Using Eq.~(\ref{eq.y3Dtwid}), we can simplify this expression into a familiar form:
\beqn
\tilde{y}_{2D}(\ell \gg 1;M,z) & \approx & \frac{a(z)}{\chi^2(z)} \int dr \, 4 \pi r^2 \frac{\sin((\ell+1/2) r/\chi)}{(\ell+1/2) r/\chi} y_{3D}(r;M) \nonumber \\
   & = & \frac{4 \pi r_s}{\ell_s^2} \int dx \, x^2 \frac{\sin((\ell+1/2) x/\ell_s)}{(\ell+1/2) x/\ell_s} y_{3D}(x;M) \,,
\label{eq.yelltwidderivation}
\eeqn
where we have performed the following change of variables in the integral over $y_{3D}$: $x \equiv a(z)r/r_s$, where $r_s$ is a characteristic scale radius of the $y_{3D}$ profile.  Finally, $\ell_s = a(z)\chi(z)/r_s = d_A(z)/r_s$ is the characteristic multipole moment associated with the scale radius, with $d_A(z)$ the angular diameter distance.  Note that the change of variables involved the scale factor because we transformed from comoving coordinates to physical coordinates.  Eq.~(\ref{eq.yelltwidderivation}) is identical to the quantity $\tilde{y}_{\ell}(M,z)$ defined in (for example) Eq.~(2) of~\cite{Komatsu-Seljak2002}, although we have explicitly used $\ell+1/2$ rather than $\ell$.  This is both technically correct and reduces the error in the approximation from $\mathcal{O}(\ell^{-1})$ to $\mathcal{O}(\ell^{-2})$~\citep{LoVerde-Afshordi2008}.  Eq.~(\ref{eq.yelltwidderivation}) is simply the flat-sky limit of Eq.~(\ref{eq.yelltwidexact1}).  Following the long-standing convention, we will use the same definition as that established in~\cite{Komatsu-Seljak2002}:
\beqn
\tilde{y}_{\ell}(M,z) & \equiv & \tilde{y}_{2D}(\ell \gg 1;M,z) \nonumber \\
   & \approx & \frac{4 \pi r_s}{\ell_s^2} \int dx \, x^2 \frac{\sin((\ell+1/2) x/\ell_s)}{(\ell+1/2) x/\ell_s} y_{3D}(x;M,z) \,.
\label{eq.yelltwid}
\eeqn
The flat-sky limit of the one-halo term given in Eq.~(\ref{eq.yCl1hexact2}) is thus given by
\beq
\label{eq.yCl1hflatsky}
C_{\ell \gg 1}^{y,1h} \approx \int dz \frac{d^2V}{dz d\Omega} \int dM \frac{dn(M,z)}{dM} \left| \tilde{y}_{\ell}(M,z) \right|^2 \,,
\eeq
as written down in (for example) Eq.~(1) of~\cite{Komatsu-Seljak2002}.

Evaluating Eq.~(\ref{eq.yCl1hexact2}) numerically is somewhat computationally expensive, as it contains five nested integrals (including the Fourier transform to obtain $\tilde{y}_{3D}$), two of which involve highly oscillatory Bessel functions.  However, the flat-sky limit in Eq.~(\ref{eq.yCl1hflatsky}) contains only three nested integrals, and involves no oscillatory functions.  Furthermore, for our fiducial cosmology, we find that the flat-sky result in Eq.~(\ref{eq.yCl1hflatsky}) only overestimates the exact result in Eq.~(\ref{eq.yCl1hexact2}) by $\approx 13$\%, $5$\%, and $3$\% at $\ell=2$, $10$, and $20$, respectively.  By $\ell=60$, the two results are identical within our numerical precision.  In addition, at $\ell=2$ where the correction is largest, the one-halo term is only $\approx 67\%$ as large as the two-halo term, and thus the total tSZ power spectrum is only overestimated by $\approx 5\%$.  Note that for non-Gaussian cosmologies this overestimate is far smaller, because the two-halo term dominates by a much larger amount at low-$\ell$ than in a Gaussian cosmology (for example, the two-halo term at $\ell=2$ is 2.2 times as large as the one-halo term for \fnl$=50$).  Given the small size of this correction and the significant computational expense required to evaluate the exact expression, we thus use the flat-sky result in Eqs.~(\ref{eq.yelltwid}) and (\ref{eq.yCl1hflatsky}) to compute the one-halo contribution to the tSZ power spectrum in this work.

\subsection{The Two-Halo Term}
We now compute the contribution of the two-halo term to the tSZ power spectrum.  We do the exact calculation first, and then consider the Limber-approximated (small-angle) limit.  The exact result is necessary for studying the signature of the scale-dependent halo bias induced by primordial non-Gaussianity on the tSZ power spectrum, since the effect is only significant at very low $\ell$.  To calculate the two-halo contribution to the angular power spectrum of the (2D) Compton-$y$ field, we project Eq.~(\ref{eq.y3D2h}) along the LOS using the projection kernel in Eq.~(\ref{eq.WtSZ}), which gives:
\beqn
C_{\ell}^{y,2h} & = & \int d\chi_1 W^{y}(\chi_1) \int d\chi_2 W^{y}(\chi_2) \int \frac{2 k^2 dk}{\pi} j_{\ell}(k\chi_1) j_{\ell}(k\chi_2) P_{y_{3D}}^{2h}(k) \nonumber \\ 
  & = & \int d\chi_1 \frac{W^{y}(\chi_1)}{\sqrt{\chi_1}} \int d\chi_2 \frac{W^{y}(\chi_2)}{\sqrt{\chi_2}} \int dk \, k \, J_{\ell+1/2}(k\chi_1) J_{\ell+1/2}(k\chi_2) P_{y_{3D}}^{2h}(k) \nonumber \\
  & = & \int d\chi_1 \frac{a(\chi_1)}{\sqrt{\chi_1}} \int d\chi_2 \frac{a(\chi_2)}{\sqrt{\chi_2}} \int dk \, k \, J_{\ell+1/2}(k\chi_1) J_{\ell+1/2}(k\chi_2) P_{y_{3D}}^{2h}(k) \nonumber \\
  & = & \int dz_1 \frac{c}{H(z_1)} \frac{a(z_1)}{\sqrt{\chi(z_1)}} \int dz_2 \frac{c}{H(z_2)} \frac{a(z_2)}{\sqrt{\chi(z_2)}} \int dk \, k \, J_{\ell+1/2}(k\chi(z_1)) J_{\ell+1/2}(k\chi(z_2)) \times \nonumber \\
  &    & \int dM_1 \frac{dn}{dM_1} b(k,M_1,z_1) \tilde{y}_{3D}(k;M_1,z_1) \int dM_2 \frac{dn}{dM_2} b(k,M_2,z_2) \tilde{y}_{3D}(k;M_2,z_2) P_{\mathrm{lin}}(k;z_1,z_2) \nonumber \\
  & = & \int \frac{dz_1}{\sqrt{\chi(z_1)}} \frac{d^2V}{dz_1 d\Omega} \int \frac{dz_2}{\sqrt{\chi(z_2)}} \frac{d^2V}{dz_2 d\Omega} \int dk \, k \, J_{\ell+1/2}(k\chi(z_1)) J_{\ell+1/2}(k\chi(z_2)) P_{\mathrm{lin}}(k;z_1,z_2) \times \nonumber \\
  &    & \int dM_1 \frac{dn}{dM_1} b(k,M_1,z_1) \tilde{y}_{k\chi_1}(M_1,z_1) \int dM_2 \frac{dn}{dM_2} b(k,M_2,z_2) \tilde{y}_{k\chi_2}(M_2,z_2) \nonumber \\
  & = & \int dk \, k \, \frac{P_{\mathrm{lin}}(k;z_{in})}{D^2(z_{in})} \left[ \int \frac{dz}{\sqrt{\chi(z)}} \frac{d^2V}{dz d\Omega} J_{\ell+1/2}(k\chi(z)) D(z) \int dM \frac{dn}{dM} b(k,M,z) \tilde{y}_{k\chi(z)}(M,z) \right]^2 \,,
\label{eq.yCl2hexact}
\eeqn
where we have again used $j_{\nu}(x) = \sqrt{\frac{\pi}{2x}} J_{\nu+1/2}(x)$ and the notation $P_{\mathrm{lin}}(k;z_1,z_2)$ refers to the re-scaling of the linear matter power spectrum by the growth factor $D(z)$:
\beq
\label{eq.Pofkrescaled}
P_{\mathrm{lin}}(k;z_1,z_2) = \frac{D(z_1) D(z_2)}{D^2(z_{in})} P_{\mathrm{lin}}(k;z_{in}) \,,
\eeq
where $z_{in}$ is a reasonable input redshift for the linear theory matter power spectrum (e.g., our choice is $z_{in}=30$).  Also, note that we have explicitly included the possible scale-dependence of the bias, $b(k,M,z)$, as arises in cosmologies with local primordial non-Gaussianity.  Finally, the notation $\tilde{y}_{k\chi}(M,z)$ in Eq.~(\ref{eq.yCl2hexact}) refers to the expression for $\tilde{y}_{\ell}(M,z)$ given in Eq.~(\ref{eq.yelltwid}) evaluated with $\ell+1/2=k\chi$.  This notation is simply a mathematical convenience; no flat-sky or Limber approximation was used in deriving Eq.~(\ref{eq.yCl2hexact}), and no $\ell$ appears in $\tilde{y}_{k\chi}(M,z)$.  Note that this expression only requires the evaluation of four nested integrals (whereas the exact one-halo term required five), although the redshift integrand is highly oscillatory due to the Bessel function.

In order to recover the Limber-approximated (i.e., small-angle) limit of Eq.~(\ref{eq.yCl2hexact}), we again use Eq.~(\ref{eq.sphBessiden}) given above.  This step is most easily accomplished starting from the first line of the derivation that led to Eq.~(\ref{eq.yCl2hexact}), which yields:
\beqn
C_{\ell \gg 1}^{y,2h} & \approx & \int d\chi_1 W^{y}(\chi_1) \int d\chi_2 W^{y}(\chi_2) \int \frac{k^2 dk}{\ell+1/2} \delta_D(\ell+1/2-k\chi_1) \delta_D(\ell+1/2-k\chi_2) P_{y_{3D}}^{2h}(k) \nonumber \\
   & = & \int \frac{d\chi_1}{\chi_1} a(\chi_1) \int d\chi_2 a(\chi_2) \int \frac{k^2 dk}{\ell+1/2} \delta_D\left(k-\frac{\ell+1/2}{\chi_1}\right) \delta_D(\ell+1/2-k\chi_2) P_{y_{3D}}^{2h}(k) \nonumber \\
   & = & \int \frac{d\chi_1}{\chi_1^3} a(\chi_1) \int d\chi_2 a(\chi_2) (\ell+1/2) \delta_D\left(\ell+1/2-\frac{\ell+1/2}{\chi_1}\chi_2\right) P_{y_{3D}}^{2h}\left(\frac{\ell+1/2}{\chi_1}\right) \nonumber \\
   & = & \int \frac{d\chi_1}{\chi_1^2} a(\chi_1) \int d\chi_2 a(\chi_2) \delta_D\left( \chi_2-\chi_1 \right) P_{y_{3D}}^{2h}\left(\frac{\ell+1/2}{\chi_1}\right) \nonumber \\
   & = & \int d\chi \left( \frac{a(\chi)}{\chi} \right)^2 P_{y_{3D}}^{2h}\left(\frac{\ell+1/2}{\chi}\right) \nonumber \\ 
  & = & \int d\chi \left( \frac{a(\chi)}{\chi} \right)^2 \left[ \int dM \frac{dn}{dM} b(M) \tilde{y}_{3D}\left(\frac{\ell+1/2}{\chi};M\right) \right]^2 P_{\mathrm{lin}}\left(\frac{\ell+1/2}{\chi}\right) \nonumber \\
  & = & \int dz \frac{c}{H(z)} \frac{a^2}{\chi^2} \left[ \int dM \frac{dn}{dM} b(M) \int dr \, 4 \pi r^2 \frac{\sin((\ell+1/2) r/\chi)}{(\ell+1/2) r/\chi} y_{3D}(r;M) \right]^2 P_{\mathrm{lin}}\left(\frac{\ell+1/2}{\chi}\right) \nonumber \\
  & = & \int dz \frac{d^2V}{dz d\Omega} \frac{a^2}{\chi^4} \left[ \int dM \frac{dn}{dM} b(M) \int dr \, 4 \pi r^2 \frac{\sin((\ell+1/2) r/\chi)}{(\ell+1/2) r/\chi} y_{3D}(r;M) \right]^2 P_{\mathrm{lin}}\left(\frac{\ell+1/2}{\chi}\right) \nonumber \\
  & = & \int dz \frac{d^2V}{dz d\Omega} \left[ \int dM \frac{dn}{dM} b(M) \frac{4 \pi r_s}{\ell_s^2} \int dx \, x^2 \frac{\sin((\ell+1/2) x/\ell_s)}{(\ell+1/2) x/\ell_s} y_{3D}(x;M) \right]^2 P_{\mathrm{lin}}\left(\frac{\ell+1/2}{\chi}\right) \nonumber \\
  & = & \int dz \frac{d^2V}{dz d\Omega} \left[ \int dM \frac{dn(M,z)}{dM} b(k,M,z) \tilde{y}_{\ell}(M,z) \right]^2 P_{\mathrm{lin}}\left(\frac{\ell+1/2}{\chi(z)};z\right) \,,
\label{eq.yCl2hLimber}
\eeqn
where we have restored all of the mass, redshift, and scale dependences in the final expression, and $\tilde{y}_{\ell}(M,z)$ is given by Eq.~(\ref{eq.yelltwid}).  Eq.~(\ref{eq.yCl2hLimber}) precisely matches the result written down for the Limber-approximated two-halo term in~\cite{Komatsu-Kitayama1999}, although again we
have explicitly used $\ell+1/2$ in the Limber approximation (rather than $\ell$), as this choice increases the accuracy of the calculation (and is formally correct).

As noted above, the exact expression for the two-halo term in Eq.~(\ref{eq.yCl2hexact}) requires the evaluation of four nested integrals; the Limber-approximated result in Eq.~(\ref{eq.yCl2hLimber}) requires three.  Thus, the computational expense is not vastly different, although the Limber case is roughly an order of magnitude faster.  For our fiducial cosmology, we find that the Limber result in Eq.~(\ref{eq.yCl2hLimber}) overestimates the exact result in Eq.~(\ref{eq.yCl2hexact}) by $\approx 7$\%, $2$\%, and $1$\% at $\ell=2$, $4$, and $20$, respectively.  By $\ell=30$, the two results are identical within our numerical precision.  Note that although the fractional difference between the exact and flat-sky results at low-$\ell$ is smaller for the two-halo term than for the one-halo term, the two-halo term dominates in this regime, and thus greater precision is required in its computation in order to predict the total $C_{\ell}^y$ precisely.  Note that using the exact result for the two-halo term is more important for \fnl$\neq 0$ cosmologies, for which the Limber approximation has been found to be less accurate~\cite{Pillepichetal2012}.  For a cosmology with \fnl$=100$, we find that the Limber result in Eq.~(\ref{eq.yCl2hLimber}) overestimates the exact result in Eq.~(\ref{eq.yCl2hexact}) by $\approx 18$\%, $5$\%, and $1$\% at $\ell=2$, $4$, and $20$, respectively.  To be conservative, we thus use the exact result for the two-halo term for all calculations at $\ell < 50$, while we use the Limber-approximated result at higher multipoles.  We note that the fairly small size of the correction to the Limber approximation, even at $\ell=2$, can be explained using arguments from~\cite{LoVerde-Afshordi2008} regarding the width of the tSZ projection kernel, which is very broad (see Eq.~(\ref{eq.WtSZ})).  In particular, their results imply that the Limber approximation is reliable when $\ell+1/2 \gtrsim \bar{r}/\sigma_r$, where $\bar{r}$ is the distance at which the projection kernel peaks and $\sigma_r$ is the width of the projection kernel, which are effectively comparable for the tSZ signal.  Thus the Limber approximation is reliable for $\ell+1/2 \gtrsim 1$, which our numerical calculations verify.


\subsection{The Covariance Matrix}
In order to obtain a complete expression for the covariance matrix of the tSZ power spectrum, we need to compute the tSZ angular trispectrum.  Trispectrum configurations are quadrilaterals in $\ell$-space, characterized by four sides and one diagonal.  The configurations that contribute to the power spectrum covariance matrix are of a ``collapsed'' shape characterized by two lines of length $\ell$ and $\ell'$ with zero diagonal~\cite{Komatsu-Seljak2002}.  Analogous derivations to those that led to Eqs.~(\ref{eq.yCl1hexact2}) and (\ref{eq.yCl1hflatsky}) lead to the exact and flat sky-approximated expressions for the one-halo contribution to these configurations of the tSZ trispectrum:
\beqn
T_{\ell\ell'}^{y,1h} & = & \int \frac{dz}{\chi^2(z)} \frac{d^2V}{dz d\Omega} \int dM \frac{dn}{dM} \left| \int k \, dk \, J_{\ell+1/2}(k\chi(z)) \tilde{y}_{3D}(k;M,z) \int \frac{c \, dz'}{H(z')(1+z')\sqrt{\chi(z')}} J_{\ell+1/2}(k\chi(z')) \right|^2 \times \nonumber \\
   & & \left| \int k' \, dk' \, J_{\ell'+1/2}(k'\chi(z)) \tilde{y}_{3D}(k';M,z) \int \frac{c \, dz''}{H(z'')(1+z'')\sqrt{\chi(z'')}} J_{\ell'+1/2}(k'\chi(z'')) \right|^2 \,\, \mathrm{(exact)}
\label{eq.yTl1hexact} \\
T_{\ell\ell' \gg 1}^{y,1h} & \approx & \int dz \frac{d^2V}{dz d\Omega} \int dM \frac{dn}{dM} \left| \tilde{y}_{\ell}(M,z) \right|^2 \left| \tilde{y}_{\ell'}(M,z) \right|^2 \,\, \mathrm{(flat \,\, sky)} \,.
\label{eq.yTl1hflatsky}
\eeqn
For computational efficiency, we choose to implement the flat-sky result at all $\ell$ values in our calculations.  Based on the errors discussed earlier for the flat-sky version of the one-halo contribution to the power spectrum compared to the exact result, we estimate that the error in the trispectrum due to this approximation may be $\sim 25-30$\% at $\ell=2$ (where the discrepancy would be maximal).  However, the only parameter forecast that would likely be affected is the \fnl constraint (due to the necessity of measuring the influence of the scale-dependent bias in order to constrain this parameter), for which we do not find competitive results compared to other probes.  If our forecasts for \fnl were in need of percent-level precision, we would certainly want to use the exact trispectrum; however, this is clearly not the case, and thus we neglect this small error in our results (the constraints on all other parameters are insensitive to moderate changes in the errors at the lowest few $\ell$ values).  Moreover, in the masked cases (which present the greatest promise for cosmological constraints), the trispectrum contribution is heavily suppressed at low-$\ell$ (see Figs.~\ref{fig.covcomps},~\ref{fig.covcompsmasked}, and~\ref{fig.covcompssupermasked}), and the total errors are dominated by the Gaussian term.  Thus, for the masked cases, the exact vs.\ flat-sky correction should be vanishingly small.

Note that we neglect the two-halo, three-halo and four-halo contributions to the trispectrum, as it is dominated even more heavily by the Poisson term than the power spectrum is~\cite{Cooray2001}.  The two-halo term will contribute to some extent at low-$\ell$, but it is unlikely that higher-order terms will be significant even in this regime.  For the masked calculations, the two-halo term may be somewhat important, though likely only at very low-$\ell$, where, as we have argued above, it seems we do not need percent-level accuracy on the errors (since the forecasts for \fnl are not particularly promising, and it is the only parameter very sensitive to this region of the power spectrum).

The full covariance matrix of the tSZ power spectrum, $M^{y}_{\ell\ell'}$ is then given by~\cite{Komatsu-Seljak2002}:
\beqn
M^y_{\ell\ell'} & \equiv & \langle (C^{y,\mathrm{obs}}_{\ell} - C^y_{\ell}) (C^{y,\mathrm{obs}}_{\ell'} - C^y_{\ell'}) \rangle \nonumber \\
   & = & \frac{1}{4 \pi f_{\mathrm{sky}}} \left( \frac{4 \pi (C^y_{\ell} + N_{\ell})^2}{\ell+1/2} \delta_{\ell\ell'} + T^y_{\ell\ell'} \right) \,,
\label{eq.yClcov}
\eeqn
where the angular brackets denote an ensemble average, $f_{\mathrm{sky}}$ is the sky fraction covered by a given experiment (we assume $f_{\mathrm{sky}}=0.7$ throughout this paper), $N_{\ell}$ is the power spectrum due to instrumental noise after multifrequency subtraction (computed for Planck and PIXIE in Section~\ref{sec:exp}), and we approximate $T^y_{\ell\ell'} \approx T^{y,1h}_{\ell\ell' \gg 1}$.  Note that Eq.~(\ref{eq.yClcov}) does not include the so-called ``halo sample variance'' term, as discussed in Section~\ref{sec:covhalomodel}, as this term is negligible for a (nearly) full-sky survey.

We can then compute the covariance matrix $\mathrm{Cov}(p_i,p_j)$ for the cosmological and astrophysical parameters of interest $p_i = \left\{ \Omega_b h^2, \Omega_c h^2, \Omega_{\Lambda}, \sigma_8, n_s, f_{\mathrm{NL}}, M_{\nu}, C_{P_0}, C_{\beta} \right\}$:
\beq
\mathrm{Cov}(p_i,p_j) = \left[ \frac{\partial C^y_{\ell}}{\partial p_i} \left(M^y_{\ell\ell'}\right)^{-1} \frac{\partial C^y_{\ell'}}{\partial p_j} \right]^{-1} \,,
\label{eq.paramCov}
\eeq
where summation over the repeated indices is implied.  The Fisher matrix for these parameters is then simply given by the inverse of their covariance matrix:
\beq
F_{ij} = \mathrm{Cov}^{-1}(p_i,p_j) \,.
\label{eq.paramFisher}
\eeq
The Fisher matrix encodes the constraining power of the tSZ power spectrum on the cosmological and astrophysical parameters.

\section{Comparison with Planck Results}
\label{appendixPlanck}
While this manuscript was under review, the Planck team released its initial set of cosmological results, including the construction of a Compton-$y$ map and estimation of the tSZ power spectrum from this map~\cite{Plancketal2013b}.  In this Appendix, we provide a brief comparison of the publicly released Planck results with the forecasts in our work.  Based on the analysis presented in Fig.~\ref{fig.fNLMnuwerrs}, the Planck tSZ power spectrum should be signal-dominated over roughly the multipole range $100 \lesssim \ell \lesssim 1500$.  Comparing with Fig. 15 in~\cite{Plancketal2013b}, this prediction is in very good agreement.  The bandpowers and associated error bars presented in Table 3 of~\cite{Plancketal2013b} imply a detection of the tSZ power spectrum with SNR $\approx 12.3$, assuming a diagonal covariance matrix (no off-diagonal terms are presented in the Planck results).  Our basic Planck forecast predicts a SNR $\approx 35$.  There are several reasons behind the difference in SNR between our forecast and the initial Planck result:
\begin{itemize}
\item The usable fraction of sky in the Planck analysis ($f_{sky} \approx 0.5$) is found to be somewhat lower than that used in our analysis ($f_{sky} = 0.7$) --- this is primarily due to heavier masking of Galactic dust contamination than we anticipated;
\item The number of frequency channels used in the Planck analysis (six HFI channels only) is lower than that used in our analysis (all nine of the HFI and LFI channels);
\item The Planck analysis explicitly accounts for uncertainties in the contributions of the relevant foreground components (clustered CIB, IR point sources, and radio point sources) to the derived tSZ power spectrum, and finds that these uncertainties dominate the overall errors on the power spectrum; in our analysis, we have used reasonable models for the foregrounds to compute their contributions to the tSZ power spectrum, but have not explicitly propagated through uncertainties in these models to the final error bars.  Our choice on this issue is partly driven by the fact that it is hard to quantify these uncertainties --- in the Planck analysis, simulations are used to provide an estimate of the amplitude of each residual spectrum in the derived Compton-$y$ power spectrum, but a $50$\% uncertainty remains.  This uncertainty dominates the derived errors on the tSZ bandpowers, which is likely the main reason the Planck analysis SNR is significantly lower than our forecast.
\end{itemize}
In addition to these differences, we also note that the Planck analysis does not consider the possibility of masking nearby, massive clusters to reduce the sample variance in the tSZ power spectrum --- however, it appears that the angular trispectrum contribution to the covariance matrix may not have been included in the Planck analysis at all, in which case masking would not be relevant.  Regardless, this is another difference between our forecasts and the Planck results.  Ultimately, the Planck analysis constrains $\sigma_8 \left( \Omega_m / 0.28 \right)^{3.2/8.1} = 0.784 \pm 0.016$.  These constraints are obtained in a $\Lambda$CDM framework with all other cosmological parameters fixed, using the pressure profile of~\cite{Arnaudetal2010} with a hydrostatic mass bias of $20$\%, and without including the Planck constraints from the primordial CMB temperature power spectrum.  In addition, the amplitudes of the foreground contributions to the tSZ power spectrum are allowed to vary, and are included as nuisance parameters.  Given that this framework is rather different from ours, it is difficult to compare directly our forecasted parameter constraints with those obtained in the Planck analysis.  Regardless, it is clear that the tSZ power spectrum is a useful cosmological probe, especially of the low-redshift amplitude of fluctuations, provided that uncertainties related to cluster gas physics and foreground contamination are treated with care.


\end{document}